\documentclass[aps,pra,amssymb,amsmath,eqsecnum,nofootinbib]{revtex4}
\usepackage{graphicx}

\newtheorem{definition}{Definition}[section]
\newtheorem{theorem}{Theorem}[section]
\newtheorem{proposition}{Proposition}[section]

\newtheorem{corollary}{Corollary}[section]
\newtheorem{conjecture}{Conjecture}[section]
\newtheorem{remark}{Remark}[section]

\newenvironment{hypothesis}{HP: \begin{center}} {\end{center}}
\newenvironment{thesis}{TH: \begin{center}} {\end{center}}
\newenvironment{proof}{\begin{center}PROOF: \end{center}} {$ \blacksquare $}
\newtheorem{example}{Example}[section]
\begin{document}
\title{Quantum substitutions of Pisot type, their quantum topological entropy and their use for optimal spacing}
\author{Gavriel Segre}
\date{11-15-2008}
\homepage{http://www.gavrielsegre.com}
\begin{abstract}
Quantum substitutions of Pisot type and their topological entropy
are introduced.

Their utility as algorithms for optimal spacing is analyzed.
\end{abstract}
\maketitle
\newpage
\tableofcontents
\newpage
\section{Acknowledgements}

I would like first of all to thank Marjorie Senechal for precious
teachings concerning Voronoi tilings.

Then I would like to thank Vittorio de Alfaro for his friendship
and his moral support, without which I would have already given
up.

Finally I  would like to thank Andrei Khrennikov and the whole
team at the International Center of Mathematical Modelling in
Physics and Cognitive Sciences of V\"{a}xj\"{o} for their very
generous informatics' support.

Of course nobody among the mentioned people has responsibilities
as to any (eventual) error contained in these pages.
\newpage
\section{Introduction}
The mathematical Theory of Formal Systems, the Theoretical
Physics' Theory of Dynamical Systems and the Theoretical Computer
Science's Theory of Computation may be seen as three different
perspectives from which to investigate a same object:

actually, as it has been strongly remarked by John. L. Casti in
the $ 9^{th} $ chapter "The Mystique of Mechanics: Computation,
Complexity and the Limits to Reason" of \cite{Casti-92}, one can
pass from one perspective to an other through the following
translation's diagram:

\bigskip

\begin{tabular}{|c|c|c|} \hline
  MATHEMATICS & THEORETICAL PHYSICS & THEORETICAL COMPUTER SCIENCE \\
  \hline
  formal system & dynamical system & computational device \\
  axioms & initial conditions & input \\
  logical inference & dynamical evolution & computation \\
  theorems & final state  & output \\ \hline
\end{tabular}

\bigskip

that is particularly fruitful as soon as one looks at the
phenomenon of Incompleteness.

G\"{o}del's Theorem \cite{Godel-65} stating the existence of
propositions undecidable (id est that cannot neither be proved nor
disproved) in suitable formal systems is seen with indifference,
if not with suspicion, by Mathematicians not working on
foundational issues, the involved undecidable propositions being
seen as self-referential intellectual masturbations (such as
G\"{odel}'s sentence being a numerical translation, realized
through G\"{o}del numbering, of a statement asserting it own non
provability).

Even more, the same occurs as to Theoretical Physicists whose
formalisms so often prefer aesthetical elegance to the
mathematical rigor, the translation of these formalisms in
consistent mathematical terms being seen as a not particularly
interesting bureaucratic  task left to mathematical physicists.

Contrary, the phenomenon of incompleteness pervades different
fields of Mathematics and might have astonishing consequences also
in Theoretical Physics:
\begin{itemize}
    \item as it has been suggested by Geroch and Hartle \cite{Geroch-Hartle-86}, Markov's Theorem stating the undecidability of the Homeomorphism
    Problem for four-manifolds (see for instance \cite{Collins-Zieschang-98}) might compromise our
    possibility of quantizing gravity;
    \item as it has been suggested by Agnes and Rasetti \cite{Agnes-Rasetti-91}, the  theorems by Boone and Novikov (and their
    consequences such as the theorem stating the recursive undecidability of Dehn's isomorphism problem) stating the  unsolvability of the
    Word Problem of suitable finitely presented groups might
    affect the predictability in suitable classical dynamical
    systems;
    \item Berger's Theorem (see \cite{Grunbaum-Shephard-87}) stating the undecidability of
    the Tiling Problem might have consequences in the theory of
    quasicrystals \cite{Senechal-95a};
    \item Matyasevich's proof of the undecidability of Hilbert
    Tenth Problem \cite{Matiyasevich-93}, \cite{Matiyasevich-99} is a Damocles' sword above the head as to all
    the diophantine equations occurring in Physics.
\end{itemize}

From the other side Thoretical Computer Science has affected
Mathematics, allowing to look at the phenomenon of Undecidability
from a different perspective:

from the viewpoint of Gregory Chaitin's two Information-Theoretic
Incompleteness Theorems \cite{Chaitin-87}, \cite{Calude-02} the
existence of limitations in the predictive power of formal systems
appears quite reasonable and its role in Physics appears  natural
\cite{Svozil-93}, in particular as far as Chaos Theory is
concerned, as we will show in a specifically dedicated section of
this paper.

It would seem rather natural to think that all this classical
stuff could be generalized to the quantum case.

Unfortunately, the problem  of extending  Algorithmic Information
Theory to the quantum domain is a not yet settled issue
\cite{Svozil-95}, \cite{Svozil-96}, \cite{Manin-99},
\cite{Vitanyi-99}, \cite{Gacs-01}, \cite{Vitanyi-01},
\cite{Berthiaume-van-Dam-Laplante-01}.

In particular the problem, first proposed by Karl Svozil, that
almost a decade ago raised our interest in the subject during the
Phd-studies \cite{Segre-02}, id est that of formulating and
proving quantum analogues of Chaitin's Information-theoretic
Incompleteness Theorems, is still open.

Since there exists also a combinatorial approach to Information
Theory coarser than the algorithmic approach but incomparably
easier, it appears natural, as a first step in such a direction,
to introduce and study the quantum combinatorial information of
the quantum analogues of the simplest formal systems, id est those
whose inference rules are simple substitution rules on the
involved symbols \cite{Fogg-02}.

In this way we have  been naturally led to the first result of
this paper, namely the definition of the quantum topological
entropy for quantum substitutions.

Among all the substitutions, furthermore, there exists a
particular class, the substitutions of Pisot type, whose very
useful properties leading to their role in the Theory of
quasicrystals and of nonperiodic tilings
\cite{Bombieri-Taylor-86}, \cite{Bombieri-Taylor-87},
\cite{Hof-95}, \cite{Hof-97}  are naturally derived from the
mathematical properties of a remarkable set of numbers, the
Pisot-Vijayaraghavan numbers
\cite{Bertin-Decomps-Guilloux-Grandet-Hugot-Pathiaux-Delefosse-Schreiber-92},
underlying much of the idolatrous mystique of the \emph{golden
number} $ \tau$ (for this reason we have emphasized how the same
properties are shared with the poor relative of $ \tau $, the so
called \emph{plastic number} $ \rho $.)

This has led us to the second result of this paper, id est the
formulation of a quantum algorithm for optimal spacing.

\newpage
\section{Topological entropy of substitutions} \label{sec:Topological entropy of substitutions}
Given a finite alphabet $ A = \{ a_{1} , \cdots a_{|A|} \} $ (id
est a set A such that the cardinality  $ | A | $ of A  is such
that $ 2 \leq | A| < \aleph_{0} $) let us denote, following
\cite{Calude-02}, \cite{Fogg-02}, \cite{Allouche-Shallit-03} and
\cite{Lothaire-05}, by $ A^{+} := \cup_{n \in \mathbb{N}_{+}} A
^{n}$ the free semi-group generated by A, id est the set of all
the finite words over A.

Given $ \vec{x} , \vec{y} \in A^{+} $ let us denote by $ \vec{x}
\cdot \vec{y} $ the concatenation of $ \vec{x} $ and $ \vec{y} $,
id est the string $ ( x_{1} , \cdots , x_{|\vec{x} |}, y_{1} ,
\cdots ,  y_{| \vec{y} |} ) $ where $ | \vec{x} | $ denotes the
length of the strings $ \vec{x} $.

Introduced the set $ A^{\mathbb{N}_{+}}$ of the \emph{sequences
over A} let us endow A with the discrete topology, and let us
endow $ A^{+} $  and $ A^{\mathbb{N}_{+}}$  with the induced
product topology.

Such a topology over  $ A^{\mathbb{N}_{+}}$ is the metric topology induced by the following distance:
\begin{equation} \label{eq:distance between sequences}
    d ( \bar{x} , \bar{y} ) \; := \; \left\{%
\begin{array}{ll}
    0, & \hbox{if $ \bar{x} = \bar{y} $;} \\
    \frac{1}{2^{ \min \{ n \in \mathbb{N}_{+} \, : \, x_{n} \neq y_{n} \} }  }, & \hbox{otherwise.} \\
\end{array}%
\right.
\end{equation}

In the following we will denote sequences through a bar, id est we
will  adopt the following notation:
\begin{equation}
    \bar{x} \; := \; \{ x_{n} \}_{n \in \mathbb{N}_{+}} \in A^{\mathbb{N}_{+}}
\end{equation}
Given a word $ \vec{w} = ( w_{1} , \cdots , w_{r} ) \in A^{+} $
and a sequence $ \bar{x} \in A^{\mathbb{N}_{+}} $:

\begin{definition}
\end{definition}
\emph{ $ \vec{w} $ occurs in  $ \bar{x} $ at the position $ n \in
\mathbb{N}_{+}$:}
\begin{equation}
    x_{n} =  w_{1} \; \wedge \; \cdots  \; \wedge \;  x_{n+r-1} =  w_{r}
\end{equation}
\begin{definition}
\end{definition}
\emph{$\vec{w}$ is a factor of $ \bar{x} $:}
\begin{equation}
  \vec{w} \leq_{occ} \bar{x} \; := \; \exists n \in \mathbb{N}_{+}
   \: : \: \text{$ \vec{w} $ occurs in  $ \bar{x} $ at the position $ n \in
\mathbb{N}_{+}$}
\end{equation}

In an analogous way, given two words $  \vec{x}, \vec{y} \in A^{+}
$:
\begin{definition}
\end{definition}
\emph{$ \vec{x}$ is a factor of $ \vec{y} $}:
\begin{equation}
 \vec{x} \leq_{occ} \vec{y} \; := \;  \exists n \in \mathbb{N}_{+}
 \, : \, y_{n} = x_{1} \wedge \cdots \wedge y_{n+|\vec{x}| } = x_{|\vec{x}| }
\end{equation}

Given two words $ \vec{x} , \vec{y} \in A^{+} $ let us denote by $
| \vec{x} |_{\vec{y}} $ the number of occurences of $ \vec{y} $ in
$ \vec{x} $.

In a similar manner, given a sequence $ \bar{x} \in
A^{\mathbb{N}_{+}} $ and a word $ \vec{y}  \in A^{+} $, let us
denote by $ | \bar{x} |_{\vec{y}} $ the number (eventually
infinite) of occurences of $ \vec{y} $ in $ \bar{x} $.

\begin{definition} \label{def:language of length n of a string}
\end{definition}
\emph{language of length n of $ \bar{x}$:}
\begin{equation}
    \mathcal{L}_{n} ( \bar{x} ) \; := \; \{ \vec{y} \in A^{n} \, :
     \vec{y} \text{ is a factor of } \bar{x} \}
\end{equation}

\begin{definition}
\end{definition}
\emph{language of $ \bar{x}$:}
\begin{equation}
    \mathcal{L} ( \bar{x} ) \; := \; \cup_{n \in \mathbb{N}_{+}}  \mathcal{L}_{n} ( \bar{x} )
\end{equation}

Given a factor $ \vec{y} \in A^{+}  $ of the sequence $ \bar{x}
\in A^{\mathbb{N}_{+}} $:
\begin{definition}
\end{definition}
\emph{frequency of $ \vec{y} $ in $ \bar{x}$:}
\begin{equation}
    f_{\vec{y}} (  \bar{x} ) \; := \; \lim_{n \rightarrow + \infty} \frac{ | \vec{x}(n)|_{\vec{y}}}{n}
\end{equation}

Let us introduce some remarkable classes of sequences:
\begin{definition} \label{def:periodic sequences}
\end{definition}
\emph{set of the periodic sequences:}
\begin{equation}
    PERIODIC( A^{\mathbb{N}_{+}} ) \; := \; \{  \bar{x} \in A^{\mathbb{N}_{+}} : ( \exists T \in \mathbb{N}_{+} : ( x_{n+T} = x_{n} \; \; \forall n \in \mathbb{N}_{+} ) ) \}
\end{equation}
\begin{definition}
\end{definition}
\emph{set of the ultimately periodic sequences:}
\begin{equation}
    PERIODIC_{ult}( A^{\mathbb{N}_{+}} ) \; := \; \{  \bar{x} \in A^{\mathbb{N}_{+}} : ( \exists T  \in \mathbb{N}_{+} , \exists N \in \mathbb{N}_{+} : ( x_{n+T} = x_{n} \; \; \forall n \in \mathbb{N}_{+} : n \geq N ) )  \}
\end{equation}

\begin{definition}
\end{definition}
\emph{set of the Borel-normal sequences over A:}
\begin{equation}
  NORMAL ( A^{\mathbb{N}_{+}} ) \; := \; \{ \bar{x} \in A^{\mathbb{N}_{+}} \: : \: f_{\vec{y}} (  \bar{x} ) = \frac{1}{|A|^{|\vec{y}|}} \; \; \forall \vec{y} \in A^{+} \}
\end{equation}

Let us now introduce the following:
\begin{definition} \label{def:combinatorial information function}
\end{definition}
\emph{combinatorial information function of $ \bar{x} $: }
\begin{equation}
  p_{n} (  \bar{x} )  \; = \; |  \mathcal{L}_{n} ( \bar{x} )|
\end{equation}

Obviously:
\begin{equation}
 1  \; \leq  \;  p_{n}  ( \bar{x} ) \;  \leq |A|^{n} \; \;
 \forall \bar{x} \in A^{\mathbb{N}_{+}} , \forall n \in \mathbb{N}_{+}
\end{equation}

\begin{proposition}
\end{proposition}
\emph{Properties of the periodic and ultimately periodic sequences:}
\begin{enumerate}
    \item
\begin{equation}
     PERIODIC( A^{\mathbb{N}_{+}} ) \; \subset \;  PERIODIC_{ult}( A^{\mathbb{N}_{+}} )
\end{equation}
    \item
\begin{equation}
    \exists n \in \mathbb{N}_{+} : p_{n} ( \bar{x} ) \leq n \; \Rightarrow \; \bar{x} \in PERIODIC_{ult}( A^{\mathbb{N}_{+}} )
\end{equation}
\end{enumerate}

\begin{definition} \label{def:topological entropy of a sequence}
\end{definition}
\emph{topological entropy of $  \bar{x} $:}
\begin{equation}
    H_{top} ( \bar{x} ) \; := \; \lim_{n \rightarrow + \infty} \frac{ \log_{|A|}(  p_{n} (  \bar{x}) )   }{n}
\end{equation}
where the existence of the limit follows from the subadditivity of
the function $ n \mapsto \log_{|A|}(  p_{n} (  \bar{x}) ) $.

\bigskip

\begin{remark}
\end{remark}
Clearly the topological entropy of a sequence is different from
zero only if $ p_{n} (  \bar{x}) $ grows exponentially with n.

For instance introduced the following:
\begin{definition} \label{def:Sturmian sequences}
\end{definition}
\emph{Sturmian sequences:}
\begin{equation}
    STURMIAN ( A^{\mathbb{N}_{+}} ) \; := \; \{ \bar{x} \in
 A^{\mathbb{N}_{+}} \, : \, p_{n} ( \bar{x} ) = n +1 \; \; \forall
    n \in \mathbb{N}_{+} \}
\end{equation}
it may be easily proved that:
\begin{proposition}
\end{proposition}
\begin{equation}
    H_{top} ( \bar{x} ) \; = \; 0 \; \; \forall \bar{x} \in  STURMIAN ( A^{\mathbb{N}_{+}} )
\end{equation}
\begin{proof}
Applying the definition \ref{def:topological entropy of a
sequence} and the definition \ref{def:Sturmian sequences} it
follows that:
\begin{equation}
    H_{top} ( \bar{x} ) \; = \; \lim_{n \rightarrow + \infty} \frac{ \log_{|A|} ( n+1)
    }{n} \; = \; 0 \; \; \forall \bar{x} \in  STURMIAN ( A^{\mathbb{N}_{+}} )
\end{equation}
\end{proof}

\bigskip

\begin{remark} \label{rem:positive and negative feature of the combinatorial approach to information}
\end{remark}
Let us recall how Andrei Nikolaevic Kolmogorov, in his seminal paper on Algorithmic Information Theory \cite{Kolmogorov-65},
showed with his usual intellectual clearness that there exist three possible approaches to Information Theory:
\begin{itemize}
    \item the \emph{combinatorial approach} in which the information of an object is defined relatively to
    a context of objects distributed uniformly
    \item the \emph{probabilistic approach} in which the information of an object is defined relatively to
    a context whose objects are weighted according to a non necessarily uniform probability distribution
    \item  the \emph{algorithmic approach} furnishing an absolute measure of the information of an object, id est not relative to  any context of different objects
    of which it is thought to be a member
\end{itemize}

As we will now show quantitatively the topological entropy
furnishes a measure of the information contained in a sequence
coarser than the one given by Algorithmic Information Theory.

From the other side it has the positive feature of being  purely
of combinatorial nature, id est not to require the introduction of
concepts from Computability Theory.

\bigskip

Let us then introduce the following useful partial ordering over $
A^{+} $:
\begin{definition}
\end{definition}
\emph{$ \vec{x} $ is a prefix of $ \vec{y} $:}
\begin{equation}
  \vec{x} <_{p}  \vec{y} \; := \; \exists \vec{z} \in A^{+} \;
  : \; \vec{y} = \vec{x} \cdot \vec{z}
\end{equation}

Given a set $ S \subset A^{+} $:
\begin{definition}
\end{definition}
\emph{S is prefix-free:}
\begin{equation}
 \vec{x} \nless_{p} \vec{y} \; \; \forall \vec{x} , \vec{y} \in S
\end{equation}

Let us now briefly recall some necessary rudiments of
Computability Theory \cite{Cutland-80}, \cite{Rogers-87},
\cite{Odifreddi-89}, \cite{Odifreddi-99a}.

Given a partial function $ C : A^{+} \stackrel{\circ}{\mapsto}
A^{+} $ we can look at it as a total function  adding a
"non-halting" symbol" (that following \cite{Calude-02} we will
assume to be the infinity symbol $ \infty $) and posing $ C (
\vec{x} ) := \infty $ whether C doesn't halt on $ \vec{x} $.

We can then introduce the following:
\begin{definition}
\end{definition}
\emph{halting set of C:}
\begin{equation}
    HALT_{C} \; := \; \{ \vec{x} \in A^{+} : C(\vec{x} ) \neq
    \infty \}
\end{equation}

We will say that:
\begin{definition}
\end{definition}
\emph{C is a partial computable function:}
\begin{equation}
 \exists f :  \mathbb{N} \stackrel{\circ}{\mapsto}  \mathbb{N} \;
 \text{ partial recursive } : C(\vec{x}) =
 string(f(string^{-1}(\vec{x}))
\end{equation}
where $ string : \mathbb{N} \mapsto A^{+} $ is the map associating
to an integer n the $ n^{th} $ word of $ A^{+} $ in lexicographic
order and  where we demand to the mentioned literature as to the
definition of a partial recursive function on natural numbers.

Given a set $ X \subset A^{+} $:
\begin{definition}
\end{definition}
\emph{X is computable:}
\begin{center}
 the characteristic function $ \chi_{X} $ of X is computable
\end{center}
A notion weaker than computability is the following:
\begin{definition}
\end{definition}
\emph{X is computably enumerable:}
\begin{equation}
    \exists C : A^{+}  \stackrel{\circ}{\mapsto} A^{+} \text{ partial computable
    function} \; : \; X = HALT_{C}
\end{equation}

\begin{definition}
\end{definition}
\emph{Chaitin computer:}

\begin{center}
a partial recursive function $ C : A^{+} \stackrel{\circ}{\mapsto}
A^{+} $ such that $ HALT_{C} $  is prefix free.
\end{center}

\begin{definition}
\end{definition}
\emph{Universal Chaitin computer:}

a Chaitin computer U such that for every Chaitin computer C there
exist a constant $ c_{U,C} \in \mathbb{R}_{+} $ such that:
\begin{equation}
 \forall \vec{x} \in HALT_{C} , \exists \vec{y} \in  A^{+} \; :
 \; U ( \vec{y} ) = C ( \vec{x} ) \, \wedge \, | \vec{y} | \leq |
 \vec{x}| + c_{U,C}
\end{equation}

Given a universal Chaitin computer U and a word $ \vec{x} \in
A^{+} $ :
\begin{definition} \label{def:algorithmic information}
\end{definition}
\emph{algorithmic information of $ \vec{x} $ with respect to U:}
\begin{equation}
    I_{U} ( \vec{x} ) \; := \; \left\{%
\begin{array}{ll}
    \min \{ | \vec{y} | :  \vec{y} \in A^{+} , U( \vec{y} ) = \vec{x} \}, & \hbox{if $ \{ \vec{y} \in A^{+} : U( \vec{y} ) = \vec{x} \} \neq \emptyset $;} \\
    + \infty , & \hbox{otherwise.} \\
\end{array}%
\right.
\end{equation}

Given a universal Chaitin computer U and  $ \vec{x} \in A^{+} $:
\begin{definition} \label{def:algorithmic probability}
\end{definition}
\emph{algorithmic probability of $ \vec{x} $ with respect to U:}
\begin{equation}
    P_{U} ( \vec{x} ) \; := \; \sum_{ \vec{y} \in U^{-1} ( \vec{x} ) }  \frac{1}{ |A|^{| \vec{y} |}}
\end{equation}

\smallskip

\begin{definition} \label{def:halting probability}
\end{definition}
\emph{Chaitin's halting probability with respect to U:}
\begin{equation}
    \Omega_{U} \; := \; \sum_{ \vec{x} \in A^{+} } P_{U} ( \vec{x} )
\end{equation}

In order to show the conceptual importance of $ \Omega_{U} $ let
us recall first of all the following Alan Turing's celebrated
result \cite{Turing-65}:
\begin{theorem} \label{th:undecidability of the halting problem}
\end{theorem}
\emph{Undecidability of the Halting Problem:}
\begin{center}
  $HALT_{U}$ is not computable
\end{center}

Chaitin's halting probability codifies the undecidability stated
by the theorem \ref{th:undecidability of the halting problem} in a
very useful form owing to the following:

\begin{theorem} \label{th:Chaitin's Theorem on the Halting Problem}
\end{theorem}
\emph{Chaitin's Theorem on the Halting Problem:}
\begin{center}
 Given the first n cbits of $ r_{\{0,1 \}}( \Omega_{U} )$ one can
 decide whether $ U ( \vec{x} ) < + \infty $ for every $ \vec{x} \in \cup_{k=1}^{n} \{ 0 , 1 \}^{k} $
\end{center}
where $ r_{\{0,1 \}}$ is the nonterminating symbolic
representation with respect to the alphabet $ \{ 0 , 1 \} $  of
the definition \ref{def:nonterminating symbolic representation}.

\bigskip

Let us now introduce the concept of \emph{algorithmic randomness}
(introduced in independent but equivalent ways by Martin-L\"{o}f,
by Solovay and by Chaitin) by defining the algorihmically random
sequences as those sequences with maximal algorithmic information
(and that hence cannot be algorithmically compressed):

\begin{definition} \label{def:algorithmic randomness}
\end{definition}
\emph{algorithmically random sequences on A:}
\begin{equation}
    RANDOM ( A^{\mathbb{N}_{+}} ) \; := \; \{ \bar{x} \in
 A^{\mathbb{N}_{+}} \, : \, \forall U \, \text{universal Chaitin
    computer} \, \exists c_{U} \in ( 0 , + \infty ) \; :
    \;  I ( \vec{x}(n) ) \geq n - c_{U} \, \, \forall n \in
    \mathbb{N}_{+} \}
\end{equation}
where $ \vec{x}(n) $ is the prefix of length n of the sequence $
\bar{x}$.

Such a notion doesn't depend upon the chosen alphabet in the
following sense:
\begin{theorem}  \label{th:Independence of algorithmic randomness from the adopted
alphabet}
\end{theorem}
\emph{Independence of algorithmic randomness from the adopted
alphabet:}
\begin{equation}
    RANDOM ( A_{2}^{\mathbb{N}_{+}} ) \; = \; T_{A_{1},A_{2}} [ RANDOM (  A_{1}^{\mathbb{N}_{+}}
    ) ] \; \; \forall A_{1}, A_{2} : \max \{ | A_{1} | ,  | A_{2}
    | \} \in \mathbb{N}
\end{equation}
where $  T_{A_{1},A_{2}} $ is the alphabet's transition map of the
definition \ref{def:alphabet's transition map}.

\begin{definition} \label{def:random real in the unit interval}
\end{definition}
\emph{algorithmically random elements of the interval $ ( 0 ,1
)$:}
\begin{equation}
 RANDOM[(0,1)] \; := \; v_{A} ( RANDOM ( A^{\mathbb{N}_{+}} ))  \;
 , \; 2 \leq | A | < \aleph_{0}
\end{equation}
where $ v_{A} $ is the numerical value map of the definition
\ref{def:numerical value of a sequence}.

\bigskip

\begin{remark}
\end{remark}
Let us remark that, owing to the theorem \ref{th:Independence of
algorithmic randomness from the adopted alphabet}, the definition
\ref{def:random real in the unit interval} doesn't depend on the
adopted alphabet A.

\bigskip

The paradigmatic example of a random real is given by Chaitin's
halting probabilities:
\begin{proposition}
\end{proposition}
\emph{Halting probabilities are random:}
\begin{equation}
    \Omega_{U} \in RANDOM [ (0 , 1) ] \; \; \forall U
    \text{ universal Chaitin computer}
\end{equation}

\bigskip

Let us now observe that clearly:
\begin{proposition}
\end{proposition}
\emph{Periodic and ultimately periodic sequences are not random:}
\begin{equation}
    PERIODIC_{ult}( \mathbb{A}^{\mathbb{N}_{+}} ) \cap RANDOM ( \mathbb{A}^{\mathbb{N}_{+}}
    ) \; = \; \emptyset
\end{equation}
\begin{proof}
Let $ \bar{x} $ be a sequence periodic of period $ T \in
\mathbb{N}_{+} $ after a prefix of  $ N \in \mathbb{N} $ digits.

By the subadditivity of algorithmic information:
\begin{equation}
    I ( \vec{x}(n T) ) \; \leq \; I (  \vec{x}(N) ) + I ( \{ x_{N+1} , \cdots , x_{T}
    \} ) + O(1)  \; \; \forall n \in \mathbb{N}_{+}
\end{equation}
and is hence bounded.

\end{proof}

Let us now introduce the following:
\begin{definition}
\end{definition}
\emph{recurrent sequences on A:}
\begin{equation}
    RECURRENT( A^{\mathbb{N}_{+}} ) \; = \; \{ \bar{x} \in
    \mathbb{A}^{\mathbb{N}_{+}} :  | \bar{x} |_{\vec{y}} \; = \; + \infty \; \; \forall \vec{y} \in
    A^{+} \}
\end{equation}

\begin{definition}
\end{definition}
\emph{uniformly recurrent sequences on A:}
\begin{equation}
     RECURRENT_{un}( A^{\mathbb{N}_{+}} ) \; := \; \{ \bar{x} \in
      RECURRENT( A^{\mathbb{N}_{+}} )  \; : \; \forall \vec{y} \in A^{+}
      , \exists r \in \mathbb{N}_{+} \:  \vec{y} \leq_{occ} \{
      x_{n}, \cdots , x_{n+r} \} \, \forall n \in \mathbb{N}_{+}  \}
\end{equation}

Demanding to \cite{Calude-02} for the proofs we can now state some
remarkable property of the algorithmically random sequences:
\begin{theorem} \label{th:Calude-Chitescu}
\end{theorem}
\emph{Calude-Chitescu Theorem:}
\begin{equation}
    RANDOM ( A^{\mathbb{N}_{+}} ) \; \subset \;  RECURRENT( A^{\mathbb{N}_{+}} )
\end{equation}

\bigskip

\begin{remark} \label{rem:link to Cantor sets}
\end{remark}
The Calude-Chitescu Theorem shows how theorem \ref{th:Independence
of algorithmic randomness from the adopted alphabet} has not be
equivocated in the following sense.

Let us consider two finite alphabets $ A_{1} $ and $ A_{2} $ such
that $ A_{1} \subset A_{2} $; then clearly:
\begin{equation}
    | \bar{x}|_{a} \; = \; 0 \; \; \forall \bar{x} \in RANDOM (
    A_{1}^{\mathbb{N}_{+}} ) , \forall a \in A_{2} - A_{1}
\end{equation}
and hence, according to the theorem \ref{th:Calude-Chitescu}:
\begin{equation}
    \bar{x} \notin RANDOM(  A_{2}^{\mathbb{N}_{+}} ) \; \; \forall
    \bar{x} \in  RANDOM(  A_{1}^{\mathbb{N}_{+}} )
\end{equation}
as it can be appreciated considering the alphabet $ A_{1} := \{ 0
,  2 \} $ and $ A_{2} := \{ 0 , 1 , 2 \} $ involved in the
construction of Cantor's middle third set extensively analyzed in
the section \ref{sec:A brief information theoretic analysis of
singular Lebesgue-Stieltjes measures supported on Cantor sets and
almost periodicity}.

\bigskip

\begin{remark} \label{rem:link to almost periodicity}
\end{remark}
Can the Calude-Chitescu Theorem (id est the theorem
\ref{th:Calude-Chitescu}) be strengthened by substituting the set
$ RECURRENT(  A^{\mathbb{N}+} ) $ with the set $ RECURRENT_{un}(
A^{\mathbb{N}+} ) $ ?

Up to our knowledge such a question is still open.

Furthermore, since uniformly recurrent sequences are deeply linked
with the notion of almost periodicity, the issue is deeply linked
with the problem of estimating the algorithmic information's
content of an almost periodic function, a problem discussed in the
section \ref{sec:A brief information theoretic analysis of
singular Lebesgue-Stieltjes measures supported on Cantor sets and
almost periodicity}.

\begin{theorem} \label{th:Borel normality versus algorithmic randomness}
\end{theorem}
\begin{equation}
    NORMAL ( A^{\mathbb{N}_{+}} ) \; \subset \; RANDOM ( A^{\mathbb{N}_{+}} )
\end{equation}

\begin{remark}
\end{remark}
The strict inclusion of theorem \ref{th:Borel normality versus algorithmic randomness} implies that though an algorithmically random sequence is always Borel normal, there exist
Borel normal sequences that are not algorithmically random.

The classical counterexample is the following:
\begin{definition} \label{def:Champerknowne sequence}
\end{definition}
\emph{Champerknowne sequence:}
\begin{equation}
    \bar{x}_{Champerknowne} \; := \; \cdot_{n \in \mathbb{N_{+}}} string(n)
\end{equation}
where $ string(n) $ is the $ n^{th} $ string in lexicographical ordering.

By construction:
\begin{equation}
    \bar{x}_{Champerknowne} \in  NORMAL ( A^{\mathbb{N}_{+}} )
\end{equation}
Anyway the Champerknowne sequence is obviously highly algorithmically compressible, being defined through
a very short algorithm:
\begin{equation}
   \bar{x}_{Champerknowne} \; \notin \; RANDOM ( A^{\mathbb{N}_{+}} )
\end{equation}

\bigskip

 Let us remark  that:
\begin{proposition} \label{prop:link between topological entropy and algorithmic
randomness}
\end{proposition}
\emph{Link between topological entropy and algorithmic
randomness:}
\begin{enumerate}
    \item
\begin{equation}
    H_{top} ( \bar{x} ) \, = \, 1 \; \; \forall \bar{x} \in RANDOM ( A^{\mathbb{N}_{+}} )
\end{equation}
    \item
\begin{equation}
  H_{top} ( \bar{x} ) \, = \, 1  \; \nRightarrow \; \bar{x} \in RANDOM ( A^{\mathbb{N}_{+}} )
\end{equation}
\end{enumerate}
\begin{proof}
\begin{enumerate}
    \item
The theorem \ref{th:Calude-Chitescu} implies that:
\begin{equation}
    p_{n} ( \bar{x} ) \; = \; |A|^{n} \; \; \forall \bar{x} \in RANDOM ( A^{\mathbb{N}_{+}}
    ) , \forall n \in \mathbb{N}_{+}
\end{equation}
and hence:
\begin{equation}
     H_{top} ( \bar{x} ) \; = \; \lim_{n \rightarrow \infty}
     \frac{n}{n} \; = \; 1 \; \; \forall \bar{x} \in RANDOM ( A^{\mathbb{N}_{+}} )
\end{equation}
    \item the fact that the Champernowne sequence of the definition \ref{def:Champerknowne sequence} is Borel normal implies that:
\begin{equation}
    p_{n} (  \bar{x}_{Champerknowne} ) \; = \; |
    A |^{n} \; \; \forall n \in \mathbb{N}_{+}
\end{equation}
and hence:
\begin{equation}
     H_{top} (  \bar{x}_{Champerknowne}) \; = \; \lim_{n \rightarrow \infty}
     \frac{n}{n} \; = \; 1
\end{equation}
though, as we have already remarked, $ \bar{x} $ is not algorithmically random.
\end{enumerate}
\end{proof}

\bigskip

Let us now suppose to have a probability distribution f over the
alphabet A, id est a map $ f : A \mapsto [ 0,1] $ such that:
\begin{equation}
    \sum_{a \in A} f(a) \; = \; 1
\end{equation}
and let us introduce the following:

\begin{definition}
\end{definition}
\emph{probabilistic information of f:}
\begin{equation}
    S_{prob} ( f ) \; := \; < - \log_{|A|} f > \; = \; - \sum_{a \in A} f(a) \log_{|A|} f(a)
\end{equation}

As it is natural for different approaches devoted to formalize
from different perspectives the same concept, the algorithmic
information and the probabilistic information are deeply linked.

Let us consider, at this purpose, a stochastic process in which,
at the $ n^{th}$ temporal step, a letter $ x_{n} $  from the
alphabet A is chosen at random according to the distribution f.

Supposed to have chosen n letters, the random variables $ x_{1} ,
\cdots , x_{n} $ are independent and identical distributed so that
the  random vector $ ( x_{1} , \cdots x_{n} ) $ has probability
distribution:
\begin{equation}
    f^{(n)}( x_{1} , \cdots ,  x_{n} ) \; := \; \prod_{i=1}^{n} f( x_{i} )
\end{equation}

Then:
\begin{theorem} \label{th:link between the probabilistic information and the algorithmic information}
\end{theorem}
\emph{Link between the probabilistic information and the algorithmic information:}
\begin{equation}
    \lim_{n \rightarrow + \infty} \frac{< I( \vec{x} (n) >}{n} \; = \; S(f)
\end{equation}
where of course:
\begin{equation}
    < I( \vec{x} (n) ) > \; = \sum_{\vec{x} \in A^{n}} f^{(n)}( \vec{x} ) I( \vec{x} )
\end{equation}

\begin{remark}  \label{rem:link between the probabilistic information and the algorithmic information}
\end{remark}

A first generalization of the theorem \ref{th:link between the
probabilistic information and the algorithmic information} may be
obtained removing the hypothesis that different letters $ x_{1} ,
\cdots, x_{n} $ are identically distributed and independent, id
est considering  the case in which they form an arbitrary
stochastic process  (an arbitrary shift in the language of
Abstract Dynamical System Theory \cite{Kornfeld-Sinai-Vershik-00},
\cite{Sinai-94}).

It can be further generalized to arbitrary abstract dynamical systems through a suitable symbolic codification of its trajectories.

The resulting Brudno Theorem \cite{Brudno-78}, \cite{Brudno-83},
\cite{Alekseev-Yakobson-1981}, \cite{Segre04b} states that the
dynamical entropy of an abstract dynamical system , defined as the
asymptotic rate of production of probabilistic-information
produced by its dynamics, is equal to the asymptotic rate of
algorithmic information of almost every its orbit.

\bigskip

Let us now introduce the following basic:
\begin{definition}
\end{definition}
\emph{substitution over A:}
\begin{center}
  a map $ \sigma : A \mapsto A^{+}  $
\end{center}

A substitution $ \sigma $ over A may be extended to a morphism of
$  A^{+} $ by concatenation, id est posing:
\begin{equation}
    \sigma [ ( x_{1} , \cdots , x_{n} ) ] \; := \; \cdot_{i=1}^{n}
    \sigma( x_{i} )
\end{equation}

In the same way it may be naturally extended to a map over $
 A^{\mathbb{N}_{+}} $  by posing:
\begin{equation}
    \sigma ( \{ x_{n} \}_{n \in \mathbb{N}} ) \; := \; \cdot_{n \in
    \mathbb{N}}  \sigma (x_{n})
\end{equation}

\begin{remark}
\end{remark}
Substitutions are very efficient ways for producing sequences
owing to the following:

\begin{proposition} \label{prop:fixed points aof a substitution beginning with a given letter}
\end{proposition}
\emph{Fixed point of a substitution beginning with a given
letter:}

\begin{hypothesis}
\end{hypothesis}
\begin{equation*}
      a \in A
\end{equation*}
\begin{center}
   $ \sigma $ substitution  such that $ \sigma (a) $ begins with a
    and $ |\sigma ( a) | \geq 2 $
\end{center}
\begin{thesis}
\end{thesis}
\begin{equation*}
    \exists ! \bar{\sigma}(a) \in A^{\mathbb{N}_{+}} \; : \; \sigma (
    \bar{\sigma}(a) ) \; = \; \bar{\sigma}(a)
\end{equation*}

Given a substitution $ \sigma $ it is natural to define its topological entropy as the sum of the topological entropy of its fixed point starting with different letters:
\begin{definition} \label{def:topological entropy of a substitution}
\end{definition}
\emph{topological entropy of $\sigma $:}
\begin{equation}
    H_{top} ( \sigma ) \; := \; \sum_{a \in A : | \sigma (a) | \geq 2}  H_{top} ( \bar{\sigma}(a))
\end{equation}

Let us now briefly recall some basic notions concerning Matrices' Theory.

Given $ n \in \mathbb{N}_{+} : n \geq 2 $ and a square matrix $ A \in M_{n} ( \mathbb{Z} ) $:
\begin{definition}
\end{definition}
\emph{A is primitive:}
\begin{equation}
    ( A_{ij} \in \mathbb{N} \: \forall i,j \in \{ 1 , \cdots , n \} ) \; \wedge \; ( \exists n \in \mathbb{N}_{+} : ( A^{n})_{ij} \in \mathbb{N}_{+}  \: \forall i,j \in \{ 1 , \cdots , n \} )
\end{equation}
\begin{definition}
\end{definition}
\emph{A is irreducible:}
\begin{equation}
    \nexists S \text{ linear subspace of $ \mathbb{R}^{n}$} \; : \; (  A \vec{v}  \in S \, \; \forall \vec{v} \in S )
\end{equation}

Let us recall the following basic:
\begin{theorem} \label{th:Perron-Frobenius' Theorem}
\end{theorem}
\emph{Perron-Frobenius' Theorem:}

\begin{hypothesis}
\end{hypothesis}

\begin{center}
 A irreducible primitive
\end{center}

\begin{thesis}
\end{thesis}
\begin{center}
  A admits a strictly positive eigenvalue $ \alpha  $ (called the \emph{leading eigenvalue of A}) whose absolute value is greater than
  the absolute value of all the other eigenvalues, $ \alpha  $ is a simple eigenvalue and there exists an eigenvector with positive entries associated
  to $ \alpha $
\end{center}

\smallskip

Given a substitution $ \sigma $ let us introduce the following:
\begin{definition}
\end{definition}
\emph{$\sigma$ is primitive:}
\begin{equation}
    \exists n \in \mathbb{N}_{+} \; : \; \text{a occurs in $ \sigma^{n} (b)$} \; \; \forall a,b \in A
\end{equation}

\begin{definition}
\end{definition}
\emph{incidence matrix of $ \sigma $:}
\begin{equation}
 M_{\sigma} \in \mathbb{M}_{|A|}( \mathbb{N} ) \; :=
 \; (M_{\sigma})_{i j} \, := \, | \sigma ( a_{j}) |_{a_{i}}
\end{equation}

Then:
\begin{proposition}
\end{proposition}
\begin{center}
 $ \sigma $ is primitive $ \Leftrightarrow \; M_{\sigma} $ is primitive
\end{center}

Demanding to the appendix \ref{sec:Pisot-Vijayaraghavan numbers}
for the involved mathematical notions let us introduce the
following:
\begin{definition}
\end{definition}
\emph{$ \sigma $ is of Pisot type:}
\begin{center}
 $ M_{\sigma} $ has a leading eigenvalue $ \lambda  $ such that for every other eigenvalue $ \alpha $ one gets $
 \lambda > 1 > | \alpha |$
\end{center}

It may be proved that:
\begin{proposition} \label{prop:Basic properties of substitutions of Pisot type}
\end{proposition}
\emph{Basic properties of substitutions of Pisot type:}

\begin{hypothesis}
\end{hypothesis}
\begin{center}
 $ \sigma $ substitution of Pisot type
\end{center}

\begin{thesis}
\end{thesis}
\begin{enumerate}
    \item $ \sigma $ is primitive
    \item  the characteristic polynomial $ P_{\sigma} := det(\lambda \mathbb{I} - M_{\sigma} ) \in \mathbb{Z}[\lambda] $ is irreducible over $ \mathbb{Q} $ \cite{Prasolov-99} and hence   the leading eigenvalue $ \alpha $ of $ M_{\sigma} $ is a Pisot-Vijayaraghavan number.
     \item in the fixed point $ \bar{\sigma}(a) $ of $ \sigma $ starting with a letter a such that $ | \sigma(a) | \geq 2 $   the frequencies of the letters are given by the coordinates of the positive eigenvector associated with the leading eigenvalue
     normalized in such a way that the sum of its coordinates equal 1.
\end{enumerate}
where we demand to the appendix \ref{sec:Pisot-Vijayaraghavan numbers} for the definition and the remarkable properties of the Pisot-Vijayaraghavan numbers.

All this stuff may be concretely implemented in Mathematica 5
 through the notebook reported in the appendix \ref{sec:Mathematica implementation of this paper}.

\begin{proposition} \label{Number theoretic characterization of substitutions of Pisot type}
\end{proposition}
\emph{Number theoretic characterization of substitutions of Pisot type:}

\begin{hypothesis}
\end{hypothesis}
\begin{center}
 $ \sigma $ primitive substitution
\end{center}

\begin{thesis}
\end{thesis}
\begin{center}
 $ \sigma $ is of Pisot type if and only if it its leading eigenvalue is a  Pisot-Vijayaraghavan number.
\end{center}

\bigskip

\begin{example} \label{ex:Fibonacci substitution}
\end{example}

The mathematician \emph{Leonardo da Pisa}, son of the customer
inspector Bonaccio and hence later commonly known as Fibonacci,
considered in his \emph{Liber Abaci} (published in 1202) the
following problem:

let us suppose to have couples of rabbits such that:
\begin{enumerate}
    \item each month a couple of  baby rabbits becomes adult
    \item each month a couple of adult rabbits generates a couple
    of baby rabbits
\end{enumerate}
Such a situation may be codified representing an adult couple of rabbits with a zero, a baby couple of rabbits with a one and
introducing the following substitution over the two-letter alphabet $ \{ 0, 1 \} $:
\begin{equation}
    \sigma (0) \; := \{ 0 , 1 \}
\end{equation}
\begin{equation}
      \sigma (1) \; := \{ 0 \}
\end{equation}
from which one derives the Fibonacci sequence $
\bar{\sigma}(0) $:
\begin{verbatim}
{0}

{0,1}

{0,1,0}

{0,1,0,0,1}

{0,1,0,0,1,0,1,0}

{0,1,0,0,1,0,1,0,0,1,0,0,1}

{0,1,0,0,1,0,1,0,0,1,0,0,1,0,1,0,0,1,0,1,0}

{0,1,0,0,1,0,1,0,0,1,0,0,1,0,1,0,0,1,0,1,0,0,1,0,0,1,0,1,0,0,1,0,0,1}

{0,1,0,0,1,0,1,0,0,1,0,0,1,0,1,0,0,1,0,1,0,0,1,0,0,1,0,1,0,0,1,0,0,1,0,1,0,0,\
1,0,1,0,0,1,0,0,1,0,1,0,0,1,0,1,0}

{0,1,0,0,1,0,1,0,0,1,0,0,1,0,1,0,0,1,0,1,0,0,1,0,0,1,0,1,0,0,1,0,0,1,0,1,0,0,\
1,0,1,0,0,1,0,0,1,0,1,0,0,1,0,1,0,0,1,0,0,1,0,1,0,0,1,0,0,1,0,1,0,0,1,0,1,0,0,\
1,0,0,1,0,1,0,0,1,0,0,1}

{0,1,0,0,1,0,1,0,0,1,0,0,1,0,1,0,0,1,0,1,0,0,1,0,0,1,0,1,0,0,1,0,0,1,0,1,0,0,\
1,0,1,0,0,1,0,0,1,0,1,0,0,1,0,1,0,0,1,0,0,1,0,1,0,0,1,0,0,1,0,1,0,0,1,0,1,0,0,\
1,0,0,1,0,1,0,0,1,0,0,1,0,1,0,0,1,0,1,0,0,1,0,0,1,0,1,0,0,1,0,1,0,0,1,0,0,1,0,\
1,0,0,1,0,0,1,0,1,0,0,1,0,1,0,0,1,0,0,1,0,1,0,0,1,0,1,0}

...

\end{verbatim}
Let us remark that:
\begin{equation}
    \sigma^{n}(0) \; = \;  \sigma^{n-1}(0) \cdot   \sigma^{n-2}(0) \; \; \forall n \in \mathbb{N} : n \geq 2
\end{equation}
and hence:
\begin{equation}
    | \sigma^{n} (0) | \; = \; F_{n+2} \; \; \forall n \in \mathbb{N}
\end{equation}
where $ F_{n} $ is the $ n^{th} $ Fibonacci number (equal to the number of adult couples of rabbits at the $ n^{th} $ month) defined recursively by
\begin{equation}
    F_{n+2} \; := \; F_{n+1} + F_{n}
\end{equation}
\begin{equation}
    F_{0} \; := \; 0
\end{equation}
\begin{equation}
    F_{1} \; := \; 1
\end{equation}
Furthermore:
\begin{equation} \label{eq:number of adults couples at step n}
     | \sigma^{n} (0) |_{0} \; = \; F_{n+1} \; \; \forall n \in \mathbb{N}_{+}
\end{equation}
\begin{equation} \label{eq:number of baby couples at step n}
     | \sigma^{n} (0) |_{1} \; = \; F_{n}  \; \; \forall n \in \mathbb{N}_{+}
\end{equation}
The incidence matrix of $ \sigma $ is::
\begin{equation}
    M_{\sigma} \; = \; \left(%
\begin{array}{cc}
  1 & 1 \\
  1 & 0 \\
\end{array}%
\right)
\end{equation}
whose leading eigenvalue is the golden number $ \tau :=
\frac{1+\sqrt{5} }{2} $, id est the least Pisot-Vijayaraghavan
number  of order two (see the example \ref{ex:the golden number is
a PV-number}).

Since:
\begin{equation}
  M_{\sigma}^{n} \; = \; \left(%
\begin{array}{cc}
  F_{n+1} & F_{n} \\
  F_{n} & F_{n-1} \\
\end{array}%
\right) \; \; \forall n \in \mathbb{N}
\end{equation}

it follows that $ \sigma $ is also primitive and hence is of Pisot
type.

The eigenvector of $ M_{\sigma} $ normalized so that the sum of
its components equals to one is $ ( \frac{\tau}{1+ \tau} ,
\frac{1}{1+ \tau} ) $; hence:
\begin{equation}
    f_{0} ( \bar{\sigma} (0) ) \; = \; \frac{\tau}{1+\tau}
\end{equation}
\begin{equation}
    f_{1} ( \bar{\sigma} (1) ) \; = \; \frac{1}{1+\tau}
\end{equation}
and hence in particular:
\begin{equation}
    \frac{ f_{0} ( \bar{\sigma} (0) )}{f_{1} ( \bar{\sigma} (1) ) } \; = \; \tau
\end{equation}
that is consistent with the equation \ref{eq:number of adults couples at step n} and the equation \ref{eq:number of baby couples at step n} since:
\begin{equation}
    \lim_{n \rightarrow + \infty } \frac{F_{n+1}}{F_{n}} \; = \; \tau
\end{equation}

Let us remark that since:
\begin{equation}
    \frac{ f_{0} ( \bar{\sigma} (0) ) }{f_{1} ( \bar{\sigma} (0) ) } \; = \; \tau \neq 1
\end{equation}
it follows that the Fibonacci sequence is not Borel-normal and
hence, in particular, is not algorithmically-random, as we could
have inferred by the fact that it is Sturmian and hence:
\begin{equation}
    H_{top} ( \sigma ) \; = \; H_{top} ( \bar{\sigma} (1) ) \; = \; 0
\end{equation}

The fact that $ \frac{ f_{0} ( \bar{\sigma} (1) ) }{f_{1} ( \bar{\sigma} (0) ) } = \tau \notin \mathbb{Q} $ implies that the Fibonacci sequence is not ultimately periodic.

\bigskip

\begin{example} \label{ex:Padovan substitution}
\end{example}
Let us consider the following substitution over the alphabet of three letters $ \{ 0 , 1 , 2 \} $:
\begin{equation}
    \sigma (0) \; := \{ 1 , 2 \}
\end{equation}
\begin{equation}
  \sigma (1) \; := \{ 2 \}
\end{equation}
\begin{equation}
  \sigma (2) \; := \; \{ 0 \}
\end{equation}

from which one derives the \emph{Padovan sequence} $ \bar{\sigma}(0) $ :
\begin{verbatim}
{0}

{1,2}

{2,0}

{0,1,2}

{1,2,2,0}

{2,0,0,1,2}

{0,1,2,1,2,2,0}

{1,2,2,0,2,0,0,1,2}

{2,0,0,1,2,0,1,2,1,2,2,0}

{0,1,2,1,2,2,0,1,2,2,0,2,0,0,1,2}

{1,2,2,0,2,0,0,1,2,2,0,0,1,2,0,1,2,1,2,2,0}

{2,0,0,1,2,0,1,2,1,2,2,0,0,1,2,1,2,2,0,1,2,2,0,2,0,0,1,2}

{0,1,2,1,2,2,0,1,2,2,0,2,0,0,1,2,1,2,2,0,2,0,0,1,2,2,0,0,1,2,0,1,2,1,2,2,0}

...
\end{verbatim}

It may be easily proved that:
\begin{equation}
    | \sigma^{n}(0) | \; = \; P_{n+2} \; \; \forall n \in \mathbb{N}_{+}
\end{equation}

where $ P_{n} $ is the $ n^{th} $ \emph{Padovan number} defined
by:
\begin{equation}
    P(0) \; := \; P(1) \; := P(2) \; := 1
\end{equation}
\begin{equation}
    P_{n} \; := \; P_{n-2} + P_{n-3}
\end{equation}

Furthermore:
\begin{equation}
    \lim_{n \rightarrow + \infty} \frac{ | \sigma^{n}(0) |   }{ | \sigma^{n-1}(0) |} \; = \;  \lim_{n \rightarrow + \infty} \frac{P_{n+2}}{ P_{n+1} } \; = \;  \lim_{n \rightarrow + \infty} \frac{P_{n}}{ P_{n-1} } \; = \; \rho
\end{equation}
where  $ \rho \; := \; \frac{(9- \sqrt{69})^{\frac{1}{3}} \, + \,
(9 + \sqrt{69})^{\frac{1}{3}}  }{ 2^{\frac{1}{3}}  3^{\frac{2}{3}}
} $ is the \emph{plastic number}, id est the least
Pisot-Vijayaraghavan number (see the example \ref{ex:the plastic
number is a PV-number}).

The incidence matrix of $ \sigma $ is:
\begin{equation}
    M_{\sigma} \; = \; \left(%
\begin{array}{ccc}
  0 & 0 & 1 \\
  1 & 0 & 0 \\
  1 & 1 & 0 \\
\end{array}%
\right)
\end{equation}
whose leading  eigenvalue is $ \rho $.

Since:
\begin{equation}
 M_{\sigma}^{n} \; = \; \left(%
\begin{array}{ccc}
  P_{n-2} & P_{n-4} & P_{n-3} \\
  P_{n-3} & P_{n-5} & P_{n-4}  \\
  P_{n-1} & P_{n-3} & P_{n-2} \\
\end{array}%
\right)   \; \;  \forall n \in \mathbb{N} : n \geq 5
\end{equation}

$ \sigma $ is also primitive and hence is of Pisot type.

The eigenvector of $ M_{\sigma} $ associated to the eigenvalue $
\rho $ and normalized so that the sum of its components equals to
one is $ \{ \rho -1 , \frac{1+ \rho- \rho^{2}}{1+\rho} ,
\frac{1}{1+\rho} \} $ and hence:
\begin{equation}
    f_{0} ( \bar{x}(0) ) \; = \;  \rho -1
\end{equation}
\begin{equation}
    f_{1} ( \bar{x}(0) ) \; = \;  \frac{1+ \rho - \rho^{2}}{1+ \rho}
\end{equation}
\begin{equation}
     f_{2} ( \bar{x}(0) ) \; = \; \frac{1}{1+\rho}
\end{equation}

Since:
\begin{equation}
    \frac{f_{1} ( \bar{x}(0) )}{f_{0} ( \bar{x}(0) )} \; = \; \frac{ 1 + \rho - \rho^{2}  }{-1 + \rho^{2}}  \; \neq \; 1
\end{equation}
\begin{equation}
    \frac{f_{2} ( \bar{x}(0) )}{f_{0} ( \bar{x}(0) )} \; = \; \frac{1}{-1 + \rho^{2}}  \; \neq \; 1
\end{equation}
it follows that the Padovan sequence is not Borel-normal and hence, in particular, it is not-algorithmically random.

Since $  \frac{f_{1} ( \bar{x}(0) )}{f_{0} ( \bar{x}(0) )} \notin \mathbb{Q} $ and
$ \frac{f_{2} ( \bar{x}(0) )}{f_{0} ( \bar{x}(0) )} \; = \; \frac{1}{-1 + \rho^{2}} \notin \mathbb{Q} $ it follows that
the Padovan sequence is not ultimately periodic.

\bigskip

\begin{example} \label{ex:Pell substitution}
\end{example}
Let us consider the Pell substitution, id est the substitution $
\sigma $ over the binary alphabet $ \{ 0 , 1 \}$:
\begin{equation}
    \sigma ( 0 ) \; := \; \{ 0 , 1 \}
\end{equation}
\begin{equation}
    \sigma (1) \; := \; \{ 0 , 0 , 1 \}
\end{equation}
from which one derives the Pell sequence:
\begin{verbatim}
{0}

{0,1}

{0,1,0,0,1}

{0,1,0,0,1,0,1,0,1,0,0,1}

{0,1,0,0,1,0,1,0,1,0,0,1,0,1,0,0,1,0,1,0,0,1,0,1,0,1,0,0,1}

{0,1,0,0,1,0,1,0,1,0,0,1,0,1,0,0,1,0,1,0,0,1,0,1,0,1,0,0,1,0,1,0,0,1,0,1,0,1,\
0,0,1,0,1,0,0,1,0,1,0,1,0,0,1,0,1,0,0,1,0,1,0,0,1,0,1,0,1,0,0,1}

{0,1,0,0,1,0,1,0,1,0,0,1,0,1,0,0,1,0,1,0,0,1,0,1,0,1,0,0,1,0,1,0,0,1,0,1,0,1,\
0,0,1,0,1,0,0,1,0,1,0,1,0,0,1,0,1,0,0,1,0,1,0,0,1,0,1,0,1,0,0,1,0,1,0,0,1,0,1,\
0,1,0,0,1,0,1,0,0,1,0,1,0,0,1,0,1,0,1,0,0,1,0,1,0,0,1,0,1,0,1,0,0,1,0,1,0,0,1,\
0,1,0,0,1,0,1,0,1,0,0,1,0,1,0,0,1,0,1,0,1,0,0,1,0,1,0,0,1,0,1,0,1,0,0,1,0,1,0,\
0,1,0,1,0,0,1,0,1,0,1,0,0,1}

{0,1,0,0,1,0,1,0,1,0,0,1,0,1,0,0,1,0,1,0,0,1,0,1,0,1,0,0,1,0,1,0,0,1,0,1,0,1,\
0,0,1,0,1,0,0,1,0,1,0,1,0,0,1,0,1,0,0,1,0,1,0,0,1,0,1,0,1,0,0,1,0,1,0,0,1,0,1,\
0,1,0,0,1,0,1,0,0,1,0,1,0,0,1,0,1,0,1,0,0,1,0,1,0,0,1,0,1,0,1,0,0,1,0,1,0,0,1,\
0,1,0,0,1,0,1,0,1,0,0,1,0,1,0,0,1,0,1,0,1,0,0,1,0,1,0,0,1,0,1,0,1,0,0,1,0,1,0,\
0,1,0,1,0,0,1,0,1,0,1,0,0,1,0,1,0,0,1,0,1,0,1,0,0,1,0,1,0,0,1,0,1,0,0,1,0,1,0,\
1,0,0,1,0,1,0,0,1,0,1,0,1,0,0,1,0,1,0,0,1,0,1,0,1,0,0,1,0,1,0,0,1,0,1,0,0,1,0,\
1,0,1,0,0,1,0,1,0,0,1,0,1,0,1,0,0,1,0,1,0,0,1,0,1,0,0,1,0,1,0,1,0,0,1,0,1,0,0,\
1,0,1,0,1,0,0,1,0,1,0,0,1,0,1,0,1,0,0,1,0,1,0,0,1,0,1,0,0,1,0,1,0,1,0,0,1,0,1,\
0,0,1,0,1,0,1,0,0,1,0,1,0,0,1,0,1,0,0,1,0,1,0,1,0,0,1,0,1,0,0,1,0,1,0,1,0,0,1,\
0,1,0,0,1,0,1,0,0,1,0,1,0,1,0,0,1,0,1,0,0,1,0,1,0,1,0,0,1,0,1,0,0,1,0,1,0,1,0,\
0,1,0,1,0,0,1,0,1,0,0,1,0,1,0,1,0,0,1}

{0,1,0,0,1,0,1,0,1,0,0,1,0,1,0,0,1,0,1,0,0,1,0,1,0,1,0,0,1,0,1,0,0,1,0,1,0,1,\
0,0,1,0,1,0,0,1,0,1,0,1,0,0,1,0,1,0,0,1,0,1,0,0,1,0,1,0,1,0,0,1,0,1,0,0,1,0,1,\
0,1,0,0,1,0,1,0,0,1,0,1,0,0,1,0,1,0,1,0,0,1,0,1,0,0,1,0,1,0,1,0,0,1,0,1,0,0,1,\
0,1,0,0,1,0,1,0,1,0,0,1,0,1,0,0,1,0,1,0,1,0,0,1,0,1,0,0,1,0,1,0,1,0,0,1,0,1,0,\
0,1,0,1,0,0,1,0,1,0,1,0,0,1,0,1,0,0,1,0,1,0,1,0,0,1,0,1,0,0,1,0,1,0,0,1,0,1,0,\
1,0,0,1,0,1,0,0,1,0,1,0,1,0,0,1,0,1,0,0,1,0,1,0,1,0,0,1,0,1,0,0,1,0,1,0,0,1,0,\
1,0,1,0,0,1,0,1,0,0,1,0,1,0,1,0,0,1,0,1,0,0,1,0,1,0,0,1,0,1,0,1,0,0,1,0,1,0,0,\
1,0,1,0,1,0,0,1,0,1,0,0,1,0,1,0,1,0,0,1,0,1,0,0,1,0,1,0,0,1,0,1,0,1,0,0,1,0,1,\
0,0,1,0,1,0,1,0,0,1,0,1,0,0,1,0,1,0,0,1,0,1,0,1,0,0,1,0,1,0,0,1,0,1,0,1,0,0,1,\
0,1,0,0,1,0,1,0,0,1,0,1,0,1,0,0,1,0,1,0,0,1,0,1,0,1,0,0,1,0,1,0,0,1,0,1,0,1,0,\
0,1,0,1,0,0,1,0,1,0,0,1,0,1,0,1,0,0,1,0,1,0,0,1,0,1,0,1,0,0,1,0,1,0,0,1,0,1,0,\
0,1,0,1,0,1,0,0,1,0,1,0,0,1,0,1,0,1,0,0,1,0,1,0,0,1,0,1,0,1,0,0,1,0,1,0,0,1,0,\
1,0,0,1,0,1,0,1,0,0,1,0,1,0,0,1,0,1,0,1,0,0,1,0,1,0,0,1,0,1,0,0,1,0,1,0,1,0,0,\
1,0,1,0,0,1,0,1,0,1,0,0,1,0,1,0,0,1,0,1,0,0,1,0,1,0,1,0,0,1,0,1,0,0,1,0,1,0,1,\
0,0,1,0,1,0,0,1,0,1,0,1,0,0,1,0,1,0,0,1,0,1,0,0,1,0,1,0,1,0,0,1,0,1,0,0,1,0,1,\
0,1,0,0,1,0,1,0,0,1,0,1,0,0,1,0,1,0,1,0,0,1,0,1,0,0,1,0,1,0,1,0,0,1,0,1,0,0,1,\
0,1,0,1,0,0,1,0,1,0,0,1,0,1,0,0,1,0,1,0,1,0,0,1,0,1,0,0,1,0,1,0,1,0,0,1,0,1,0,\
0,1,0,1,0,0,1,0,1,0,1,0,0,1,0,1,0,0,1,0,1,0,1,0,0,1,0,1,0,0,1,0,1,0,0,1,0,1,0,\
1,0,0,1,0,1,0,0,1,0,1,0,1,0,0,1,0,1,0,0,1,0,1,0,1,0,0,1,0,1,0,0,1,0,1,0,0,1,0,\
1,0,1,0,0,1,0,1,0,0,1,0,1,0,1,0,0,1,0,1,0,0,1,0,1,0,0,1,0,1,0,1,0,0,1,0,1,0,0,\
1,0,1,0,1,0,0,1,0,1,0,0,1,0,1,0,1,0,0,1,0,1,0,0,1,0,1,0,0,1,0,1,0,1,0,0,1,0,1,\
0,0,1,0,1,0,1,0,0,1,0,1,0,0,1,0,1,0,0,1,0,1,0,1,0,0,1,0,1,0,0,1,0,1,0,1,0,0,1,\
0,1,0,0,1,0,1,0,1,0,0,1,0,1,0,0,1,0,1,0,0,1,0,1,0,1,0,0,1,0,1,0,0,1,0,1,0,1,0,\
0,1,0,1,0,0,1,0,1,0,0,1,0,1,0,1,0,0,1,0,1,0,0,1,0,1,0,1,0,0,1,0,1,0,0,1,0,1,0,\
0,1,0,1,0,1,0,0,1,0,1,0,0,1,0,1,0,1,0,0,1,0,1,0,0,1,0,1,0,1,0,0,1,0,1,0,0,1,0,\
1,0,0,1,0,1,0,1,0,0,1}

...
\end{verbatim}

It may be easily proved that:
\begin{equation}
  | \sigma^{n}(0) | \; = \; a_{n+1} \; \; \forall n \in \mathbb{N}_{+}
\end{equation}
where $ \{ a_{n }\} \in \mathbb{N}^{\mathbb{N}} $ are the Pell
numbers defined as:
\begin{equation}
    a_{0} \; := \; 0
\end{equation}
\begin{equation}
    a_{1} \; := \; 1
\end{equation}
\begin{equation}
    a_{n} \; := 2 a_{n-1} + a_{n-2} \; \; \forall n \in \mathbb{N}
    : n \geq 2
\end{equation}
Furthermore:
\begin{equation} \label{eq:occurence of zero in the Pell sequence}
    | \sigma^{n}(0)|_{0} \; = \; a_{n} + a_{n-1} \; \; \forall n \in \mathbb{N}_{+}
\end{equation}
\begin{equation} \label{eq:occurence of one in the Pell sequence}
    | \sigma^{n}(0)|_{1} \; = \; a_{n} \; \; \forall n \in \mathbb{N}_{+}
\end{equation}

The incidence matrix of $ \sigma $ is:
\begin{equation}
    M_{\sigma} \; = \; \left(%
\begin{array}{cc}
  1 & 2 \\
  1 & 1 \\
\end{array}%
\right)
\end{equation}
Since:
\begin{equation}
    M_{\sigma}^{n} \; = \; \left(%
\begin{array}{cc}
  a_{n}+a_{n-1} & 2 a_{n} \\
  a_{n} & a_{n}+a_{n-1} \\
\end{array}%
\right) \; \; \forall n \in \mathbb{N}_{+}
\end{equation}
it follows that $ M_{\sigma} $ is a primitive matrix and hence $
\sigma $ is a primitive substitution.

 Since furthermore the leading eigenvalue of $ M_{\sigma} $ is $
1 + \sqrt{2} $ that is a Pisot-Vijayaraghavan number (see the
example \ref{ex:the Pell number is a PV-number}) it follows that
the Pell substitution is of Pisot type.

The eigenvector of $ M_{\sigma} $ associated to the leading
eigenvalue $ 1 + \sqrt{2} $ and normalized so that the sum of its
components equals to one is $ \{ \frac{ \sqrt{2} }{ 1 + \sqrt{2} }
, \frac{1}{1 + \sqrt{2}} \} $. Therefore:
\begin{equation}
    f_{0}( \bar{\sigma} (0) ) \; = \;  \frac{ \sqrt{2} }{ 1 + \sqrt{2} }
\end{equation}
\begin{equation}
    f_{1}( \bar{\sigma} (0) ) \; = \;  \frac{ 1 }{ 1 + \sqrt{2} }
\end{equation}
consistently with the equation \ref{eq:occurence of zero in the
Pell sequence} and the equation \ref{eq:occurence of one in the
Pell sequence} since:
\begin{equation}
    \lim_{n \rightarrow + \infty} \frac{a_{n}}{a_{n-1}} \; = \; 1
    + \sqrt{2}
\end{equation}
Since:
\begin{equation}
    \frac{f_{0}( \bar{\sigma} (0) )}{ f_{1}( \bar{\sigma} (0) )}
    \; = \; \sqrt{2} \notin \mathbb{Q}
\end{equation}
it follows that the Pell sequence is neither Borel normal nor
ultimately periodic.

\newpage
\section{Substitutions of Pisot type as algorithms for optimal
spacing} \label{sec:Substitutions of Pisot type as algorithms for
optimal spacing}

Let us suppose to have to allocate $ n \in \mathbb{N}_{+} : n \geq
2 $ petals of  a flower  so that:
\begin{enumerate}
    \item the average distance between neighbors is maximal in order to maximize the exposition of each
petal to sun and rain.
    \item the distribution of the distances among neighbour petals is the more uniform one.
\end{enumerate}

Supposed all the petals have unit length so that their vertices
belong to the unit circumference of the complex plane let us
assume that their positions are $ e^{i \theta_{1} } , \cdots e^{i
\theta_{n} } $ where $ 0 \leq \theta_{1} < \theta_{2} < \cdots <
\theta_{k} < \theta_{k+1} < \cdots < \theta_{n} < 2 \pi $ and let
us as pose:
\begin{equation}
    \alpha_{k} \; := \; d_{geodesic}( \theta_{k+1} - \theta_{k}) \; \; k= 1 , \cdots , n
\end{equation}
where:
\begin{equation} \label{eq:cyclicity constraint}
    \theta_{n+1} \; := \; \theta_{1}
\end{equation}
and where the distance $ d_{geodesic} : [ 0 , 2 \pi)^{2} \mapsto \mathbb{R} $:
\begin{equation} \label{eq:geodesic distance on the circumference}
    d_{geodesic} ( \theta_{k+1} ,  \theta_{k} ) \; = \; \left\{%
\begin{array}{ll}
     \theta_{k+1} - \theta_{k} , & \hbox{ if $ \theta_{k+1} - \theta_{k}  \leq \pi $;} \\
    2 \pi - ( \theta_{k+1} - \theta_{k} )  , & \hbox{otherwise.} \\
\end{array}%
\right.
\end{equation}
is the geodesic distance between $ e^{i \theta_{k}} $ and $ e^{i \theta_{k+1}} $.

Our optimization problem consists in  maximizing the mean of $
\alpha $:
\begin{equation}
    \bar{\alpha} \; := \; \frac{ \sum_{i=1}^{n} \alpha_{i} }{n}
\end{equation}
while minimizing the variance of $ \alpha $:
\begin{equation}
    \sigma^{2} ( \alpha ) \; := \frac{ \sum_{i=1}^{n} ( \alpha_{i} - \bar{\alpha})^{2} }{n}
\end{equation}

The solution of such an optimization problem consists in locating the $ n^{th} $ petal in the $ n^{th}$ root of unit:
\begin{equation} \label{eq:roots of unity}
  e^{ i \theta_{k} } \; = \; r_{n,k} \; := \; e^{i \frac{2 \pi}{n} k } \; \; k \in \{ 1 , \cdots , n \}
\end{equation}
so that:
\begin{equation}
    \alpha_{k} \; = \; \bar{\alpha} \; = \; \frac{2 \pi}{n} \; \;
    k = 1 , \cdots , n
\end{equation}
\begin{equation}
    \sigma^{2} ( \alpha ) \; = \; 0
\end{equation}
To prove it let us observe that, by the equation \ref{eq:geodesic
distance on the circumference},  any configuration such that $
\max \{ \theta_{k} - \theta_{k-1} \} < \pi $ and $ \sum_{k}
\alpha_{k} = 2 \pi $ furnishes the maximum value of the mean of $
\alpha $:
\begin{equation}
 \bar{ \alpha } \; = \; \frac{ 2 \pi}{n}
\end{equation}
Among these configurations the one with minimal variance is, of
course, that in which  $ \alpha_{k} = \theta_{k} - \theta_{k-1} =
\frac{ 2 \pi}{n} $ for every k so that the variance of $ \alpha $
is equal to zero.

The roots of the identity have some beautiful properties; first of
all:
\begin{proposition} \label{prop:cyclotomic identity}
\end{proposition}
\emph{Cyclotomic identity:}
\begin{equation}
    \sum_{k=1}^{n} r_{n,k} \; = \; 0 \; \; \forall n \in \mathbb{N}_{+} : n \geq 2
\end{equation}
\begin{proof}
It is sufficient to observe that:
\begin{equation}
   \sum_{k=1}^{n} r_{n,k} \; = \; \sum_{k=1}^{n} r_{1,k}^{k} \; = \; \frac{ r_{1,k} ( 1 - r_{1,k}^{n}  ) }{1 - r_{1,k}} \; = \; 0
\end{equation}
where we have used the fact that:
\begin{equation}
  \sum_{k=1}^{n} z^{k} \; = \; \frac{z ( 1 - z^{n})}{1 - z } \; \; \forall z \in \mathbb{C} - \{ 1 \} \, , \,  \forall n \in \mathbb{N}_{+} : n \geq 2
\end{equation}
\end{proof}

Let us now denote by $ polygon_{n,1} $ the regular polygon having
the $ n^{th}  $ roots of unity as vertices.

Given $ i, j \in \{ 1 , \cdots ,n \}$ such that $ i \neq j , i
\neq j \pm 1 $ let us furthermore denote by $ diagonal (i, j) $
the segment connecting $ r_{n,i} $ and $ r_{n,j} $.

Let us then consider the set of the intersections of distinct
diagonals, id est the set of points $ \{  diagonal (i_{1}, j_{1})
\cap diagonal (i_{2}, j_{2}) \, , \,  i_{1},i_{2},j_{1},j_{2} \in
\{ 1 , \cdots , n \} , i_{1} \neq j_{1} ,   i_{2} \neq j_{2} ,
i_{1} \neq i_{2}, j_{1} \neq j_{2} $ and let us call $
polygon_{n,2} $ the polygon having these points as vertices.

We will say that:
\begin{definition} \label{def:roots of unity generating similarity}
\end{definition}
\emph{the $ n^{th} $ roots of unity generate self-similarity:}
\begin{center}
  $ polygon_{n,2} $ is regular
\end{center}

Then name of the definition \ref{def:roots of unity generating
similarity} is owed to the fact that if $ polygon_{n,2} $ is
regular then there exists a scaling factor $ r_{scaling} \in ( 0
,1 ) $ and a rotation angle $ \phi_{rotation} \in [ 0 , 2 \pi) $
such that:
\begin{equation}
  polygon_{n,2} \; = \; r_{scaling} e^{i \phi_{rotation}} polygon_{n,1}
\end{equation}

Clearly in this case the procedure may be iterated generating a
sequence $  \{ polygon_{n,k} , k \in \mathbb{N}_{+} \}$ defined
recursively by:
\begin{equation}
  polygon_{n,k} \; = \; r_{scaling} e^{i \phi_{rotation}} polygon_{n,k-1}
\end{equation}

\smallskip

\begin{figure}
  \includegraphics[scale=.8]{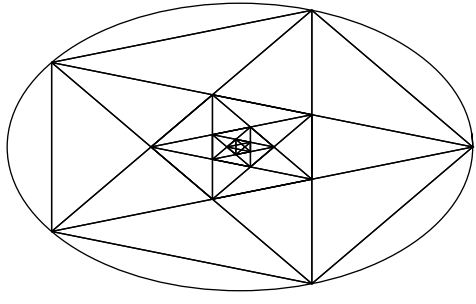}\\
  \caption{The $ 5^{th} $ roots of unity generate
  self-similarity.} \label{fig:selfsimilarity generated by fifth roots of unity}
\end{figure}

\begin{example}
\end{example}
The $ 5^{th} $ roots of unity generate self similarity, with $
r_{scaling} := \frac{\tau}{1 + 2 \tau} $ and $ \phi_{rotation} :=
\pi $ as it can be easily computed and verified looking at the
figure \ref{fig:selfsimilarity generated by fifth roots of unity}.

\bigskip

Let us now suppose to have a countable infinity of petals, so that
the uniform distribution given by the equation \ref{eq:roots of
unity} become meaningless.

Let us then follow another strategy an let us pose the petals in
the points:
\begin{equation}
    p_{\lambda,k} \; := \; \exp ( i \theta_{\lambda,k} ) \; \; k \in \mathbb{N}_{+}
\end{equation}
where:
\begin{equation} \label{eq:optimal distribution of a countable infinity of petals}
    \theta_{\lambda,k} \; := \; Mod_{2 \pi} [  ( 2 \pi \lambda)^{k}  ]  \; \; \forall k \in \mathbb{N}_{+}
\end{equation}

Let us furthermore call $ side_{\lambda,k} $ the segment
connecting $ p_{\lambda,k} $ and $ p_{\lambda,k+1} $ and  let us
then introduce the following:
\begin{definition}
\end{definition}
\emph{piece-wise linear curve associated to $ \lambda $:}
\begin{equation}
    C_{\lambda} \; := \; \cup_{k \in \mathbb{N}_{+} } side_{k}
\end{equation}
\begin{definition}
\end{definition}
\emph{$ \infty^{th} $ cusp of $  C_{\lambda} $:}
\begin{equation}
    p_{\lambda,\infty} \; := \; \lim_{k \rightarrow \infty}
    p_{\lambda,k} \: (mod \, 2 \pi)
\end{equation}

Then:
\begin{theorem} \label{th:geometrical beauty of Pisot-Vijayaraghavan numbers}
\end{theorem}
\emph{Geometrical beauty of Pisot-Vijayaraghavan numbers:}

\begin{hypothesis}
\end{hypothesis}
\begin{equation}
    \lambda \in PV( \mathbb{A})
\end{equation}

\begin{thesis}
\end{thesis}
\begin{equation}
    p_{\lambda, \infty} \; = \; p_{\lambda,1}
\end{equation}
\begin{proof}
The proposition \ref{prop:fundamental property of the
Pisot-Vijayaraghavan numbers} implies that if $ \lambda $ is
chosen to be a Pisot-Vijayaraghavan number, then:
\begin{equation} \label{eq:nonperiodic cyclotomic identity}
  \lim_{k \rightarrow + \infty}  \theta_{\lambda,k} \; = \; 0 \; (
  mod \, 2 \pi )
\end{equation}
from which the thesis follows.
\end{proof}

\begin{figure}
  \includegraphics[scale=.8]{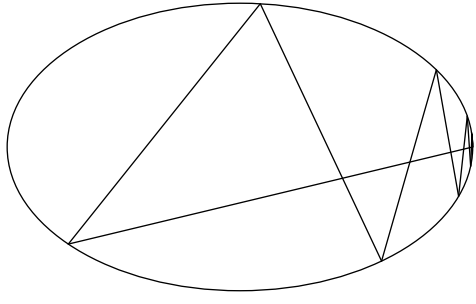}\\
  \caption{The first 10 cusps of $ C_{\tau}$.} \label{fig:golden number curve}
\end{figure}

\begin{figure}
  \includegraphics[scale=.8]{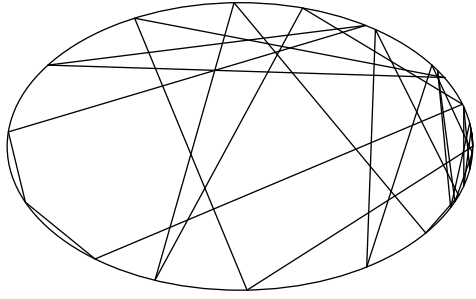}\\
  \caption{The first 30 cusps of $ C_{\rho}$.}  \label{fig:plastic number curve}
\end{figure}

The meaning of the theorem \ref{th:geometrical beauty of
Pisot-Vijayaraghavan numbers} is shown in the figure
\ref{fig:golden number curve}, in the figure \ref{fig:Pell number
curve} and in the figure \ref{fig:plastic number curve}.

\begin{figure}
  \includegraphics[scale=.8]{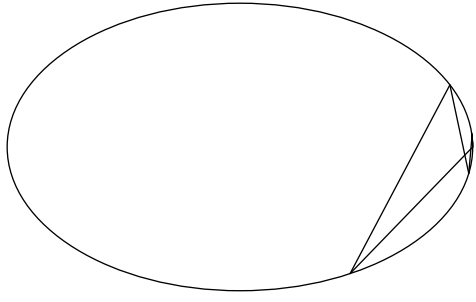}\\
  \caption{The first 5 cusps of $ C_{1+\sqrt{2}}$}  \label{fig:Pell number curve}
\end{figure}

\bigskip

\begin{remark}
\end{remark}
As for the $ n^{th}$ roots of unity one can introduce the segment
diagonal(i,j) connecting $ p_{\lambda,i} $ and $ p_{\lambda,j} $
for every $ i ,j \in \mathbb{N} : i \neq j , i \neq j \pm 1 $ and
then considerate the polygon having as vertices the intersections
of the diagonals.

\begin{figure}
  \includegraphics[scale=.8]{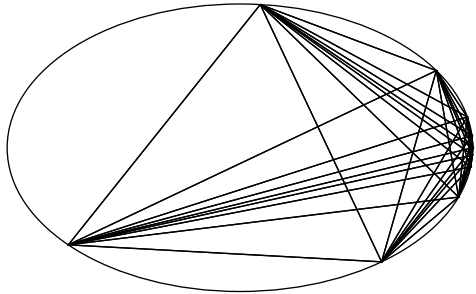}\\
  \caption{$ C_{\tau}$ doesn't give rise to self-similarity.} \label{fig:golden number pentagram}
\end{figure}

\begin{figure}
  \includegraphics[scale=.8]{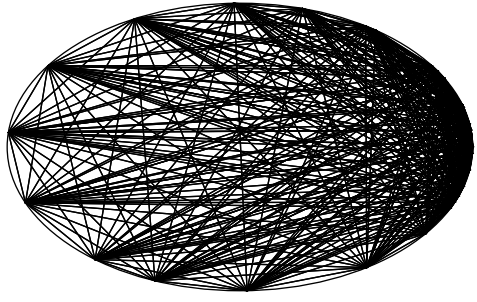}\\
  \caption{$ C_{\rho}$ doesn't give rise to self-similarity.} \label{fig:plastic number pentagram}
\end{figure}

As shown in in the figure \ref{fig:golden number pentagram} and in
the figure \ref{fig:plastic number pentagram} no self-similarity
is generated in this way.

\bigskip

As a corollary of the theorem \ref{th:geometrical beauty of
Pisot-Vijayaraghavan numbers} it follows that the cusps of $
C_{\lambda} $ are a solution of our optimization problem for a
countable infinity of petals if and only if $ \lambda $ is a
Pisot-Vijayaraghavan number.

\smallskip

Our model is a simplified version of the optimization problem that
has led Nature to fix the role of $ \lambda = \tau $ in the
phyllotaxis of many flowers (see for instance the $ 11^{th} $
chapter "The golden section and phyllotaxis" of \cite{Coxeter-89}
 and \cite{Jean-94}).

Let us now suppose to have a substitution $ \sigma $ of Pisot type over the binary alphabet $ \{ 0 , 1 \} $  such that $ | \sigma(0) | \geq 2 $  and let us call again $ \lambda $ the leading eigenvalue of its incidence matrix.

Let suppose to choose two fixed angles $ \beta_{(0)} , \beta_{(1)} \in [ 0 , 2 \pi ) $ such that:
\begin{equation}
    Mod_{2 \pi} [ \frac{\beta_{(0)}}{\beta_{(1)}} ] \; = \; \lambda
\end{equation}
If we suppose to construct our flower by following the algorithmic procedure:
\begin{equation}
    \theta_{k} \; := \;  \left\{%
\begin{array}{ll}
    Mod_{2 \pi} ( \theta_{k-1} +  \beta_{(0)} )   , & \hbox{if  $ ( \bar{\sigma}(0) )_{k} = 0 $}  \\
    Mod_{2 \pi} ( \theta_{k-1} +  \beta_{(1)} )   , & \hbox{if  $ ( \bar{\sigma}(0) )_{k} = 1 $} \\
\end{array}%
\right.
\end{equation}
(where $ (\bar{\sigma}(0))_{k} $ is the $ k^{th} $ digit of the
sequence fixed point $ \bar{\sigma}(0) $ starting with 0) the
optimal solution is reached asymptotically.

\begin{figure}
  \includegraphics[scale=0.8]{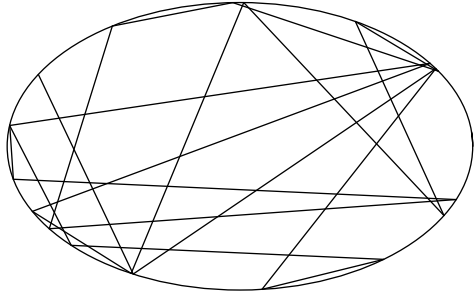}\\
  \caption{The spacing given by the first $ \sum_{k=1}^{^{20}} F_{k} +1 = 17711 $ digits of the Fibonacci sequence, associating spacing $ \tau $ to the letter 0 and spacing 1 to the
  letter 1.} \label{fig:spacing given by the fibonacci sequence}
\end{figure}

\begin{figure}
  \includegraphics[scale=0.8]{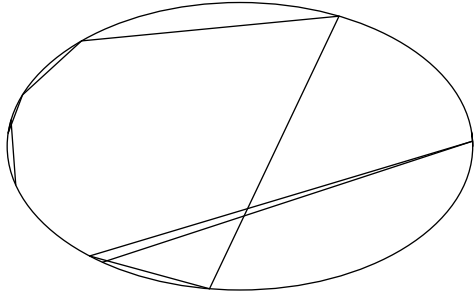}\\
  \caption{The spacing given by the first $ a_{11} = 5741 $ digits of the Pell sequence, associating spacing $ 1 + \sqrt{2} $ to the letter 0 and spacing 1 to the
  letters 1.} \label{fig:spacing given by the Pell sequence}
\end{figure}

Examples  obtained using the Fibonacci substitution and the Pell
substitution are shown, respectively, in the figure
\ref{fig:spacing given by the fibonacci sequence} and in the
figure \ref{fig:spacing given by the Pell sequence}.

\bigskip

\begin{remark}
\end{remark}
Since to build flowers may be a problem for Nature but is not
precisely a concrete problem appearing in common life, the
optimization problems discussed in this paper might appear as
mathematical curiosities with no practical application.

The whole matter appears from a different perspective as soon as
one realizes their deep link with certain searching problems, such
as the problem of localizing  with the minimum possible
uncertainty the minimum of an unknown physical observable f whose
functional dependence from an other physical observable x is
unknown, the only information available being the fact that it is
has a unique minimum on an interval (a , b), through a finite
number n  of compounds measurement $\{  x_{k} , f( x_{k} )
\}_{k=1}^{n} $, each measurement comporting a cost so that the
number of measurements n has to be minimized too (see  the $
9^{th} $ chapter "Optimal spacing and search algorithms" of
\cite{Dunlap-97},  the $ 6^{th}$ chapter "Search Techniques and
nonlinear programming" of \cite{Pierre-86}, the $ 1^{th} $ chapter
"Basic Concepts" of \cite{Knuth-97}, the $ 6^{th} $ chapter
"Searching" of \cite{Knuth-98b}, the $10^{th} $ chapter
"Minimization and Maximization of Functions" of
\cite{Press-Teukolsky-Vetterling-Flannery-07} and the $ 10^{th} $
chapter "Search and games" of \cite{Vajda-08}).

\newpage
\section{Chaos as a phenomenon of undecidability} \label{sec:Chaos as a phenomenon of undecidability}
Let us start by giving some more information concerning Chaitin's
almost mystic $ \Omega $ number.

With this regard we have, preliminarly, to introduce some basic
notion of Computability Theory over the Reals on which,
fortunately, agreement has been reached
\cite{Blum-Cucker-Shub-Smale-98}, \cite{Weihrauch-00}.

Given a sequence of rational numbers $ \{ r_{n} \} \in
\mathbb{Q}^{\mathbb{N}} $:

\begin{definition}
\end{definition}
\emph{ $  \{ r_{n} \} $ is computable:}
\begin{equation}
    \exists a,b,s : \mathbb{N} \mapsto \mathbb{N} \text{ recursive
    } \; : \; (  b(n) \neq 0 \; \wedge \; r_{n} = (-1)^{s(n)}
    \frac{a(n)}{b(n)} ) \; \; \forall n \in \mathbb{N}
\end{equation}

Given $ x \in \mathbb{R} $:
\begin{definition}
\end{definition}
\emph{$  \{ r_{n} \} $ converges computably to x:}
\begin{equation}
    comp-\lim_{n \rightarrow + \infty} r_{n} = x \; := \; \exists e
    : \mathbb{N} \mapsto \mathbb{N} \text{ recursive }  : ( | r_{k}
    - x | \leq 2^{-n} \, \, \forall k \geq e(n) ) \; \; \forall n
    \in \mathbb{N}
\end{equation}

\begin{definition}
\end{definition}
\emph{x is computable:}
\begin{equation}
    \exists  \{ r_{n} \} \in
\mathbb{Q}^{\mathbb{N}} \text{ computable } \; : \; comp-\lim_{n
\rightarrow + \infty} r_{n} = x
\end{equation}

Let us introduce also the weaker notion:
\begin{definition}
\end{definition}
\emph{x is computably enumerable:}
\begin{equation}
     \exists  \{ r_{n} \} \in
\mathbb{Q}^{\mathbb{N}} \text{ increasing ,  computable } \; : \;
\lim_{n \rightarrow + \infty} r_{n} = x
\end{equation}

We will denote the set of computable enumerable reals by $
c.e.(\mathbb{R}) $.

Then \cite{Calude-02}:
\begin{proposition} \label{prop:characterization of Chaitin's halting probabilities}
\end{proposition}
\emph{Characterization of Chaitin's halting probabilities:}
\begin{equation}
    \{ \Omega_{U} \text{ U universal Chaitin computer }      \} \;
= \; RANDOM[(0,1)] \cap c.e.(\mathbb{R})
\end{equation}

Let us now consider a formal system whose rules of inference form
a computably enumerable set of order pairs $ F:= \{ (a_{n}, T_{n}
) \}_{n \in \mathbb{N}} $, the ordered pair $ ( a_{n} , T_{n} ) $
indicating that the theorem $ T_{n} $ is deducible from the axiom
$ a_{n} $.

We will adopt the Mathematical Logic's usual notation:
\begin{definition}
\end{definition}
\emph{T is deducible in the formal system F from the axiom a:}
\begin{equation}
    a \vdash_{F} T \; := \; (a,T) \in F
\end{equation}

We can then state the following fundamental (demanding once more
to the basic reference  \cite{Calude-02} for three different
proofs):
\begin{theorem} \label{th:1th Chaitin information-theoretic incompleteness theorem}
\end{theorem}
\emph{Chaitin's $1^{th}$ Information-theoretic Incompleteness
Theorem:}

\begin{hypothesis}
\end{hypothesis}
\begin{center}
    F formal system with one axiom a
\end{center}
\begin{equation}
    a \vdash_{F} " I(x) > n" \; \Rightarrow \; I(x) > n
\end{equation}

\begin{thesis}
\end{thesis}
\begin{equation}
  \exists c_{F} \in \mathbb{R}_{+} \; : \;  a \vdash_{F} " I(x) >
  n" \: \Rightarrow \: n < I(a) + c_{F}
\end{equation}

\bigskip

\begin{example}
\end{example}
Let us look at a generic substitution $ \sigma $ over the finite
alphabet A as to a formal system $ F_{\sigma} $ having as axioms
the letters of A and as inference rules the substitution rule:
\begin{equation}
    a  \; \vdash_{F_{\sigma}} \; \sigma ( a) \; \; \forall a \in A
\end{equation}
\begin{equation}
    \vec{x} \; \vdash_{F_{\sigma}} \; \sigma( \vec{x} ) \; \;
    \forall \vec{x} \in A^{+}
\end{equation}
Let us suppose that the substitution $ \sigma $ is Sturmian (for
instance it might be the Fibonacci substitution of the example
\ref{ex:Fibonacci substitution}) so that the algorithmic
information of all the deducible theorems is rather  low.

It follows that $ F_{\sigma} $ may be used only to estimate the
information content of  objects of rather low information.

\bigskip

Let us then state also the following:

\begin{theorem} \label{th:2th Chaitin information-theoretic incompleteness theorem}
\end{theorem}
\emph{Chaitin's $2^{th}$ Information-theoretic Incompleteness
Theorem:}

\begin{hypothesis}
\end{hypothesis}
\begin{center}
    A finite alphabet
\end{center}
\begin{center}
    F formal system such that the set $ T_{F} $ of theorems deducible in F is computably enumerable and such
    that any statement of the form "the $ n^{th}$ cbit of
   $ r_{\{0,1\}}(\Omega_{U}) $ is a 0" , "the $ n^{th}$ cbit of
   $ r_{\{0,1\}}(\Omega_{U}) $ is a 1" can be represented in $ T_{F}
   $ and such a statement is a theorem of $ T_{F} $ only if it is
   true
\end{center}

\begin{thesis}
\end{thesis}
\begin{center}
 $ T_{f} $ can enable us to determine the positions and values of
 at most finitely many scattered cbits of $ r_{\{0,1\}}(\Omega_{U}) $
\end{center}

Let us now analyze how Chaitin Information-theoretic
Incompleteness Theorems are at the heart of Chaos Theory.

Let us, at this purpose, briefly recall some basic notions.

Given a finite alphabet A : $ 2 \leq | A | < \aleph_{0} $

\begin{definition} \label{def:Brudno algorithmic information of a sequence}
\end{definition}
\emph{Brudno algorithmic entropy of $ \bar{x} \in
A^{\mathbb{N}_{+}} $}:
\begin{equation}
  B( \bar{x} ) \; := \; \lim_{n \rightarrow \infty} \frac{I( \vec{x}(n))
}{n}
\end{equation}

As it has been proved by Brudno himself \cite{Brudno-78}:
\begin{proposition} \label{prop:Brudno randomness is weaker than Chaitin
randomness}
\end{proposition}
\begin{equation}
  \{ \bar{x} \in A^{\mathbb{N}_{+}} : B( \bar{x} ) > 0 \} \; \supset \;  RANDOM(  A^{\mathbb{N}_{+}}
  )
\end{equation}

given a classical probability space $ ( X \, , \, \mu ) $:
\begin{definition}  \label{def:endomorphism of a classical probability
space}
\end{definition}
\emph{endomorphism of  $ ( X \, , \, \mu ) $}:

$T \, : \, HALT_{\mu} \rightarrow HALT_{\mu} $ surjective :
\begin{equation}
  \mu ( A ) \; = \;   \mu ( T^{-1} A ) \; \; \forall A \in HALT_{\mu}
\end{equation}
where $ HALT_{\mu} $ is the halting-set of the measure $ \mu $,
namely the $ \sigma$-algebra of subsets of X on which $ \mu $ is
defined.

\begin{definition} \label{def:classical dynamical system}
\end{definition}
\emph{classical dynamical system}:

a triple $ ( X \, , \, \mu \, , \, T ) $ such that:
\begin{itemize}
  \item  $ ( X \, , \, \mu ) $ is a classical probability space
  \item  $T \, : \, HALT_{\mu} \rightarrow HALT_{\mu} $ is an
  endomorphism of  $ ( X \, , \, \mu ) $
\end{itemize}
Given a classical dynamical system  $ ( X \, , \, \mu \, , \, T )
$:

\begin{definition} \label{def:ergodic classical dynamical system}
\end{definition}
\emph{$ ( X \, , \, \mu \, , \, T ) $ is ergodic}:
\begin{equation}
lim_{n \rightarrow \infty} \frac{1}{n} \sum_{k=0}^{n-1} \, \mu ( A
\cap T^{k}(B)) \; = \; \mu(A) \,  \mu(B) \; \; \forall \, A,B \in
HALT_{\mu}
\end{equation}

Given a classical probability space $ ( X \, , \mu ) $:
\begin{definition} \label{def:partition of a classical probability space}
\end{definition}
\emph{finite measurable partition of $ ( X \, , \, \mu ) $}:
\begin{equation}
\begin{split}
  P \, &  = \; \{ \, P_{0} \, , \, \cdots \, P_{n-1} \} \; n \in
{\mathbb{N}} \, : \\
  P_{i} & \in  HALT(\mu) \; \; i \, = \, 0 \, , \, \cdots  \, , \, n-1 \\
  P_{i} & \, \cap \, P_{j} \, = \, \emptyset \; \; \forall \, i \, \neq \,
  j \\
  \mu  &  ( \, X  \,- \, \cup_{i=0}^{n-1} P_{i} \, ) \; = \; 0
\end{split}
\end{equation}

We will denote the set of all the finite measurable partitions of
$ ( X \, , \, \mu ) $ by $ \mathcal{P} ( \, X \, , \, \mu \, ) $.

Given two partitions $ P = \{ P_{i} \}_{i=0}^{n-1} \, , \,  Q  =
\{ Q_{j} \}_{j=0}^{m-1} \; \in \; \mathcal{P} ( X , \mu ) $:
\begin{definition}
\end{definition}
\emph{P is a coarse-graining of Q  $ (P \preceq Q) $}:

every atom of P is the union of atoms of Q

\smallskip

\begin{definition}
\end{definition}
\emph{coarsest refinement of  $ P = \{ P_{i} \}_{i=0}^{n-1} $ and
$ Q = \{ Q_{j} \}_{j=0}^{m-1} \in {\mathcal{P}}( \; X \, , \mu \;
) $}:
\begin{equation}
  \begin{split}
      P \, & \vee \, Q \; \in {\mathcal{P}}( X  , \mu )  \\
      P \, & \vee \, Q \; := \; \{ \, P_{i} \, \cap \, Q_{j} \, \;  i =0 ,
\cdots , n-1 \; j = 0 , \cdots , m-1  \}
  \end{split}
\end{equation}

Clearly $ \mathcal{P} ( X , \mu ) $ is closed both under coarsest
refinements and under endomorphisms of $ ( X , \mu ) $.

\begin{definition}
\end{definition}
\emph{Shannon probabilistic entropy of $ P = \{ P_{i}
\}_{i=0}^{n-1} \in \mathcal{P} ( X , \mu ) $}:
\begin{equation}
  H(P) \; := \; - \sum_{i=0}^{n-1} \mu(P_{i}) \log_{n} [ \mu(P_{i})]
\end{equation}

Given a classical dynamical system $  CDS \, := \, ( X \, , \, \mu
\, , \, T ) $ \cite{Kolmogorov-58}, \cite{AMS-LMS-00},
\cite{Sinai-76}, \cite{Kornfeld-Sinai-Vershik-00},
\cite{Sinai-94}:

\begin{definition} \label{def:Kolmogorov-Sinai entropy}
\end{definition}
\emph{Kolmogorov-Sinai entropy of CDS}:
\begin{equation}
  h_{CDS} \; := \; sup_{P \in {\mathcal{P}}(X , \mu)} \, lim_{n \rightarrow \infty} \,  \frac{1}{n} \,
H(\vee_{k=0}^{n-1} \, T^{-k} P )
\end{equation}

Let us now briefly recall how the orbits of a classical dynamical
system may be symbolically codified.

Let $ A_{n} := \{ a_{0} , \cdots , a_{n-1} \} $ be an alphabet of
n letters.

Considered a partition $ P \, = \, \{ P_{i} \}_{i = 0}^{n-1} \in
\, {\mathcal{P}}(X , \mu) $:
\begin{definition} \label{def:symbolic translator w.r.t. a partition}
\end{definition}
\emph{symbolic translator of CDS with respect to P}:

$ \psi_{P} \, : \, X  \rightarrow A_{n} $:
\begin{equation}
  \psi_{P} ( x  ) \; := \; string(i)  \, : \, x \in P_{i}
\end{equation}
where $ string(i) $ is the $i^{th}$ letter of $ A_{n} $ in
lexicographic order.

\begin{definition} \label{def:n-point symbolic translator w.r.t. a partition}
\end{definition}
\emph{k-point symbolic translator of CDS with respect to P}:

$ \psi_{P}^{(k)} \, : \, X \, \rightarrow A_{n}^{k} $:
\begin{equation}
   \psi_{P}^{(k)} ( x ) \; := \; \cdot_{j = 0}^{k-1}  \psi_{P} ( T^{j} x )
\end{equation}

\begin{definition} \label{def:orbit symbolic translator w.r.t. a partition}
\end{definition}
\emph{orbit symbolic translator of CDS with respect to P}:

$ \psi_{P}^{(\infty)} \, : \, X \,  \rightarrow \, A_{n}^{\infty}
$:
\begin{equation}
   \psi_{P}^{( \infty)} ( x ) \; := \; \cdot_{j = 0}^{\infty}  \psi_{P} ( T^{j}
x )
\end{equation}

Let us finally introduce the following basic:

\begin{definition}
\end{definition}
\emph{CDS is chaotic:}
\begin{equation}
    \exists P \in \mathcal{P}(X , \mu) \; : \;
    \psi_{P}^{(\infty)} (x) \in RANDOM(A_{|P|}^{\mathbb{N}_{+}}) \;
    \; \forall-\mu-almost \: x \in X
\end{equation}

Then:
\begin{theorem} \label{th:Brudno theorem}
\end{theorem}
\emph{Brudno Theorem:}
\begin{equation}
    CDS \text{ is chaotic } \; \Rightarrow \; h_{CDS} > 0
\end{equation}

\bigskip

\begin{remark}
\end{remark}
Let us remark that, owing essentially to the proposition
\ref{prop:Brudno randomness is weaker than Chaitin randomness},
the fact of having strictly positive Kolmogorov-Sinai entropy,
though being (according to the theorem \ref{th:Brudno theorem}) a
necessary condition for chaoticity, is not a sufficient condition
for chaoticity.

\bigskip

Let us now look at such a dynamical system from a mathematical
logic point of view, id est looking at it as a formal system.

Let us suppose that CDS is chaotic and let $ P \in \mathcal{P} ( X
, \mu ) $ be a partition such that:
\begin{equation} \label{eq:randomness of orbits}
    \psi_{P}^{(\infty)} (x) \in RANDOM(A_{|P|}^{\mathbb{N}_{+}}) \;
    \; \forall-\mu-almost \: x \in X
\end{equation}

Given the initial condition $ x \in X $, the letter $ \psi_{P}(x)
\in A_{|P|}$ is the axiom of such a formal system,

Let us now observe that the equation \ref{eq:randomness of orbits}
implies that:
\begin{equation}
    I( \psi_{P}^{(n)} (x)) \; = \; O_{n \rightarrow + \infty}(n) \; \;
    \; \forall-\mu-almost \: x \in X
\end{equation}

Chaitin's $1^{th}$ Information-theoretic Incompleteness Theorem
(id est the theorem \ref{th:1th Chaitin information-theoretic
incompleteness theorem}) implies that the eventuality  that the
divergence of $ I ( \psi_{P}^{(n)} (x)) $ is faster cannot be
decided:
\begin{corollary} \label{cor:Undecidability underlying chaotic dynamical systems}
\end{corollary}
\emph{Undecidability underlying chaotic dynamical systems:}

\begin{equation}
 \psi_{P}(x) \nvdash_{CDS}  "I( \psi_{P}^{(n)}) (x)
  = o_{n \rightarrow + \infty}(n) "  \; \;   \forall-\mu-almost \:  x \in X
\end{equation}

\bigskip

\begin{remark} \label{rem:independence od the incompleteness explication of chaos from Pesin Theorem}
\end{remark}
Let us remark that such an explanation of the mathematical logic
undecidability  underlying the phenomenon of chaos, holding for
discrete-time dynamical systems, is completely independent from
Pesin Theorem that, for  continuous-time dynamical systems, states
that the Kolmogorov-Sinai entropy is equal to the sum of the
positive Lyapunov exponents
\cite{Bunimovich-Jacobson-Pesin-Sinai-00}.

\bigskip

Also the $ 2^{th}$ Chaitin Information-theoretic Incompleteness
Theorem (id est the theorem \ref{th:2th Chaitin
information-theoretic incompleteness theorem}) affects the Theory
of Dynamical Systems.

To appreciate this fact it is enough to observe that a universal
Chaitin computer U  may be seen, from the viewpoint of the Theory
of Abstract Dynamical Systems, as a particular kind of shift (see
\cite{Wolfram-94}, \cite{Moore-98} and the references therein).

The Undecidability of the Halting Problem (id est the theorem
\ref{th:Chaitin's Theorem on the Halting Problem}) appears then as
a property of such a dynamical system ruled by $ \Omega_{U} $
owing to the theorem  \ref{th:Chaitin's Theorem on the Halting
Problem}.

\newpage
\section{Quantum topological entropy versus quantum algorithmic randomness}
Given the finite alphabet $ A := \{ a_{1} , \cdots , a_{|A|} \} $
\begin{definition}
\end{definition}
\emph{Quantum computational Hilbert space with respect to A:}
\begin{equation}
    \mathcal{H}_{A} \; := \; \mathbb{C}^{|A|}
\end{equation}
\begin{equation}
   \hat{ \mathbb{I}}_{\mathcal{H}_{A}} \; = \; \sum_{ a \in A}  | a > < a |
\end{equation}
\begin{equation}
    < a | b > \; = \; \delta_{a,b} \; \; \forall a , b \in A
\end{equation}
Given $ n \in \mathbb{N}_{+} $:
\begin{definition}
\end{definition}
\emph{n qu-$|A|$-its' Hilbert space:}
\begin{equation}
    \mathcal{H}_{A}^{\otimes n} \; := \; \otimes_{i=1}^{n} \mathcal{H}_{A} \; = \; \mathbb{C}^{|A|^{n}}
\end{equation}
\begin{equation}
    \hat{\mathbb{I}}_{\mathcal{H}_{A}^{\otimes n}} \; := \; \sum_{\vec{x} \in A^{n} } | \vec{x} > < \vec{x} |
\end{equation}
\begin{equation}
    < \vec{x} | \vec{y} > \; = \; \delta_{\vec{x} , \vec{y}} \; \; \forall \vec{x} , \vec{y} \in A^{n}
\end{equation}

The quantum analogue of the set $ A^{+} $ of all strings over A is
the following:
\begin{definition}
\end{definition}
\emph{Hilbert space of quantum strings with respect to A:}
\begin{equation}
    \mathcal{H}_{A}^{\otimes +} \; := \; \oplus_{n \in \mathbb{N}_{+}} \mathcal{H}_{A}^{\otimes n} \; := \; \oplus_{n \in \mathbb{N}_{+}} \mathbb{C}^{|A|^{n}}
\end{equation}
\begin{equation}
    \hat{\mathbb{I}}_{\mathcal{H}_{A}^{\otimes +}} \; := \; \sum_{\vec{x} \in A^{+} } | \vec{x} > < \vec{x} |
\end{equation}
\begin{equation}
    < \vec{x} | \vec{y} > \; = \; \delta_{\vec{x} , \vec{y}} \; \; \forall \vec{x} , \vec{y} \in A^{+}
\end{equation}

The quantum analogue of the set $ A^{\mathbb{N}_{+}} $ of all sequences over A is the following:
\begin{definition}
\end{definition}
\emph{Hilbert space of quantum sequences with respect to A:}
\begin{equation}
    \mathcal{H}_{A}^{\otimes \mathbb{N}_{+}} \; := \; \otimes_{n \in \mathbb{N}_{+}} \mathcal{H}_{A}
\end{equation}
\begin{equation}
    \hat{\mathbb{I}}_{\mathcal{H}_{A}^{\otimes \mathbb{N}_{+}}} \; := \; \int_{A^{\mathbb{N}_{+}}} d \bar{x} | \bar{x} > < \bar{x} |
\end{equation}
\begin{equation} \label{eq:orthonormality condition for quantum sequences}
    < \bar{x} | \bar{y} > \; = \; \delta ( \bar{x} - \bar{y} ) \; \; \forall \bar{x} , \bar{y} \in A^{\mathbb{N}_{+}}
\end{equation}
where $ d \bar{x} $ is the Lebesgue measure over $ A^{\mathbb{N}_{+}} $ and  where the $ \delta $ in the right hand side of \ref{eq:orthonormality condition for quantum sequences}
is the Dirac delta tempered distribution (defined in the appendix \ref{sec:Generalized functions on the space of all sequences over a finite alphabet}).

Given a quantum sequence $ | \psi > \in  \mathcal{H}_{A}^{\otimes
\mathbb{N}_{+}} $ the natural quantum analogue of the definition
\ref{def:combinatorial information function} is  the following:

\begin{definition} \label{def:combinatorial quantum information function}
\end{definition}
\emph{combinatorial quantum information function of $ | \psi > $: }
\begin{equation}
  p_{n} ( | \psi > )  \; := \; \int_{A^{\mathbb{N}_{+}}} d \bar{x}  | < \bar{x} |  \psi > |^{2}  p_{n} ( \bar{x} )
\end{equation}

\begin{remark}
\end{remark}
The definition \ref{def:combinatorial quantum information
function} is nothing but the expectation value of the classical
combinatorial information function when a measurement of the
sequence operator $ \hat{\bar{x}} $ \footnote{where obviously the
sequence operator is defined by:
\begin{equation}
    \hat{\bar{x}} | \bar{x} > \; := \; \bar{x}  | \bar{x} >
\end{equation}}
is performed.

The natural quantum analogue of the definition \ref{def:topological entropy of a sequence} is then the following:

\begin{definition} \label{def:topological quantum entropy of a sequence}
\end{definition}
\emph{topological quantum entropy of $  | \psi > $:}
\begin{equation}
    H_{top} ( | \psi > ) \; := \; \lim_{n \rightarrow + \infty} \frac{ \log_{|A|}(  p_{n} ( | \psi >  ))   }{n}
\end{equation}

Clearly:

\begin{proposition} \label{prop:quantum topological entropy of an element of the computational basis}
\end{proposition}
\begin{equation}
    H_{top} ( | \bar{x} > ) \; = \;  H_{top} (  \bar{x} ) \; \; \forall \bar{x} \in A^{\mathbb{N}_{+}}
\end{equation}
\begin{proof}
It is sufficient to observe that:
\begin{equation}
    p_{n} ( |\bar{x} > ) \; = \; \int_{A^{\mathbb{N}}_{+}} d \bar{y}  | < \bar{y}   | \bar{x} > |^{2}  p_{n}  ( | \bar{y} > ) \; = \;
    \int_{A^{\mathbb{N}}_{+}} d \bar{y} \delta^{2} ( \bar{x} - \bar{y} )   p_{n}  ( | \bar{y} > )  \; = \;  p_{n} ( \bar{x}  )
\end{equation}
\end{proof}

\begin{example}
\end{example}
Given the binary alphabet $ \{ 0 , 1 \} $ let us consider the quantum sequence:
\begin{equation}
    | \psi > \; := \; \frac{1}{\sqrt{2}} ( | 0000000000 \cdots > + | 1111111111 \cdots > )
\end{equation}
Then:
\begin{equation}
    p_{n} (  | \psi >  ) \; = \; \frac{ p_{n} (  |0000000000 \cdots   > ) + p_{n} ( | 1111111111 \cdots > )}{2}
\end{equation}
Since obviously:
\begin{equation}
  \mathcal{L}_{n} (  0000000000 \cdots ) \; = \; \{ \cdot_{i=1}^{n} 0 \}
\end{equation}
\begin{equation}
    p_{n} ( 0000000000 \cdots ) \; = \; 1
\end{equation}
\begin{equation}
  \mathcal{L}_{n} (  1111111111 \cdots ) \; = \; \{ \cdot_{i=1}^{n} 1 \}
\end{equation}
\begin{equation}
    p_{n} ( 1111111111 \cdots ) \; = \; 1
\end{equation}
it follows that:
\begin{equation}
    p_{n} ( | \psi > ) \; = \; 1
\end{equation}
and hence:
\begin{equation}
    H_{top} ( | \psi > ) \; = \; 0
\end{equation}

\bigskip

\begin{example}
\end{example}
Let us consider a gaussian wave-packet centered in the sequence $ | 0000000000 \cdots > $:
\begin{equation}
   | < \bar{x} | \psi > |^{2} \; = \; \exp ( - d^{2} ( \bar{x} ,  0000000000 \cdots  )
\end{equation}
where d is the metric defined by the equation \ref{eq:distance between sequences}.

Then:
\begin{equation}
    p_{n} ( | \psi > ) \; = \; \int_{A^{\mathbb{N}_{+}}} d \bar{x} \exp ( - d^{2} ( \bar{x} ,  | 0000000000 \cdots > )  p_{n} ( | \bar{x} > ) \; = \;
    \sum_{k=1}^{\infty} \int_{ \{ \bar{x} \in A^{\mathbb{N}_{+}} : \vec{x}(k+1) = ( \cdot_{i=1}^{k} 0) \cdot 1 \}  } d \bar{x} \exp ( 2^{- (k+1)^{2}} )  p_{n} ( | \bar{x} >)
\end{equation}

\bigskip

Let us now introduce the following:
\begin{definition}
\end{definition}
\emph{Coleman-Lesniewski operator:}
\begin{equation}
    \hat{\prod}_{RANDOM} \; := \; \int_{A^{\mathbb{N}_{+}}} d \bar{x} \chi_{RANDOM(A^{\mathbb{N}_{+}})} | \bar{x} > < \bar{x} |
\end{equation}
where:
\begin{equation}
    \chi_{S}(x) \; := \; \left\{%
\begin{array}{ll}
    1, & \hbox{if $ x \in S$;} \\
    0, & \hbox{otherwise.} \\
\end{array}%
\right.
\end{equation}
is the characteristic function of a set S.

It may be easily verified that $ \hat{\prod}_{RANDOM} $ is a
projection operator. It appears, then, natural to define the space
of the algorithmically random quantum sequences as the subspace on
which the Coleman-Lesniewski projects:
\begin{definition}
\end{definition}
\emph{subspace of the algorithmically random quantum sequences with respect to A:}
\begin{equation}
    RANDOM (   \mathcal{H}_{A}^{\otimes \mathbb{N}_{+}} ) \; := \;  \hat{\prod}_{RANDOM}  \mathcal{H}_{A}^{\otimes \mathbb{N}_{+}} \; = \; \{ | \psi > \in \mathcal{H}_{A}^{\otimes \mathbb{N}_{+}} :
  \hat{\prod}_{RANDOM} | \psi > = | \psi > \}
\end{equation}

We can then prove the quantum analogue of the proposition
\ref{prop:link between topological entropy and algorithmic
randomness}:

\begin{proposition} \label{prop:link between quantum topological entropy and quantum algorithmic
randomness}
\end{proposition}
\emph{Link between quantum topological entropy and quantum algorithmic
randomness:}
\begin{enumerate}
    \item
\begin{equation}
    H_{top} ( | \psi > ) \, \neq  \, 0 \; \; \forall | \psi > \in  RANDOM ( \mathcal{H}_{A}^{\otimes \mathbb{N}_{+}} )
\end{equation}
    \item
\begin{equation}
  H_{top} (  | \psi > ) \, \neq \, 0  \; \nRightarrow  \; | \psi >  \in RANDOM ( \mathcal{H}_{A}^{\otimes \mathbb{N}_{+}} )
\end{equation}
\end{enumerate}
\begin{proof}
\begin{enumerate}
    \item
Clearly:
\begin{multline}
    p_{n} ( | \psi > ) \; = \; \int_{RANDOM (A^{\mathbb{N}_{+}})} d \bar{x} | < \bar{x} | \psi > |^{2}   p_{n} ( | \bar{x} > ) \; = \\
    \int_{RANDOM (A^{\mathbb{N}_{+}})} d \bar{x} | < \bar{x} | \psi > |^{2} |A|^{n} \, > \, 0 \; \; \forall  | \psi > \in RANDOM ( \mathcal{H}_{A}^{\otimes \mathbb{N}_{+}})
\end{multline}
    \item  the quantum Champerknowne sequence is not algorithmically random:
\begin{equation}
    | \bar{x}_{Champerknowne} > \; \notin \; RANDOM ( \mathcal{H}_{A}^{\otimes \mathbb{N}_{+}})
\end{equation}
Anyway:
\begin{equation}
    H_{top} (  |  \bar{x}_{Champerknowne} > ) \; = \; H_{top} (  \bar{x}_{Champerknowne} ) \; = \; 1
\end{equation}
\end{enumerate}
\end{proof}

\bigskip

\begin{remark}
\end{remark}
The proposition  \ref{prop:link between quantum topological entropy and quantum algorithmic
randomness} shows that the considerations of the remark \ref{rem:positive and negative feature of the combinatorial approach to information} may be
thoroughly extended to the quantum case:

the quantum topological entropy furnishes a measure of the quantum
information contained in a quantum sequence coarser than the one
given by Quantum Algorithmic Information Theory  \cite{Svozil-95},
\cite{Svozil-96}, \cite{Manin-99}, \cite{Vitanyi-99},
\cite{Gacs-01}, \cite{Vitanyi-01},
\cite{Berthiaume-van-Dam-Laplante-01}.

Since no general agreement has been reached in the scientific
community as to the right quantum analogue of the definition
\ref{def:algorithmic information} (see \cite{Segre04b} for a
discussion of the involved issues), the combinatorial approach has
the advantages not to involve Quantum Computability Theory with
all its still not yet settled issues.

\bigskip

\begin{example}
\end{example}
Given $ \Omega_{1} , \Omega_{2} \in RANDOM[(0,1)] \cap
c.e.(\mathbb{R}) $ and a finite alphabet A let us consider the
quantum sequence of qu-$|A|$-its:
\begin{equation}
   |  \psi > \; := \; \frac{1}{\sqrt{2}} (| r_{A}( \Omega_{1} ) >
   + | r_{A}( \Omega_{2} ) > )
\end{equation}
where  $ r_{A}$ is the nonterminating symbolic representation with
respect to A of the definition \ref{def:nonterminating symbolic
representation}.

Obviously:
\begin{equation}
  |  \psi >  \in RANDOM ( \mathcal{H}_{A}^{\mathbb{N}_{+}})
\end{equation}
\newpage
\section{Quantum algorithms of substitution}

Given two finite alphabet A and B and a map $ f : A^{+}
\rightarrow B^{+} $ there exists a canonical way of constructing a
quantum algorithm associated to f as the linear operator $ \hat{f}
: \mathcal{H}_{A}^{\otimes + } \mapsto  \mathcal{H}_{B}^{ \otimes
+ } $  acting as f on the computational basis (about some
remarkable considerations concerning the implementation  of such
an algorithm see the section 4.9.13 "Quantum implementations" of
\cite{Calude-Paun-01}).

Applied in particular to a substitution $ \sigma $ such a strategy
leads to the following:
\begin{definition}
\end{definition}
\emph{quantum algorithm of the first kind associated to $\sigma$:}

the linear operator over $ \mathcal{H}_{A}^{ \otimes + } \cup
\mathcal{H}_{A}^{\otimes \mathbb{N}_{+}}  $ defined by the
following action on the computational bases:
\begin{equation}
    \hat{\sigma} | \vec{x} > \; := \; | \sigma ( \vec{x} ) > \; \; \vec{x} \in A^{+}
\end{equation}
\begin{equation}
    \hat{\sigma} | \bar{x} > \; := \; | \sigma ( \bar{x} ) > \; \; \bar{x} \in A^{\mathbb{N}_{+}}
\end{equation}

We will refer to $ \hat{\sigma} $ as to a \emph{quantum algorithm of  substitution} or, more concisely, as to a \emph{quantum substitution}.

Clearly:
\begin{proposition}
\end{proposition}
\begin{enumerate}
    \item
\begin{equation}
    \hat{\sigma} | \vec{x} >  \; = \; \otimes_{i=1}^{|\vec{x}| } | \sigma ( x_{i} ) > \; \; \forall \vec{x} \in A^{+}
\end{equation}
   \item
\begin{equation}
  \hat{\sigma} | \bar{x} >  \; = \;  \otimes_{n \in \mathbb{N}_{+}} | \sigma ( x_{n} )
  > \; \; \forall \bar{x} \in A^{\mathbb{N}_{+}}
\end{equation}
  \item
\begin{equation*}
 \text{$ \hat{\sigma} $ is not self-adjoint}
\end{equation*}
  \item
\begin{equation}
  \hat{\sigma} |_{\mathcal{H}_{A}^{\otimes +}} \; = \; \sum_{\vec{x} \in A^{+}
  } | \sigma ( \vec{x} ) > < \vec{x} |
\end{equation}
\end{enumerate}
\begin{proof}
\begin{enumerate}
 \item It is sufficient to observe that:
\begin{equation}
  \hat{\sigma} | \vec{x} > \; = \; | \sigma ( \vec{x} ) > \; = \; | \sigma( \cdot_{i=1}^{ | \vec{x} | } x_{i} ) > \; = \;
   | \cdot_{i=1}^{ | \vec{x} | } \sigma ( x_{i} ) > \; = \; \otimes_{i=1}^{|\vec{x}| } | \sigma ( x_{i} )
   > \; \; \forall \vec{x} \in A^{+}
\end{equation}
  \item In an analogous way:
\begin{equation}
  \hat{\sigma} | \bar{x} > \; = \; | \sigma ( \bar{x} ) > \; = \; | \sigma( \cdot_{n \in \mathbb{N}} x_{n} ) > \; = \;
   | \cdot_{n \in \mathbb{N}_{+} } \sigma ( x_{n} ) > \; = \; \otimes_{ n \in \mathbb{N}_{+}  } | \sigma ( x_{n} )
   > \; \; \forall \bar{x} \in A^{\mathbb{N}_{+}}
\end{equation}
 \item It is sufficient to observe that:
 \begin{equation}
    < \vec{x} | \hat{\sigma} | \bar{y} > \; = \; \delta_{\vec{x}, \sigma(
    \vec{y})} \; \neq \; < \vec{y} |  \hat{\sigma} | \vec{x} > \;
    = \; \delta_{\vec{y}, \sigma ( \vec{x} ) }
\end{equation}
 \item Clearly:
 \begin{multline}
   \hat{\sigma}|_{\mathcal{H}_{A}^{\otimes +}}  \; = \; \sum_{\vec{x} \in A^{+}} | \vec{x} > <
   \vec{x} | \hat{\sigma}  \sum_{\vec{y} \in A^{+}} | \vec{y} > <
   \vec{y} | \; = \\
    \sum_{\vec{x} \in A^{+}}  \sum_{\vec{y} \in
   A^{+}} < \vec{x} |  \hat{\sigma} | \vec{y} > | \vec{x} > <
   \vec{y} | \; = \;  \sum_{\vec{x} \in A^{+}}  \sum_{\vec{y} \in
   A^{+}} \delta_{\vec{x}, \sigma ( \vec{y}) } | \vec{x} > <
   \vec{y} |
\end{multline}
from which the thesis follows.
\end{enumerate}
\end{proof}

\smallskip

Let us now suppose that $ a \in A $ is such that $ | \sigma (a) | \geq 2 $. Denoted as usual with $ \bar{\sigma} (a)   $ the fixed point of
of $ \sigma $ starting with a let us observe that:
\begin{proposition}
\end{proposition}
\begin{equation}
    \hat{\sigma} |  \bar{\sigma} (a) > \; = \;  |  \bar{\sigma} (a) >
\end{equation}
\begin{proof}
  It is sufficient to observe that:
\begin{equation}
  \hat{\sigma} |  \bar{\sigma} (a) > \; = \; | \sigma [ \bar{\sigma} (a) ] > \; = \;  |  \bar{\sigma} (a) >
\end{equation}
\end{proof}

Let us introduce also the following:

\begin{definition}
\end{definition}
\emph{symmetric state of n qu-$|A|$-its:}
\begin{equation}
    | S , n > \; := \; \frac{1}{ \sqrt{ | A |^{n}  } } \sum_{\vec{x} \in
    A^{n}} | \vec{x} >
\end{equation}

The Mathematica 5  notebook of the section \ref{sec:Mathematica
implementation of this paper} may be used to compute the action of
quantum substitutions on symmetric states.

\begin{example}
\end{example}
Let $ \sigma $ denote the Fibonacci substitution of the example
\ref{ex:Fibonacci substitution}. Then:
\begin{equation}
   \hat{ \sigma } | S , 1 > \; = \; \hat{ \sigma } [
   \frac{1}{\sqrt{2}} ( | 0 > + | 1 > ) ] \; = \;
   \frac{1}{\sqrt{2}} ( | 0 1 > + | 0 > )
\end{equation}
\begin{equation}
  \hat{ \sigma }^{2} | S , 1 > \; = \; \frac{1}{\sqrt{2}} ( | 0 1
  0 > \, + \, | 0 1 > )
\end{equation}
\begin{equation}
    \hat{ \sigma }^{3} | S , 1 > \; = \; \frac{1}{\sqrt{2}} ( | 0
    1 0 0 1 > \, + \,  |0 1 0 > )
\end{equation}
and in general:
\begin{equation}
  \hat{ \sigma }^{n} | S , 1 > \; = \; \frac{1}{\sqrt{2}} ( |
  \sigma^{n}(0) > \, + \, |
  \sigma^{n-1}(0) >  ) \; \; \forall n \in \mathbb{N}_{+}
\end{equation}
and hence:
\begin{equation}
    \lim_{n \rightarrow + \infty} \hat{ \sigma }^{n} | S , 1 > \; =
    \;  \frac{1}{\sqrt{2}} ( | \bar{\sigma}(0) > +  | \bar{\sigma}(0)
    > ) \; = \; \sqrt{2} | \bar{\sigma}(0) >
\end{equation}

 Furthermore:
\begin{equation}
  \hat{ \sigma } | S , 2 > \; = \; \hat{ \sigma } \frac{1}{2} ( |
  0 0 > + | 0 1 > + | 1 0 > + | 1 1 > ) \; = \; \frac{1}{2} ( | 0 1
  0 1 > + | 0 1 0 > + | 0 0 1 > + | 0 0 > )
\end{equation}
\begin{equation}
    \hat{ \sigma }^{2} | S , 2 > \; = \; \frac{1}{2} ( |0 1 0 0 1
    0 > + | 0 1 0 0 1 > + | 0 1 0 1 0 > + | 0 1 0 1 > )
\end{equation}
\begin{equation}
   \hat{ \sigma }^{3} | S , 2 > \; = \;  \frac{1}{2} ( |0 1 0 0 1
   0 1 0 0 1 > + | 0 1 0 0 1 0 1 0 > + | 0 1 0 0 1 0 0 1 > + |0 1
   0 0 1 0 > )
\end{equation}

\bigskip

\begin{remark}
\end{remark}
The quantum algorithms of substitution of the first kind have the
great disadvantage of loosing the connection with the powerful and
beautiful Mathematics of Pisot-Vijayaraghavan numbers underlying
the incidence matrix and are of doubtful practical utility.

This suggests the introduction of quantum algorithms of
substitution of the second kind.

\bigskip

Given a substitution $ \sigma $ :

\begin{definition}
\end{definition}
\emph{quantum algorithm of the second kind associated to $
\sigma$:}
\begin{center}
 the linear operator $ \hat{\sigma} $ represented by $ M_{\sigma}
 $ in the computational basis of $ \mathcal{H}_{A} $.
\end{center}

We will refer again to $ \hat{\sigma} $ as to a quantum algorithm
of substitution or, more concisely, as to a quantum substitution.

Then:
\begin{proposition}
\end{proposition}

\begin{hypothesis}
\end{hypothesis}
\begin{center}
    $ \sigma $ Pisot substitution with leading eigenvalue $
    \lambda$
\end{center}
\begin{equation}
    a \in A \; : \; | \sigma (a) | \geq 2
\end{equation}
\begin{center}
  $ \hat{\sigma} $ quantum algorithm of the second kind associated to $
  \sigma$
\end{center}
\begin{thesis}
\end{thesis}
 \begin{equation}
   \lim_{n \rightarrow + \infty} \hat{\sigma}^{n}  \; = |
   e_{\lambda} > < e_{\lambda} |
\end{equation}
where:
\begin{equation}
\hat{\sigma}  | e_{\lambda} > \; = \; \lambda | e_{\lambda} >
\end{equation}
\begin{equation}
    < e_{\lambda} |  e_{\lambda} > \; = \; 1
\end{equation}
\begin{proof}
Since:
\begin{equation}
    \hat{\sigma} \; = \; \lambda | e_{\lambda} > < e_{\lambda} | +
    \sum_{y \in Con(\lambda) } y | e_{y} > < e_{y} |
\end{equation}
it follows that:
\begin{equation}
    \hat{\sigma}^{n} \; = \; \lambda^{n}  | e_{\lambda} > < e_{\lambda}
    | + \sum_{y \in Con(\lambda) } y^{n} | e_{y} > < e_{y} |
\end{equation}
Since $ \lambda \in PV(\mathbb{A}) $:
\begin{equation}
  \lim_{n \rightarrow + \infty}  \sum_{y \in Con(\lambda) } y^{n} | e_{y} > < e_{y}
  | \; = \; 0
\end{equation}
from which the thesis follows.
\end{proof}

\bigskip

\begin{example} \label{ex:quantum Fibonacci substitution}
\end{example}
The quantum algorithm of the second kind associated to the
Fibonacci substitution is clearly the operator $ \hat{\sigma } :
\mathcal{H}_{\{0,1\}} \mapsto \mathcal{H}_{\{0,1\}} $ such that:
\begin{equation}
    \hat{\sigma} | 0 > \; = \; | 0 > +  | 1 >
\end{equation}
\begin{equation}
    \hat{\sigma} | 1 > \; = \;  | 0 >
\end{equation}
From the example \ref{ex:Fibonacci substitution} we may
immediately infer that:
\begin{equation}
  \hat{\sigma} \; = \; \tau | e_{\tau} > < e_{\tau} | + ( -
  \frac{1}{\tau} )   | e_{\frac{-1}{\tau}} > < e_{\frac{-1}{\tau}} |
\end{equation}
\begin{equation}
  | e_{\tau} > \; = \; \frac{\tau }{\sqrt{\tau+2} } | 0 > + \frac{1 }{\sqrt{\tau +2}
  } | 1 >
\end{equation}
\begin{equation}
    | e_{- \frac{1}{\tau}} > \; = \; - \frac{1}{\tau} \sqrt{ \frac{ \tau +1 }{ \tau +2 }
    } | 0 > + \sqrt{ \frac{ \tau +1 }{ \tau +2 }
    } | 1 >
\end{equation}
(where we have used the fact that $ \tau^{2} \; = \; \tau + 1 $)
so that the probability that a measurement of the qubit operator
in the state $ \lim_{n \rightarrow + \infty} \hat{\sigma}^{n} | 0
> = | e_{\tau} > $  gives as result zero is:
\begin{equation}
    Pr_{ \lim_{n \rightarrow + \infty} \hat{\sigma}^{n} | 0 > }( 0 ) \; = \; \frac{\tau^{2} } {  \tau + 2 }
\end{equation}
while the probability that such a measurement gives as result one
is:
\begin{equation}
    Pr_{ \lim_{n \rightarrow + \infty} \hat{\sigma}^{n} | 0 >  }( 1 ) \; = \; \frac{1} { \tau + 2 }
\end{equation}

\smallskip

\begin{example} \label{ex:the quantum Padovan substitution}
\end{example}

The quantum algorithm of second kind associated to the Padovan
substitution $ \sigma $ is clearly the linear operator $
\hat{\sigma} : \mathcal{H}_{\{0,1,2\}} \mapsto
\mathcal{H}_{\{0,1,2\}} $ such that:
\begin{equation}
   \hat{ \sigma} |0 > \; := | 1 > + | 2 >
\end{equation}
\begin{equation}
 \hat{ \sigma}  |1 > \; := | 2 >
\end{equation}
\begin{equation}
  \hat{\sigma} |2 > \; := \; | 0 >
\end{equation}

From the example \ref{ex:Fibonacci substitution} we may
immediately infer that:
\begin{equation}
    \hat{\sigma} | e_{\rho} > \; = \; \rho | e_{\rho} >
\end{equation}
\begin{equation}
  | e_{\rho} > \; = \; \frac{\rho^{2}-1}{\sqrt{( \rho^{2}-1)^{2}+(1+\rho - \rho^{2})^{2}+1 } } |0 >
  + \frac{1 + \rho - \rho^2}{\sqrt{( \rho^{2}-1)^{2}+(1+\rho - \rho^{2})^{2}+1 } } | 1 > + \frac{1}{\sqrt{( \rho^{2}-1)^{2}+(1+\rho - \rho^{2})^{2}+1 } } | 2 >
\end{equation}
(where we have used the fact that $ \rho^{3} \; = \; \rho + 1 $)
so that the probability distribution of the outcome of a
measurement of the qutrit operator in the state $  \lim_{n
\rightarrow + \infty} \hat{\sigma}^{n} | 0 >  \; = \;  | e_{\rho}
> $ is:
\begin{equation}
    Pr_{ \lim_{n
\rightarrow + \infty} \hat{\sigma}^{n} | 0 >  }( 0 ) \; = \;
\frac{(\rho^{2}-1)^{2}}{( \rho^{2}-1)^{2}+(1+\rho -
\rho^{2})^{2}+1  }
\end{equation}
\begin{equation}
    Pr_{\lim_{n
\rightarrow + \infty} \hat{\sigma}^{n} | 0 >   }( 1 ) \; = \;
\frac{(1 + \rho - \rho^2)^{2}}{( \rho^{2}-1)^{2}+(1+\rho -
\rho^{2})^{2}+1 }
\end{equation}
\begin{equation}
  Pr_{\lim_{n
\rightarrow + \infty} \hat{\sigma}^{n} | 0 >  }( 2 ) \; = \;
\frac{1}{( \rho^{2}-1)^{2}+(1+\rho - \rho^{2})^{2}+1 }
\end{equation}

\bigskip

\begin{example} \label{ex:quantum Pell substitution}
\end{example}
The quantum algorithm of the second kind associated to the Pell
substitution is clearly the operator $ \hat{\sigma } :
\mathcal{H}_{\{0,1\}} \mapsto \mathcal{H}_{\{0,1\}} $ such that:
\begin{equation}
    \hat{\sigma} | 0 > \; := \; | 0 > + | 1 >
\end{equation}
\begin{equation}
    \hat{\sigma} | 1 > \; := \; 2 | 0 > + | 1 >
\end{equation}
From the example \ref{ex:Pell substitution} we immediately infer
that:
\begin{equation}
    \hat{\sigma} \; = \; ( 1 + \sqrt{2} ) | e_{1+\sqrt{2}} > < e_{1+\sqrt{2}} | + ( 1 - \sqrt{2} ) | e_{1-\sqrt{2}}
    > < e_{1-\sqrt{2}} |
\end{equation}
\begin{equation}
     | e_{1+\sqrt{2}} > \; = \; \sqrt{\frac{2}{3}} | 0 > +
     \frac{1}{\sqrt{3}} | 1 >
\end{equation}
\begin{equation}
     | e_{1 - \sqrt{2}} > \; = \; - \sqrt{\frac{2}{3}} | 0 > +
     \frac{1}{\sqrt{3}} | 1 >
\end{equation}
so that the probability distribution of the outcome of a
measurement of the qubit operator in the state $ \lim_{n
\rightarrow + \infty} \hat{\sigma}^{n} | 0 > \, = \, |
e_{1+\sqrt{2}} >   $ is:
\begin{equation}
    Pr_{\lim_{n
\rightarrow + \infty} \hat{\sigma}^{n} | 0 > } (0) \; = \;
\frac{2}{3}
\end{equation}
\begin{equation}
    Pr_{\lim_{n
\rightarrow + \infty} \hat{\sigma}^{n} | 0 > } (1) \; = \;
\frac{1}{3}
\end{equation}

\bigskip

\begin{remark}
\end{remark}
Quantum algorithms mimicking the recursive structure of Fibonacci
numbers have been introduced in
\cite{Elias-Fernandez-Mor-Weinstein-07}. In our modest opinion it
doesn't seem, anyway, to be any connection between those
remarkable algorithms and the quantum substitutions of Pisot type
discussed in this paper.
\newpage

\bigskip

\begin{remark}
\end{remark}
The idea of taking into account  suitable (specifically almost
periodic Schr\"{o}dinger)  operators associated  to substitutions
has been already considered in the Mathematical-Physics'
literature concerning quasicrystals  (see for instance
\cite{Damanik-00}, \cite{Damanik-07} and references therein).

In order to allow the reader to appreciate the possible conceptual
connections a brief review of the basic notions is presented in
the section \ref{sec:Quasicrystals}.

From the mathematical side the great achievement obtained by Barry
Simon and coworkers, namely the discovery that almost periodic
Schr\"{o}dinger  operators has commonly continuous spectral
measures supported on Cantor sets \cite{Last-07} has never been
looked from an information theoretic perspective.

As  we have already observed  in the remark \ref{rem:link to
almost periodicity}, this would be a necessary step in order to
generalize the Calude-Chitescu Theorem (id est the theorem
\ref{th:Calude-Chitescu}) clarifying the algorithmic information
theoretic  status of uniformly recurrent sequences.

The attempt to make some step forward in this direction  is
presented in the section \ref{sec:A brief information theoretic
analysis of singular Lebesgue-Stieltjes measures supported on
Cantor sets and almost periodicity}.

Remembering, furthermore, that an almost periodic function $ f:
\mathbb{R}^{D} \mapsto \mathbb{C} $ can be seen as an ergodic
stationary random function over the probability space $ ( \Gamma ,
Hull(\Gamma) , \mu_{Haar}( \Gamma ))$, $ \Gamma := \overline{
\mathcal{T}_{\vec{x}} f}  $ being the closure of all the shifts of
f (where of course $ \mathcal{T}_{\vec{x}}f(\vec{y}) := f(
\vec{x}+ \vec{y})$) (see the section 7.2 "Periodic and Almost
Periodic Potentials" of \cite{Carmona-Lacroix-90}) such a task
could be part of a more ambitious ompletely unexplored research
project investigating algorithmically-random Schr\"{o}dinger
operators that, anyway, is far beyond the purposes of this paper.

\bigskip

The natural analogue of the definition \ref{def:topological
entropy of a substitution} for the quantum algorithm of first kind
associated to a substitution is the following:
\begin{definition} \label{def:quantum topological entropy of a substitution}
\end{definition}
\emph{quantum topological entropy of $ \hat{\sigma}$:}
\begin{equation}
    H_{top} ( \hat{\sigma} ) \; := \; \sum_{a \in A \, : \, | \sigma(a) | \geq 2} \max \{ H_{top} ( | \psi > ) \: , \: | \psi > \in \mathcal{H}_{A}^{\otimes \mathbb{N}_{+}} \, : \, < \bar{\sigma}(a) | \psi > \neq 0 \}
\end{equation}

\bigskip

\begin{remark} \label{rem:about the Quantum Brudno Theorem}
\end{remark}
Quantum analogues of the theorem \ref{th:link between the
probabilistic information and the algorithmic information}
expressing the link between the quantum probabilistic information
(id est Von Neumann entropy) and the quantum algorithmic
information have been established in the same papers in which this
latter notion has been introduced \cite{Svozil-96},
\cite{Manin-99}, \cite{Vitanyi-99}, \cite{Gacs-01},
\cite{Vitanyi-01}, \cite{Berthiaume-van-Dam-Laplante-01}.

After having defined chaotic a quantum dynamical system whose
orbits, simbolically codified in a suitable way, are
algorithmically-random, as we have observed in the remark
\ref{rem:link between the probabilistic information and the
algorithmic information} the next conceptual step would then
consists in its generalization to arbitrary quantum stochastic
processes and then, through a suitable symbolic codification, to
arbitrary quantum dynamical systems resulting in a quantum
analogue of the Brudno Theorem (id est the theorem \ref{th:Brudno
theorem}) stating that the vanishing of the quantum dynamical
entropy of a quantum dynamical system (defined as the asymptotic
rate of production of probabilistic information of such a
dynamical system) is a necessary condition for chaoticity also at
the quantum level.

The first problem, with this regard, consists in the fact that, according to whether or not one assumes that measurements are performed on the quantum dynamical system
during its dynamical evolution, one results in different notions of quantum dynamical entropy: respectively the \emph{Connes-Narnhofer-Thirring entropy} \cite{Connes-Narnhofer-Thirring-87}, \cite{Benatti-93}, \cite{Stormer-02}, \cite{Neshveyev-Stormer-06}
and the \emph{Alicki-Lindblad-Fannes entropy} \cite{Alicki-Fannes-01}.

A  remarkable result going in the right direction has been
obtained in
\cite{Benatti-Kruger-Muller-Siegmund-Schultze-Szkola-06} though,
in our modest opinion, it is not yet a Quantum Brudno Theorem,
since it in no way takes into account individual trajectories and
their eventual quantum algorithmic randomness that, according to
the analysis presented in the section \ref{sec:Chaos as a
phenomenon of undecidability}, is the key point as to the
underlying undecidability phenomenon.

The definition \ref{def:quantum topological entropy of a
substitution}, involving only the combinatorial approach to
quantum information, might be a tool useful in order to proceed in
the indicated direction.

\newpage
\section{A quantum algorithm for optimal spacing}
In the section \ref{sec:Substitutions of Pisot type as algorithms
for optimal spacing} we have seen that the optimization problem
consisting in  allocating $ n \in \mathbb{N}_{+} : n \geq 2 $
petals of a flower so that:
\begin{enumerate}
    \item the average distance between neighbors is maximal in order to maximize the exposition of each
petal to sun and rain.
    \item the distribution of the distances among neighbour petals is the more uniform one.
\end{enumerate}
has as solutions the $n^{th}$-roots of unity:
\begin{equation}
    r_{n,k} \; := \; e^{ i \frac{2 \pi}{n} k } \; \; k = 1 ,
    \cdots , n
\end{equation}

Given a generic finite alphabet A and $ n \in \mathbb{N}_{+} $
\cite{Nielsen-Chuang-00}:
\begin{definition}
\end{definition}
\emph{quantum Fourier transform on n qu-$|A|$-its:}

the linear map $ \hat{F}  $  over $ \mathcal{H}_{A}^{\otimes n} $
defined by its matrix's elements in the computational basis:
\begin{equation}
   < \vec{y}| \hat{F} | \vec{x} > \; := \; \frac{1}{\sqrt{| A|^{n}}
    } \, r_{|A|^{n} , l_{loc}(\vec{x}) l_{loc}(\vec{y} )}  \; \;
    \forall \vec{x} , \vec{y} \in A^{n} , \forall n \in \mathbb{N}_{+}
\end{equation}

where $ l_{loc} $ denotes the \emph{local lexicographic number} on
$ A^{n} $ defined as:
\begin{equation}
  l_{loc} ( \cdot_{k=1}^{n} a_{1} ) \; = \; 0
\end{equation}
\begin{equation*}
    \vdots \vdots \vdots \vdots
\end{equation*}
\begin{equation*}
    \vdots \vdots \vdots \vdots
\end{equation*}
\begin{equation}
 l_{loc} ( \cdot_{k=1}^{n} a_{|A|} ) \; = \; |A|^{n} - 1
\end{equation}

Let us face, anyway, the more interesting optimization problem in
which the number of petals is countable.

Let us now suppose to have a substitution $ \sigma $ of Pisot type
over the binary alphabet $ \{ 0 , 1 \} $  such that $ | \sigma(0)
| \geq 2 $  and let us call again $ \lambda $ the leading
eigenvalue of its incidence matrix.

Let suppose to choose two fixed angles $ \beta_{(0)} , \beta_{(1)}
\in [ 0 , 2 \pi ) $ such that:
\begin{equation}
    Mod_{2 \pi} [ \frac{\beta_{(0)}}{\beta_{(1)}} ] \; = \; \lambda
\end{equation}

Let us then construct our flower according to the following
algorithmic procedure:
\begin{enumerate}
     \item set n=0
     \item label[start]
     \item  prepare the state $ | 0 > $ on the $1^{th}$ register
     \item apply  n times the operator $ \hat{\sigma} $ on the $1^{th}$ register
     \item copy the status of the $1^{th}$ register on the $2^{th}$ register
    \item  perform on the $2^{th}$ register  a measurement of the
    qubit operator
    \item pose the $ n^{th}$ petal at the angle:
\begin{equation}
    \theta_{n} =  \left\{%
\begin{array}{ll}
    Mod_{2 \pi} ( \theta_{n-1} +  \beta_{(0)} )   , & \hbox{if the result of the measurement was 0}  \\
    Mod_{2 \pi} ( \theta_{n-1} +  \beta_{(1)} )   , & \hbox{if the result of the measurement was 1} \\
\end{array}%
\right.
\end{equation}
  \item increase n of one unit
   \item goto the label[start]
\end{enumerate}

It is important to stress that the fifth step of such an algorithm
doesn't violate the No Cloning Theorem as one can infer taking
into account the precise formulation of such a theorem
\cite{Nielsen-Chuang-00}:

\begin{theorem} \label{th:No Cloning Theorem}
\end{theorem}
\emph{No Cloning Thorem:}

\begin{hypothesis}
\end{hypothesis}
\begin{center}
 $ \mathcal{H} $ Hilbert space
\end{center}

\begin{thesis}
\end{thesis}
\begin{equation}
    \nexists \hat{U} : \mathcal{H} \mapsto \mathcal{H} \otimes
    \mathcal{H} \text{ unitary } \; : \; \hat{U} | \psi_{1} > = |
    \psi_{1}> \otimes | \psi_{1}> \, \wedge \, \hat{U} | \psi_{2} > = |
    \psi_{2}> \otimes | \psi_{2}> \, \wedge \, | \psi_{1} > \neq | \psi_{2} > \, \wedge \ < \psi_{1} |
    \psi_{2} > \neq 0
\end{equation}
\begin{proof}
Let us suppose ad absurdum that such a unitary operator exists.
Then:
\begin{equation}
    < \psi_{1} | \hat{U}^{\dag} \hat{U} |  \psi_{2} > \; = (  < \psi_{1}
    | \otimes < \psi_{1}
    |) ( | \psi_{2} > \otimes | \psi_{2} > ) \; = ( < \psi_{1} |
    \psi_{2} > )^{2}  \; = \; < \psi_{1} | \psi_{2} >
\end{equation}
and hence:
\begin{equation}
   < \psi_{1} | \psi_{2} > \; = \; 0
\end{equation}
contradicting the hypothesis that the states $ | \psi_{1} > $ and
$ | \psi_{2} > $ are non-orthogonal.
\end{proof}

Theorem \ref{th:No Cloning Theorem} forbids the simultaneous
cloning of non-orthogonal states, but it doesn't forbids the
simultaneous cloning of distinct orthogonal states.

Given a finite alphabet A and considered the cloning function $
f_{clone} : A^{+} \mapsto  A^{+} $ one can immediately introduce
the quantum algorithm $ \hat{f}_{clone} : \mathcal{H}_{A}^{\otimes
+} \mapsto \mathcal{H}_{A}^{\otimes +} $ acting as $ f_{clone} $
on the computational basis:
\begin{equation}
  \hat{f}_{clone} | \vec{x} > \; := \; | \vec{x} \cdot \vec{x} > \; \;
  \vec{x} \in A^{+}
\end{equation}
perfectly compatible with the theorem \ref{th:No Cloning Theorem}.

\bigskip

The presented algorithm reaches asymptotically the optimal
solution.

\newpage
\appendix
\section{Generalized functions on the space of all sequences over a finite alphabet} \label{sec:Generalized functions on the space of all sequences over a finite alphabet}
Given a topological space $ ( X , \mathcal{T} ) $  \cite{Bourbaki-89} and a subset of its $ S \subset X $:

\begin{definition}
\end{definition}
\emph{set of the limit points of S:}
\begin{equation}
    LP ( S) \; := \; \{ x \in X \; : \; ( O \in \mathcal{T} \wedge x \in O) \; \Rightarrow \; O \cap S \neq \emptyset \}
\end{equation}

\begin{definition} \label{def:Cantor set}
\end{definition}
\emph{S is a Cantor set:}
\begin{equation*}
   ( \bar{S} = S ) \; \wedge \; ( S^{\circ} = \emptyset ) \; \wedge \; ( LP(S) \; = \; S )
\end{equation*}
where $ \bar{S} $ is the closure of S while $  S^{\circ} $ is its interior.

Given a finite alphabet $ A = \{ a_{1} , \cdots a_{|A|} \} $ (id
est a set A such that the cardinality of A $ | A| \in
\mathbb{N}_{+} $) we will denote, following the notation of
\cite{Calude-02}, \cite{Fogg-02}, \cite{Allouche-Shallit-03} and
\cite{Lothaire-05}, by $ A^{+} := \cup_{n \in \mathbb{N}_{+}}
A^{n} $ the free semi-group generated by A, id est the set of all
the finite words over A.

Given $ \vec{x} , \vec{y} \in A^{+} $ let us denote by $ \vec{x}
\cdot \vec{y} $ the concatenation of $ \vec{x} $ and $ \vec{y} $,
id est the string $ ( x_{1} , \cdots , x_{|\vec{x} |}, y_{1} ,
\cdots ,  y_{| \vec{y} |} ) $ where $ | \vec{x} | $ denotes the
length of the strings $ \vec{x} $.

Introduced the set $ A^{\mathbb{N}_{+}}$ of the \emph{sequences
over A} let us endow A with the discrete topology, and let us
endow $ A^{+} $  and $ A^{\mathbb{N}_{+}}$  with the product
topology.

Such a topology over  $ A^{\mathbb{N}_{+}}$ is the metric topology induced by the following distance:
\begin{equation}
    d ( \bar{x} , \bar{y} ) \; = \; \left\{%
\begin{array}{ll}
    0, & \hbox{if $ \bar{x} = \bar{y} $;} \\
    \frac{1}{2^{ \min \{ n \in \mathbb{N}_{+} \, : \, x_{n} \neq y_{n} \} }  }, & \hbox{otherwise.} \\
\end{array}%
\right.
\end{equation}

The Borel-$\sigma$ algebra $ \mathcal{B} ( A^{\mathbb{N}_{+}} ) $  associated to such a topology
is the $ \sigma$-algebra generated by the cylinder sets:
\begin{equation}
    C_{\vec{x}} \; := \; \{ \bar{y} \in A^{\mathbb{N}_{+}} \, : \, \vec{y}(| \vec{x} |) = \vec{x} \} \; \; \vec{x} \in A^{+}
\end{equation}
where $ \vec{y}( n) $ denotes the prefix of length n of the sequence $ \bar{y} $.

We can then introduce the following:
\begin{definition} \label{def:Lebesgue measure on sequences}
\end{definition}
\emph{Lebesgue measure associated to A:}

the probability measure $ \mu_{Lebesgue_{A}} $ over the measurable
space $ ( A^{\mathbb{N}_{+}} , \mathcal{B} ( A^{\mathbb{N}_{+}} )
) $ such that:
\begin{equation}
  \mu_{Lebesgue,A} ( C_{\vec{x}} ) \; := \; \frac{1}{|A|^{|\vec{x} | }}
\end{equation}

Given $ f \in L^{1} ( A^{\mathbb{N}_{+}} , d \mu_{Lebesgue_{A}} )
$ let us pose $ d \bar{x} : = d \mu_{Lebesgue,A} $ and hence:
\begin{equation}
    \int_{A^{\mathbb{N}_{+}}} d \bar{x} f ( \bar{x} ) \; := \;  \int_{A^{\mathbb{N}_{+}}}  d \mu_{Lebesgue_{A}} ( \bar{x}) f ( \bar{x} )
\end{equation}

The applicability of Probability Theory to the real world lies,
from a foundational perspective, on the following:
\begin{theorem} \label{th:On the Foundation of Probability Theory}
\end{theorem}
\emph{On the Foundation of Probability Theory:}
\begin{equation}
    \mu_{Lebegue,A} [ RANDOM ( A^{\mathbb{N}_{+}} ) ] \; = \; 1 \;
    \; \forall A :  2 \leq  |A | < \aleph_{0}
\end{equation}

\bigskip

Given a sequence $ \bar{x} \in A^{\mathbb{N}_{+}}$:
\begin{definition} \label{def:numerical value of a sequence}
\end{definition}
\emph{numerical value of $ \bar{x} $:}
\begin{equation}
    v_{A} ( \bar{x} ) \; := \; \sum_{n=1}^{\infty} \frac{lex( x_{n}) }{ |A|^{n} }
\end{equation}

\begin{remark}
\end{remark}
Let us remark that the map $ v : A^{\mathbb{N}_{+}} \mapsto [ 0 , 1 ] $ is surjective but it is not injective since:

\begin{proposition} \label{prop:non injectivity of the numerical value map}
\end{proposition}
\begin{equation}
 v_{A} ( \vec{x} \cdot a_{i} \cdot_{k= | \vec{x} |+2}^{\infty} a_{|A|}) \; = \;
 v_{A} ( \vec{x} \cdot a_{i+1} \cdot_{k= | \vec{x} |+2}^{\infty}  a_{1} )  \; \; \forall \vec{x} \in A^{+} , \forall i \{ 1 , \cdots, |A| -1 \}
\end{equation}

\begin{example}
\end{example}
Let us suppose that $ A := \{ 0 ,1,2,3,4,5,6,7,8,9 \} $. Then we are taught from childhood that:
\begin{equation}
    0. \bar{9} \; = \; 1
\end{equation}
\begin{equation}
    0. 1 \bar{9} \; = \; 0.2
\end{equation}
\begin{equation}
    0. 2 \bar{9} \; = \; 0.3
\end{equation}
and so on.

\smallskip

 Let us then introduce the following:
\begin{definition} \label{def:nonterminating symbolic representation}
\end{definition}
\emph{nonterminating symbolic representation with respect to A:}

the map $ r_{A} : [ 0 , 1 ] \mapsto  A^{\mathbb{N}_{+}} $:
\begin{equation}
    r_{A} ( \sum_{n=1}^{\infty} \frac{ lex(x_{n}) }{ | A|^{n} } ) \; :=
    \; \cdot_{n \in \mathbb{N}_{+}} x_{n}
\end{equation}
with the nonterminating choice in the cases of the  proposition
\ref{prop:non injectivity of the numerical value map}.

Given two finite alphabets $ A_{1} $ and  $ A_{2} $ we can then
introduce the following:
\begin{definition} \label{def:alphabet's transition map}
\end{definition}
\emph{alphabet's transition map  from $ A_{1}$ to $ A_{2}$:}

the map $ T_{A_{1},A_{2}} :  A_{1}^{\mathbb{N}_{+}} \mapsto
A_{2}^{\mathbb{N}_{+}} $:
\begin{equation}
  T_{A_{1},A_{2}}  ( \bar{x} ) \; = \; r_{A_{2}} [  v_{A_{1}} ( \bar{x}
  ) ]
\end{equation}

\bigskip

The structure of topological space that we have given to $
A^{\mathbb{N}_{+}} $ is sufficient, as it is well known, to define
limits and derivatives.

Given a smooth (id est infinitely differentiable) map $ f :
A^{\mathbb{N}_{+}} \mapsto \mathbb{C} $ and  $ n,m \in \mathbb{N}
$:
\begin{definition}
\end{definition}
\begin{equation}
    \| f \|_{n,m} \; := \; \sup_{ \bar{x} \in A^{\mathbb{N}_{+}}  } | (v_{A}(\bar{x}))^{n} \frac{d^{m}}{d x^{m}   } f ( v_{A}(\bar{x})) |
\end{equation}

\begin{definition}
\end{definition}
\emph{Schwartz space of rapid decrease functions over $ A^{\mathbb{N}_{+}} $:}
\begin{equation}
    \mathcal{S} ( A^{\mathbb{N}_{+}} ) \; := \; \{ f : A^{\mathbb{N}_{+}} \mapsto \mathbb{C} \text{ smooth } \, : \, \| f \|_{n,m} < +
    \infty \: \forall n,m \in \mathbb{N} \}
\end{equation}
The family of seminorms  $  \| f \|_{n,m} $ induces a natural topology over $ \mathcal{S} ( A^{\mathbb{N}_{+}} ) $ that can be used to introduce the following:
\begin{definition}
\end{definition}
\emph{space of tempered distributions over $ A^{\mathbb{N}_{+}} $:}
\begin{center}
   $ \mathcal{S}' ( A^{\mathbb{N}_{+}} ) \; := $ the topological dual of $ \mathcal{S} ( A^{\mathbb{N}_{+}} )$.
\end{center}

Given $ \bar{x} \in A^{\mathbb{N}_{+}} $:
\begin{definition}
\end{definition}
\emph{Dirac delta in $ \bar{x} $:}
\begin{center}
 $ \delta_{\bar{x}} \; := $ the linear functional $ \delta_{\bar{x}} ( f) \; := \; f ( \bar{x} ) $
\end{center}

It may be easily proved that:
\begin{proposition}
\end{proposition}
\begin{equation}
  \delta_{\bar{x}} \in  \mathcal{S}' ( A^{\mathbb{N}_{+}} ) \; \; \forall \bar{x} \in A^{\mathbb{N}_{+}}
\end{equation}

In this paper we will adopt the usual notation:
\begin{equation}
    \int_{A^{\mathbb{N}_{+}}} d \bar{x}  f( \bar{x} ) \delta ( \bar{x} - \bar{y} ) \; := \; \delta_{\bar{y}} ( f)
\end{equation}

\newpage
\section{Pisot-Vijayaraghavan numbers} \label{sec:Pisot-Vijayaraghavan numbers}

Given $ x \in \mathbb{C} $:
\begin{definition}
\end{definition}
\emph{$x $ is algebraic:}
\begin{equation}
    \exists P \in \mathcal{P}[\mathbb{Q}] \; : \; P(x) \, = \, 0
\end{equation}
where we have denoted with $ \mathcal{P}[\mathbb{K}] $ the linear space of
all the monic polynomials with coefficients on the generic ring $
\mathbb{K} $.

We will denote the set of all the algebraic numbers by $
\mathbb{A}$.

\begin{definition}
\end{definition}
\emph{x is algebraic integer:}
\begin{equation}
    \exists P \in \mathcal{P}[\mathbb{Z}] \; : \; P(x) \, = \, 0
\end{equation}

We will denote the set of all the algebraic numbers by $
\mathbb{A}_{\mathbb{Z}}$.

Given $ n \in \mathbb{N}_{+} $ let us denote with $
\mathcal{P}_{n}[\mathbb{K}] $ the set of all the monic polynomials
of degree n with coefficients belonging to the generic ring $
\mathbb{K} $.

Given $ x \in \mathbb{A} $:
\begin{definition}
\end{definition}
\emph{degree of x:}
\begin{equation}
    deg( x ) \; := \; \min \{ n \in \mathbb{N}_{+} \, : \, P( x) = 0 \; P \in \mathcal{P}_{n}[\mathbb{Q}] \}
\end{equation}
\begin{definition}
\end{definition}
\emph{minimal polynomial of $ x $:}
\begin{equation}
    P_{x-minimal} \in \mathcal{P}_{deg( x )}[\mathbb{Q}] \; : \; \  P_{x-minimal} ( x ) = 0
\end{equation}
\begin{definition}
\end{definition}
\emph{set of the conjugates of $ x $:}
\begin{equation}
    Con( x) \; := \; \{ y  \in \mathbb{C} \, : \, y \neq x \, \wedge \,  P_{
    x-minimal}(y) = 0 \}
\end{equation}

We can then state the following \cite{Weiss-98},
\cite{Narkiewicz-04} :
\begin{proposition} \label{prop:fundamental property of algebraic integers}
\end{proposition}
\emph{Fundamental property of algebraic integers:}

\begin{hypothesis}
\end{hypothesis}
\begin{equation*}
    x \in \mathbb{A}_{\mathbb{Z}}
\end{equation*}
\begin{thesis}
\end{thesis}
\begin{equation*}
    x^{n} \, + \sum_{y \in Con(x)} y^{n} \; \in \; \mathbb{Z} \;
    \; \forall n \in \mathbb{N}_{+} : n \geq deg(x)
\end{equation*}

\bigskip

Let us now finally introduce the main actor of this paper:

\begin{definition}
\end{definition}
\emph{set of the Pisot-Vijayaraghavan numbers:}
\begin{equation}
    PV( \mathbb{A}) \; := \; \{ x \in \mathbb{A}_{\mathbb{Z}} \; :  \; x > 1 \; \wedge \; (  | y | < 1 \, \forall y \in Con( x ) )     \}
\end{equation}

Let us recall that given $ a , b \in \mathbb{R} $:
\begin{definition}
\end{definition}
\emph{floor of a:}
\begin{equation}
    \llcorner a \lrcorner \; := \; \max \{ n \in \mathbb{N} : n \leq a \}
\end{equation}
\begin{definition}
\end{definition}
\begin{equation}
    Mod_{b}( a) \; := \; a - \llcorner \frac{a}{b} \lrcorner b
\end{equation}

Given $ a \in \mathbb{R}_{+} $ let us endow the interval $ [ 0 , a
) $ with the topology of the circle, id est by the topology
induced by the following metric:
\begin{definition}
\end{definition}
\emph{ $ S^{1} $-geodesic distance over $ [ 0 , a ) $:}
\begin{equation}
    d_{geodesic} ( x ,y ) : = \left\{%
\begin{array}{ll}
    | x - y|  , & \hbox{ if $  | x - y| \geq \frac{a}{2} $;} \\
    a - | x - y|, & \hbox{otherwise.} \\
\end{array}%
\right.
\end{equation}

Given a sequence $ \{ x_{n} \}_{n \in \mathbb{N}_{+}} \in [0 , a
)^{\mathbb{N}_{+}}$ and $ y \in [ 0 ,a ) $:
\begin{definition}
\end{definition}
\emph{convergence modulo a:}
\begin{equation}
    \lim_{n \rightarrow + \infty} x_{n} \, = \, y \; (mod \, a) \;
    := \;
    \forall \epsilon > 0 , \exists N \in \mathbb{N} \, : \,(
    d_{geodesic} ( x_{n} , y ) < \epsilon \: \forall n
    \in \mathbb{N} : n > N )
\end{equation}

Then
\cite{Bertin-Decomps-Guilloux-Grandet-Hugot-Pathiaux-Delefosse-Schreiber-92},
\cite{Finch-03}:

\begin{proposition} \label{prop:fundamental property of the Pisot-Vijayaraghavan numbers}
\end{proposition}
\emph{Fundamental property of the Pisot-Vijayaraghavan numbers:}

\begin{hypothesis}
\end{hypothesis}
\begin{equation}
    x \in PV ( \mathbb{A} )
\end{equation}

\begin{thesis}
\end{thesis}
\begin{equation}
     \lim_{n \rightarrow + \infty}  Mod_{1}( x^{n})  = 0
\end{equation}
\begin{proof}
By the proposition \ref{prop:fundamental property of algebraic
integers}:
\begin{equation} \label{eq:1th auxiliary property}
    x^{n} \, + \sum_{y \in Con(x)} y^{n} \; \in \; \mathbb{Z} \;
    \; \forall n \in \mathbb{N}_{+} : n \geq deg(x)
\end{equation}
\begin{equation}
    Mod_{1} (  x^{n} \, + \sum_{y \in Con(x)} y^{n} ) \; = \; 0 \;
    \; \forall n \in \mathbb{N}_{+} : n \geq deg(x)
\end{equation}
 let us observe that since $ x \in PV( \mathbb{A} ) $:
\begin{equation}
     \sum_{y \in Con(x)} y^{n} \; =
    \; O_{n \rightarrow \infty} ( \frac{1}{n})
\end{equation}
so that:
\begin{equation}
   Mod_{1} [ x^{n} \, + \,  O_{n \rightarrow \infty} ( \frac{1}{n}) ] \; =
   \; 0 \; \; \forall n \in \mathbb{N}_{+} : n \geq deg(x)
\end{equation}
from which the thesis follows
\end{proof}

\bigskip

Another important property of Pisot--Vijayaraghavan numbers
concerns their link with suitable linear recurrence numeric
sequences, according to the following:
\begin{definition} \label{def:linear recurrence numeric sequence}
\end{definition}
\emph{linear recurrence numeric sequence:}
\begin{equation}
   \{ f_{n} \}_{n \in \mathbb{N}} \in
\mathbb{Z}^{\mathbb{N}} \; : \; ( \exists \, d(f) \in
\mathbb{N}_{+} , \exists  c_{1}, \cdots , c_{d(f)} \in
\mathbb{Z}^{d(f)} \; : \; f_{n} \; = \; \sum_{k=1}^{d(f)} c_{k}
f_{n-k} \; \; \forall n \in N : n \geq d(f) )
\end{equation}
We will call the integer d(f) the \emph{degree} of the linear
recurrence numeric sequence f.

\bigskip

\begin{remark}
\end{remark}
It is important to remark that a linear recurrence numeric
sequence is completely defined by primitive recursion once the
values of $ f_{0} , \cdots , f_{d(f)-1} $ have been assigned.

\bigskip

Then (see the $13^{th}$ chapter "Pisot sequences, Boyd sequences
and linear recurrence" of
\cite{Bertin-Decomps-Guilloux-Grandet-Hugot-Pathiaux-Delefosse-Schreiber-92}):
\begin{proposition} \label{prop:Linear recurrence numeric sequence associated to a
Pisot--Vijayaraghavan number}
\end{proposition}
\emph{Linear recurrence numeric sequence associated to a
Pisot--Vijayaraghavan number:}

\begin{hypothesis}
\end{hypothesis}
\begin{equation}
    \lambda \in PV ( \mathbb{A} )
\end{equation}

\begin{thesis}
\end{thesis}
\begin{equation}
    \exists \{ f_{n} \}_{n \in \mathbb{N}} \in
\mathbb{Z}^{\mathbb{N}}  \text{ linear recurrence numeric sequence
} \; : \; \lim_{n \rightarrow + \infty} \frac{f_{n}}{f_{n-1}} \; =
\; \lambda
\end{equation}

\bigskip

\begin{remark}
\end{remark}
Let us remark that, in in general, there might exist many
primitive recursive numeric sequences associated to a
Pisot--Vijayaraghavan number.

\bigskip

Let us observe furthermore that:
\begin{proposition} \label{prop:asymtotic exponential behaviour}
\end{proposition}

\begin{hypothesis}
\end{hypothesis}
\begin{equation*}
    \lambda \in PV( \mathbb{A}) : ord( \lambda ) = 2
\end{equation*}
\begin{equation*}
   \{ f_{n} \}_{n \in \mathbb{N}} \in
\mathbb{Z}^{\mathbb{N}}  \text{ linear recurrence numeric sequence
associated to } \lambda
\end{equation*}

\begin{thesis}
\end{thesis}
\begin{equation}
    \exists \{ c_{n} \} \in \mathbb{R}^{n} \; : \; f_{n} \; = \; c_{n}
    \lambda^{n} + O_{n \rightarrow + \infty} ( \frac{1}{n} ) \;
    \wedge \; \lim_{n \rightarrow + \infty} \frac{c_{n}}{c_{n-1}}
    \, = \, 1
\end{equation}
\begin{proof}
Let $ y \in Con ( \lambda ) : y \neq \lambda $ be the conjugate of
$ \lambda $.

Posed:
\begin{equation}
     c_{n} \; := \; \frac{f_{n}}{( \lambda^{n} - y^{n})}
\end{equation}
one obtains that:
\begin{equation}
    \frac{f_{n}}{f_{n-1}} \; = \; \frac{ c_{n} \lambda^{n} + O_{n \rightarrow + \infty}( \frac{1}{n})   }{c_{n-1} \lambda^{n-1} + O_{n \rightarrow + \infty}( \frac{1}{n})}
\end{equation}
Using the fact that $ \{ f_{n} \} $ is associated to $ \lambda $:
\begin{equation}
    \lambda + O_{n \rightarrow + \infty}( \frac{1}{n}) \; = \;
    \frac{c_{n}}{c_{n-1}} \lambda
\end{equation}
from which the thesis follows
\end{proof}

\bigskip

\begin{remark}
\end{remark}
If the linear recurrence numeric sequence is of the form $ f_{n} =
m f_{n-1} + f_{n-2} $, with $ f_{0} = 0 $ and $ f_{1} = 1 $ Binet
has proved that the sequence $ c_{n} $ of the proposition
\ref{prop:asymtotic exponential behaviour} is constant not only
asymptotically:
\begin{equation}
    c_{n} \; = \; \frac{1}{ \sqrt{m^{2}+4} } \; \; \forall n \in
    \mathbb{N}
\end{equation}

\bigskip

\begin{example} \label{ex:the golden number is a PV-number}
\end{example}
Given the \emph{golden number} $ \tau := \frac{1 + \sqrt{5}}{2} $:
\begin{equation}
    P_{ \tau-minimal} ( x ) \; = \; x^{2} - x - 1
\end{equation}
and hence $  P_{ \tau-minimal} \in \mathcal{P} [ \mathbb{Z} ]$.

Furthermore $ \tau > 1 $ while:
\begin{equation}
    Con(\tau) \; = \; \{ - \frac{1}{\tau} \}
\end{equation}
where $ | - \frac{1}{\tau} | < 1 $.

Hence $ \tau \in  PV( \mathbb{A}) $.

The \emph{golden number} is the least Pisot-Vijayaraghavan number
of order two.

Furthermore it is the least limit point of Pisot-Vijayaraghavan
numbers:
\begin{equation}
    \tau \; = \; \min LP[ PV( \mathbb{A})]
\end{equation}

A linear recurrence numeric sequence (of degree 2) associated to $
\tau $ is the sequence of Fibonacci numbers:
\begin{equation}
    F_{n+2} \; := \; F_{n+1} + F_{n}
\end{equation}
\begin{equation}
    F_{0} \; := \; 0
\end{equation}
\begin{equation}
    F_{1} \; := \; 1
\end{equation}
\bigskip

\begin{example} \label{ex:the plastic number is a PV-number}
\end{example}
Given the \emph{plastic number} $ \rho \; := \; \frac{(9- \sqrt{69})^{\frac{1}{3}} \, + \, (9 + \sqrt{69})^{\frac{1}{3}}  }{ 2^{\frac{1}{3}}  3^{\frac{2}{3}} } $ :
\begin{equation}
    P_{\rho-minimal}(x) \; = \; x^{3}-x-1
\end{equation}
and hence $  P_{ \rho-minimal} \in \mathcal{P} [ \mathbb{Z} ]$.

Furthermore $ \rho > 1 $ while:
\begin{equation}
    Con(\rho) \; = \; \{  \frac{  i ( i + \sqrt{3}) (9- \sqrt{69})^{\frac{1}{3}} \, + \, (-1 -i \sqrt{3})  (9 + \sqrt{69})^{\frac{1}{3}}  }{2 \cdot  2^{\frac{1}{3}}  3^{\frac{2}{3}} } \: , \: \frac{(-1-i\sqrt{3})(9- \sqrt{69})^{\frac{1}{3}} \, + \, i( i + \sqrt{3})(9 + \sqrt{69})^{\frac{1}{3}}  }{ 2 \cdot 2^{\frac{1}{3}}  3^{\frac{2}{3}} } \, \}
\end{equation}
with $ | \frac{  i ( i + \sqrt{3}) (9- \sqrt{69})^{\frac{1}{3}} \, + \, (-1 -i \sqrt{3})  (9 + \sqrt{69})^{\frac{1}{3}}  }{2 \cdot  2^{\frac{1}{3}}  3^{\frac{2}{3}} }   | < 1 $ and $ | \frac{(-1-i\sqrt{3})(9- \sqrt{69})^{\frac{1}{3}} \, + \, i( i + \sqrt{3})(9 + \sqrt{69})^{\frac{1}{3}}  }{ 2 \cdot 2^{\frac{1}{3}}  3^{\frac{2}{3}} } | < 1$.

Hence $ \rho \in PV( \mathbb{A}) $.

The \emph{plastic number} is the least Pisot-Vijayaraghavan
number.

Furthermore, introduced the notion of the \emph{height} $ h( x )$
of an algebraic number $ x \in \mathbb{A} $ (a sort of measure of
the algebraic complication needed to describe $ x $ for whose
definition and properties we demand to the $ 3^{th} $ chapter
"Pisot and Salem Numbers" of  \cite{Borwain-02}, to the section
3.5 "Smyth's theorem" of
\cite{{Bertin-Decomps-Guilloux-Grandet-Hugot-Pathiaux-Delefosse-Schreiber-92}}
and to the $ 4^{th} $ chapter "Small points" of
\cite{Bombieri-Gubler-06}), the plastic number appears in the
following:
\begin{theorem}
\end{theorem}
\emph{Smith Theorem}:

\begin{hypothesis}
\end{hypothesis}
\begin{equation}
    x \in \mathbb{A}_{\mathbb{Z}} \; : \; x \neq 0 \,
    \wedge \, \frac{1}{x} \notin Con(x)
\end{equation}

\begin{thesis}
\end{thesis}
\begin{equation}
    h ( x ) \; \geq \; \frac{ \log \rho  }{ deg(x) }
\end{equation}

A linear recurrence numeric sequence (of degree 3) associated to $
\rho $ is the sequence of the Padovan numbers:
\begin{equation}
    P(0) \; := \; P(1) \; := P(2) \; := 1
\end{equation}
\begin{equation}
    P_{n} \; := \; P_{n-2} + P_{n-3}
\end{equation}

\bigskip

\begin{example} \label{ex:the Pell number is a PV-number}
\end{example}
The number $ 1 + \sqrt{2} > 1 $ has as minimal polynomial:
\begin{equation}
    P_{1+\sqrt{2}-mimimal} ( x) \; = \; x^{2} -2 x -1
\end{equation}
and is hence an algebraic integer.

Furthermore:
\begin{equation}
    Con (  1 + \sqrt{2} ) \; = \; \{ 1 - \sqrt{2} \}
\end{equation}
where obviously:
\begin{equation}
    |  1 - \sqrt{2} | \; < \; 1
\end{equation}
Hence $ 1 + \sqrt{2} \in PV( \mathbb{A} ) $.

A linear recurrence numeric sequence (of degree 2) associated to $
1 + \sqrt{2} $ is the sequence of Pell numbers:
\begin{equation}
    a_{0} \; := \; 0
\end{equation}
\begin{equation}
    a_{1} \; := \; 1
\end{equation}
\begin{equation}
    a_{n} \; := 2 a_{n-1} + a_{n-2} \; \; \forall n \in \mathbb{N}
    : n \geq 2
\end{equation}

\bigskip

We want to conclude this Number Theoretic section with a
celebrated well known result, the recursive undecidability of
Hilbert $ 10^{th}$ problem showing its deep link with Theoretical
Computer Science \cite{Matiyasevich-93}, \cite{Matiyasevich-99}.

Given a set $ S \subseteq \mathbb{N}$:
\begin{definition}
\end{definition}
\emph{S is Diophantine:}
\begin{equation}
    \exists P \in \mathcal{P}[\mathbb{Z}] \; : \; S = \{ x \in
    \mathbb{N} \, : \, P(x) = 0 \}
\end{equation}

\begin{theorem} \label{th:Matiyasevich Theorem}
\end{theorem}
\emph{Matiyasevich Theorem:}

\begin{hypothesis}
\end{hypothesis}
\begin{equation}
     S \subseteq \mathbb{N}
\end{equation}

\begin{thesis}
\end{thesis}
\begin{center}
    S is Diophantine if and only if it is recursively enumerable
\end{center}

Theorem \ref{th:Matiyasevich Theorem} implies that it doesn't
exist an algorithm that, receiving as input a polynomial $ P \in
\mathcal{P} ( \mathbb{Z}) $ outputs "yes" if its has non-negative
integer solutions while it outputs "no" if it doesn't.

\newpage
\section{A brief information theoretic analysis of singular Lebesgue-Stieltjes
measures supported on Cantor sets and almost periodicity}
\label{sec:A brief information theoretic analysis of singular
Lebesgue-Stieltjes measures supported on Cantor sets and almost
periodicity}

Demanding to \cite{Reed-Simon-80}, \cite{Ash-Doleans-Dade-00},
\cite{Bogachev-07} for any further information let us  recall that
given a continuous function $ f : \mathbb{R} \mapsto \mathbb{R}$
non decreasing, id est such that:
\begin{equation} \label{eq:non-decreasing condition}
    f( x) \geq f(y) \; \; \forall x , y \in \mathbb{R} \, : \, x >
    y
\end{equation}
we can introduce the following:
\begin{definition}
\end{definition}
\emph{Lebsegue-Stieltjes measure associated to the f:}

the measure $ \mu_{f} : \mathcal{B}( \mathbb{R} ) \mapsto [ 0 , +
\infty ) $ such that:
\begin{equation}
    \mu_{f} [ ( a , b) ] \; := \; \lim_{\epsilon \rightarrow
    0^{-}} f ( b - \epsilon ) \, - \, \lim_{\epsilon \rightarrow
    0^{+}} f ( a + \epsilon ) \; \; \forall a , b \in \mathbb{R} :
     a < b
\end{equation}
where $  \mathcal{B}( \mathbb{R} ) $ is the Borel-$\sigma$-algebra
on $ \mathbb{R}$, id est the $ \sigma$-algebra induced by the
topology induced by the usual metric $ d( x,y ) := | x - y |$.

For instance, choosing the identity function $ Id : \mathbb{R}
\rightarrow \mathbb{R} $:
\begin{equation}
    Id(x) := x
\end{equation}
the Lebesgue-Stieltjes measure $ \mu_{Id} $ is nothing but the
Lebesgue measure on numbers $ v_{A} ( \mu_{Lebesgue} ) $  (where $
\mu_{Lebesgue} $ is the Lebesgue measure on sequences of the
definition \ref{def:Lebesgue measure on sequences}).

Given a measure $ \mu : \mathcal{B} \rightarrow [ 0 , + \infty ) $
let us recall that:

\begin{definition}
\end{definition}
\emph{set of pure points of $ \mu $ }:
\begin{equation}
    PP( \mu ) \; := \; \{ x \in \mathbb{R} \; : \;
    \mu ( \{ x \} ) \neq 0 \}
\end{equation}

\begin{definition}
\end{definition}
\emph{$\mu $ is pure point:}
\begin{equation}
    \mu( X) \; = \; \sum_{x \in X} \mu ( \{ x \} ) \; \; \forall X
    \in \mathcal{B}( \mathbb{R} )
\end{equation}

\begin{definition}
\end{definition}
\emph{$\mu $ is absolutely continuous (with respect to the
Lebesgue measure):}
\begin{equation}
    \mu \prec \mu_{Id} \; := \;  \mu ( B ) \, = \, 0 \; \Rightarrow \;  \mu_{Id}( B ) \, = \, 0
\end{equation}

Let us recall the basic:
\begin{theorem} \label{th:Radon-Nikodym Theorem}
\end{theorem}
\emph{Radon-Nikodym Theorem:}

\begin{hypothesis}
\end{hypothesis}
\begin{equation*}
   \mu \prec \mu_{Id}
\end{equation*}

\begin{thesis}
\end{thesis}
\begin{center}
 There exists a $ \mu_{Id} $-measurable function, called the
 Radon-Nikodym derivative of $ \mu $ with respect to  $ \mu_{Id} $
 and consequentially denoted as $ \frac{d \mu }{d \mu_{Id}} $, such
 that $ \int_{\mathbb{R}} f (x) d \mu (x) \; = \; \int_{\mathbb{R}} f (x) \frac{d \mu }{d \mu_{Id}}(x)  d \mu_{Id}
 (x)$ for every $ \mu$-measurable function f
\end{center}

\begin{definition}
\end{definition}
\emph{$\mu$ is singular (with respect to the Lebesgue measure):}
\begin{equation}
   \mu \perp \mu_{Id}  \; := \; \exists S \; : \; \mu (S) = 0 \; \wedge \; \mu_{Id} (
    \mathbb{R} - S) \, = \, 0
\end{equation}

Let us then recall the following:

\begin{theorem} \label{th:Lebesgue Decomposition Theorem}
\end{theorem}
\emph{Lebesgue Decomposition Theorem:}

\begin{hypothesis}
\end{hypothesis}
\begin{equation*}
    \mu : \mathcal{B} ( \mathbb{R} ) \mapsto [ 0 , + \infty )
    \text{ measure }
\end{equation*}

\begin{thesis}
\end{thesis}
\begin{equation*}
    \exists ! ( \mu_{p.p} , \mu_{a.c.} , \mu_{sing} ) \text{ triple of measures
    } \; : \; \mu_{p.p}  \text{ is pure point } \: \wedge \:   \mu \prec \mu_{a.c.} \: \wedge \: \mu \perp \mu_{sing}
\end{equation*}

Given a map $ f : [ a , b ] \rightarrow \mathbb{R} $:
\begin{definition}
\end{definition}
\emph{f is absolutely continuous:}
\begin{multline}
    \forall \epsilon > 0 \, , \, \exists \delta > 0 :
    \sum_{i=1}^{n} | f( b_{i}) - f (a_{i})| \leq \epsilon \\
    \forall \{ (a_{i} , b_{i} ) \}_{i=1}^{n}  : (  a_{i} , b_{i} )
    \subset (a , b ) \wedge ( a_{i} , b_{i} ) \cap ( a_{j} , b_{j}
    ) \, = \, \emptyset \; \wedge \sum_{i=1}^{n} \mu_{Id} [ (
    a_{i}, b_{i} ) ] < \delta \; \forall n \in \mathbb{N}_{+}
\end{multline}

A remarkable feature of Lebesgue-Stieltjes measures is the
following:

\begin{theorem} \label{th:On the singularity of Lebesgue-Stieltjes measures}
\end{theorem}
\emph{On the singularity of Lebesgue-Stieltjes measures:}

\begin{hypothesis}
\end{hypothesis}
\begin{center}
  $ \mu_{f} $ Lebesgue-Stieltjes measure
\end{center}

\begin{thesis}
\end{thesis}
\begin{center}
  $ \mu_{f} $ is an absolutely continuous measure if and only if f
  is an absolutely continuous function
\end{center}

\bigskip

Let us now show how Algorithmic Information Theory enters in the
game.

Given a trinary alphabet $ A := \{ a_{1} ,a_{2} , a_{3} \} $ let
us consider the set of all sequences over A not containing the
letter $ a_{2} $:
\begin{equation}
    NO_{A,a_{2}} \; := \; \{ \bar{x} \in A^{\mathbb{N}_{+}} \; : \; |  \bar{x} |_{a_{2}} = 0  \}
\end{equation}
As we have already observed in the remark \ref{rem:link to Cantor
sets},  the theorem \ref{th:Calude-Chitescu} implies that:
\begin{equation} \label{eq:1th condition}
   NO_{A,a_{2}} \cap RANDOM( A^{\mathbb{N}_{+}} )\; = \; \emptyset
\end{equation}
Let us now consider the binary alphabet $ B := \{ a_{1} , a_{3} \}
$.

Clearly:
\begin{equation} \label{eq:2th condition}
  NO_{A,a_{2}} \; = \; B^{\mathbb{N}_{+}} \; \supset \; RANDOM(
  B^{\mathbb{N}_{+}})
\end{equation}

At a first sight the equation \ref{eq:1th condition} and the
equation \ref{eq:2th condition} might appear incompatible.

A deeper analysis allows, anyway, to understand that everything is
absolutely consistent:

simply we have to remember that the correct way of passing from
the alphabet A to the alphabet B is through the transition map $
T_{A,B} $ of the definition \ref{def:alphabet's transition map}.

Actually the theorem \ref{th:Independence of algorithmic
randomness from the adopted alphabet} guarantees that:
\begin{equation}
    RANDOM ( B^{\mathbb{N}_{+}} )  \; = \; T_{A,B} [ RANDOM (  A^{\mathbb{N}_{+}}
    ) ]
\end{equation}

\bigskip

\begin{remark}
\end{remark}
In a very interesting paper \cite{Calude-Staiger-Svozil-05} that
has greatly inspired this section, Cristian Calude, Ludwig Staiger
and Karl Svozil have introduced a generalization of the notion of
algorithmic randomness, id est of the definition
\ref{def:algorithmic randomness}, to sequences $ \{ x_{n} \} $
each letter of which belongs to a different alphabet $ A_{n} $.
The analysis of the probably deep link existing between such a
generalization and what we are going to discuss here is something
that, at present, we are not able to formalize and is left as an
interesting challenge for the reader.

\bigskip

Let us now introduce the set:
\begin{definition}
\end{definition}
\emph{Cantor's middle third set:}
\begin{equation}
    \mathcal{C}_{A,a_{2}} \; := \; v_{A} ( NO_{A,a_{2}} )
\end{equation}

It may be easily proved that:
\begin{proposition} \label{pr:properties of Cantor's middle third set}
\end{proposition}
\begin{enumerate}
    \item
\begin{equation}
  \mu_{Id} (  \mathcal{C}_{A,a_{2}} ) \; = \; 0
\end{equation}
    \item
\begin{equation}
    |  \mathcal{C}_{A,a_{2}} | \; > \; \aleph_{0}
\end{equation}
  \item
\begin{equation}
    dim_{Hausdorff} ( \mathcal{C}_{A,a_{2}} ) \; = \; \frac{ \log |B| }{  \log |A| }
\end{equation}
\end{enumerate}

Given $ n \in \mathbb{N}_{+}$ let us introduce the following:
\begin{definition}
\end{definition}
\emph{Cantor function:}

the map $ f_{A,a} : B^{+} \mapsto \mathbb{R} $:
\begin{equation}
  f_{A,a}( \vec{x} ) \; := \; v_{B}( \vec{x} ) + \frac{|B|}{|B|^{|\vec{x}|+1 }}
  \forall \vec{x} \in B^{+} \; : \; v_{B}( \vec{x} ) \in (
  v_{A}( \vec{x}) + \frac{1}{|A|^{| \vec{x} | + 1}} , v_{A}( \vec{x}) +
  \frac{|B|}{|A|^{| \vec{x} | +1 }} )
\end{equation}

It may be easily proved that the map $ \tilde{f} : [ 0 ,1] \mapsto
\mathbb{R} $:
\begin{equation}
  \tilde{f} (x) \; := \; f ( r_{B}(x) )
\end{equation}
is a continuous non-decreasing function so that it induces a
Lebesgue-Stieltjes measure $ \mu_{\tilde{f}} $.

Since $ \tilde{f} $ is not absolutely continuous it follows that $
\mu_{\tilde{f}} $  is singular with respect to the Lebesgue
measure;

actually:
\begin{equation}
    \mu_{\tilde{f}} [  \mathcal{C}_{A,a_{2}} ] \; = \; 1
\end{equation}
while we know by the proposition \ref{pr:properties of Cantor's
middle third set} that $ \mathcal{C}_{A,a_{2}} $ has vanishing
Lebesgue measure.

Let us remark that $ \mathcal{C}_{A,a_{2}} $ is a Cantor set in
the sense of the definition \ref{def:Cantor set}.

Now the measure $ \mu_{\tilde{f}} : \mathcal{B} [ [ 0 ,1 ) ]
\mapsto [ 0 , + \infty )  $ induces naturally the measure $
\mu_{f} :  \mathcal{B} [ A^{\mathbb{N}_{+}} ] \mapsto [ 0 , +
\infty ) $ defined by:
\begin{equation}
    \mu_{f} ( X ) \; := \;  \mu_{\tilde{f}} ( v_{A} (X) )
\end{equation}

Then:
\begin{equation}
  \mu_{Lebesgue,B} [  NO_{A,a_{2}} ] \; = \: \mu_{Lebesgue,B}  [
  B^{\mathbb{N}_{+}} ] \; = \; 1
\end{equation}
\begin{equation}
    \mu_{Lebesgue,B} [  NO_{A,a_{2}} \cap RANDOM (
    B^{\mathbb{N}_{+}}) ] \; = \;  \mu_{Lebesgue,B} [ RANDOM (
    B^{\mathbb{N}_{+}}) ] \; = \; 1
\end{equation}
\begin{equation}
    \mu_{Lebesgue,A} [  NO_{A,a_{2}} ] \; = \; 0
\end{equation}
\begin{equation}
    \mu_{f} [ NO_{A,a_{2}} ] \; = \; 1
\end{equation}
where we have used the theorem \ref{th:On the Foundation of
Probability Theory}.

\bigskip

The whole story may be easily generalized in the following way:

given a finite alphabet $ A : 2 \leq | A| < \aleph_{0} $ and a
letter $ a \in A $ let us introduce the set of all the sequences
on  A not containing the digit a:
\begin{equation}
    NO_{A,a} \; := \; \{ \bar{x} \in A^{\mathbb{N}_{+}} \, : \, |
    \bar{x}|_{a} \, = \, 0 \}
\end{equation}
and the new alphabet B obtained from A excluding the letter a:
\begin{equation}
  B \; := \; A - \{ a \}
\end{equation}
Then one can introduce the set:
\begin{definition}
\end{definition}
\emph{generalized Cantor set associated to $ (A,a)$:}
\begin{equation}
    \mathcal{C}_{A,a} \; := \; v_{A} ( NO_{A,a} )
\end{equation}

 Given $ n \in \mathbb{N}_{+}$ let us introduce the
following:
\begin{definition}
\end{definition}
\emph{generalized Cantor function associated to $ (A,a)$:}

the map $ f_{A,a} : B^{+} \mapsto \mathbb{R} $:
\begin{equation}
  f_{A,a}( \vec{x} ) \; := \; v_{B}( \vec{x} ) + \frac{|B|}{|B|^{|\vec{x}|+1 }}
  \forall \vec{x} \in B^{+} \; : \; v_{B}( \vec{x} ) \in (
  v_{A}( \vec{x}) + \frac{1}{|A|^{| \vec{x} | + 1}} , v_{A}( \vec{x}) +
  \frac{|B|}{|A|^{| \vec{x} | +1 }} )
\end{equation}

and the associated  map $ \tilde{f} : [ 0 ,1] \mapsto \mathbb{R}
$:
\begin{equation}
  \tilde{f}_{A,a} (x) \; := \; f ( r_{B}(x) )
\end{equation}

Then:
\begin{proposition} \label{pr:on generalized Cantor functions}
\end{proposition}
\emph{Properties of the generalized Cantor functions:}

\begin{hypothesis}
\end{hypothesis}
\begin{equation}
   0 < lex(a) < |A| -1
\end{equation}
\begin{thesis}
\end{thesis}
\begin{enumerate}
    \item $ \tilde{f}_{A,a} $ is a  continuous non decreasing
    function
    \item $ \tilde{f}_{A,a} $ is not absolutely continuous
\end{enumerate}
\begin{proof}
It is sufficient to follow step by step the analogous proof
holding in the particular case of the Cantor middle-third set and
generalize in the natural way
\end{proof}

The proposition \ref{pr:on generalized Cantor functions} implies
that one can introduce the Lebesgue-Stieltjes measure $
\mu_{\tilde{f}} $. Then:
\begin{corollary}
\end{corollary}
\begin{center}
  $ \mu_{\tilde{f}_{A,a}} $  is  singular
\end{center}
\begin{proof}
The thesis follows by the proposition \ref{th:On the singularity
of Lebesgue-Stieltjes measures} and the proposition \ref{pr:on
generalized Cantor functions}.
\end{proof}

\smallskip

Furthermore:
\begin{equation}
  \mu_{Lebesgue,B} [  NO_{A,a} ] \; = \: \mu_{Lebesgue,B}  [
  B^{\mathbb{N}_{+}} ] \; = \; 1
\end{equation}
\begin{equation}
    \mu_{Lebesgue,B} [  NO_{A,a} \cap RANDOM (
    B^{\mathbb{N}_{+}}) ] \; = \;  \mu_{Lebesgue,B} [ RANDOM (
    B^{\mathbb{N}_{+}}) ] \; = \; 1
\end{equation}
\begin{equation}
    \mu_{Lebesgue,A} [  NO_{A,a} ] \; = \; 0
\end{equation}
\begin{equation}
    \mu_{f} [ NO_{A,a} ] \; = \; 1
\end{equation}

\bigskip

We will now briefly outline the deep  link existing between the
singular Lebesgue-Stieltjes measures associated to generalized
Cantor functions and the theory of almost periodic functions,
sequences and measures \cite{Amerio-Prouse-71},
\cite{Corduneanu-89}, \cite{de-Lamadrid-Argabright-90}.

Given a map $ f : \mathbb{R} \mapsto \mathbb{C} $:

\begin{definition}
\end{definition}
\emph{f is periodic:}
\begin{equation}
    \exists T \in (0, + \infty ) \; : \; f( t+T) = f(t) \; \;
    \forall t \in \mathbb{R}
\end{equation}
Given a periodic function f:
\begin{definition}
\end{definition}
\emph{fundamental period of f:}
\begin{equation}
 Period_{fund}[f] \; := \; \min \{  T \in (0, + \infty ) \; : \; f( t+T) = f(t) \; \;
    \forall t \in \mathbb{R}  \}
\end{equation}

We will denote by $ PER_{T} ( \mathbb{R} , \mathbb{C} ) $  the set
of all the periodic functions.

\begin{remark}
\end{remark}
In the section \ref{sec:Topological entropy of substitutions} we
saw that the \emph{combinatorial information} and the
\emph{algorithmic information} of a periodic sequence (according
to the definition \ref{def:periodic sequences}) are very low since
a periodic sequence is completely specified assigning the values
it takes over a single period.

The same thing can clearly be said of periodic functions.

\bigskip

Clearly:
\begin{proposition}
\end{proposition}
\begin{equation}
    PER_{T} ( \mathbb{R} , \mathbb{C} ) \text{ is a complex linear space }
     \; \; \forall T \in (0, + \infty  )
\end{equation}

\smallskip

Given $ n \in \mathbb{N}_{+} $, $ \omega_{1} , \cdots, \omega_{n}
\in ( 0 , + \infty ) $ and a map $ f : \mathbb{R} \mapsto
\mathbb{C} $:
\begin{definition}
\end{definition}
\emph{f is quasiperiodic with pulsations $ \omega_{1} , \cdots,
\omega_{n} $:}
\begin{multline}
    ( \exists \phi : \mathbb{R}^{n} \mapsto \mathbb{C} : \phi ( x_{1} ,
     \cdots , x_{i} + 2 \pi , \cdots , x_{n} ) \, = \, \phi ( x_{1} ,
     \cdots , x_{i} , \cdots , x_{n} ) \; \forall i \in \{ 1 ,  \cdots , n \} ) \;
     \wedge \\
      ( f( t) = \phi ( \omega_{1} t , \cdots , \omega_{n} t) \; \;
     \forall t \in \mathbb{R} )
\end{multline}

\begin{definition}
\end{definition}
\emph{$\omega_{1}, \cdots , \omega_{n} $ are rationally
independent:}
\begin{equation}
   ( \sum_{i=1}^{n} k_{i} \omega_{i} \, = \, 0 \; \Rightarrow \;
    k_{1}= \cdots = k_{n} = 0 ) \; \; \forall k_{1} , \cdots ,
    k_{n} \in \mathbb{Q}
\end{equation}

Then:
\begin{proposition}
\end{proposition}
\emph{On the relation between periodicity and quasiperiodicity:}
\begin{enumerate}
    \item f is periodic $ \Rightarrow $ f is quasiperiodic
    \item f is periodic $ \Leftrightarrow $ f is quasiperiodic
    with rationally dependent pulsations
\end{enumerate}

\smallskip

\begin{definition} \label{def:complex trigonometric polynomial}
\end{definition}
\emph{f is a complex trigonometric polynomial:}
\begin{equation}
    \exists n \in \mathbb{N}_{+} , \exists A_{1} ,  ,
    \cdots , A_{n} \in \mathbb{C} , \exists \omega_{1} , \cdots ,
    \omega_{n} \in ( 0 , + \infty ) \; : \; f(t) \, = \,
    \sum_{k=1}^{n} A_{k} \exp ( i \omega_{k} t) \; \; \forall t \in \mathbb{R}
\end{equation}

\begin{proposition}
\end{proposition}
\begin{equation}
    f \text{ complex trigonometric polynomial } \; \Rightarrow \;
    f \text{ is quasi-periodic}
\end{equation}
\begin{proof}
Given the complex trigonometric polynomial:
\begin{equation}
    f(t) \; := \; \sum_{k=1}^{n} A_{k} \exp ( i \omega_{k} t)
\end{equation}
let us introduce the following  map $ \phi : \mathbb{R}^{n}
\mapsto \mathbb{C} $:
\begin{equation}
    \phi ( x_{1} , \cdots , x_{n} ) \; := \;  \sum_{k=1}^{n} A_{k} \exp ( i x_{k})
\end{equation}
Obviously:
\begin{equation}
    \phi( x_{1}+2 \pi , \cdots, x_{n} ) \; = \; \cdots \; = \;
    \phi ( x_{1} , \cdots , x_{n} + 2 \pi ) \; \; \forall x_{1} ,
    \cdots, x_{n} \in \mathbb{R}
\end{equation}
and:
\begin{equation}
     f( t) \; =  \; \phi ( \omega_{1} t , \cdots , \omega_{n} t) \; \;
     \forall t \in \mathbb{R}
\end{equation}
\end{proof}

\bigskip

Let us now introduce Harald Bohr's notion of almost periodic
function.

\begin{definition} \label{def:almost periodic function}
\end{definition}
\emph{almost periodic function:}

a map $ f : \mathbb{R} \mapsto \mathbb{C} $ :
\begin{equation}
    \forall \epsilon > 0 \, \exists \, T_{\epsilon} \text{ complex trigonometric polynomial
    } \; : \; | f(x) - T_{\epsilon} (x) | < \epsilon \; \forall x
    \in \mathbb{R}
\end{equation}

We will denote by $ A-PER ( \mathbb{R} , \mathbb{C} ) $  the set
of all the almost periodic functions.

To appreciate the meaning of the definition \ref{def:almost
periodic function} let us recall that \cite{Reed-Simon-80}  given
a metric space $ ( M , d) $ and a set $ S \subset M $:
\begin{definition}
\end{definition}
\emph{S is dense in $( M , d)$:}
\begin{equation}
 \forall y \in S \; \exists \{ x_{n} \}_{n \in \mathbb{N}} \in
 M^{\mathbb{N}} \; : \; \lim_{n \rightarrow + \infty} d(x_{n}, y )
 = 0
\end{equation}

Then definition \ref{def:almost periodic function} implies that:
\begin{proposition}
\end{proposition}
\begin{center}
  $ \{ \text{ complex trigonometric polynomials } \} $ is dense in the metric space $
  ( A-PER ( \mathbb{R} , \mathbb{C} ) , d_{1}) $
\end{center}
where:
\begin{equation}
    d_{1} ( f , g ) \; := \; \sup_{t \in \mathbb{R}} | f(t) -g(t) |
\end{equation}

\smallskip

Furthermore:
\begin{proposition}
\end{proposition}
\emph{On the relation between quasiperiodicy and almost
periodicity:}
\begin{enumerate}
    \item f is quasiperiodic $ \Rightarrow $ f is almost periodic
    \item f is almost periodic $ \nRightarrow $ f is quasiperiodic
\end{enumerate}

\smallskip

The basic properties of almost periodic functions are encoded in
the following:
\begin{proposition} \label{prop:basic properties of almost periodic functions}
\end{proposition}
\emph{Basic properties of almost periodic functions:}
\begin{enumerate}
     \item
\begin{multline}
    f \text{ is almost periodic } \; \Leftrightarrow \; \forall \epsilon \in ( 0 , + \infty
    )  \, \exists l( \epsilon ) \in ( 0 , + \infty )
     \; : \\
     (  \forall a \in \mathbb{R} , \exists \xi \in
    \mathbb{R} \, : \, | f( x+ \xi ) - f(x) | < \epsilon \,  \forall x \in ( a , a + l( \epsilon
    ) )
\end{multline}
    \item f complex trigonometric polynomial $ \Rightarrow $ f
    almost periodic
    \item  $ A-PER( \mathbb{R} , \mathbb{C} ) $ is a complex linear space
    \item
\begin{equation}
    f_{1} \cdot f_{2} \text{ is almost periodic} \; \; \forall f_{1} ,
    f_{2} \text{ almost periodic}
\end{equation}
    \item
\begin{equation}
  \frac{d}{d a} \lim_{T \rightarrow + \infty} \int_{a}^{a+T} f(x)
  dx \; = \; 0
\end{equation}
\end{enumerate}

Given an almost periodic function f:
\begin{definition}
\end{definition}
\emph{mean value of f:}
\begin{equation}
    M[f] \; := \; \lim_{T \rightarrow + \infty} \int_{a}^{a+T} f(x)
  dx
\end{equation}

Then:
\begin{proposition} \label{prop:property of the mean value}
\end{proposition}
\emph{Basic property of the mean value of an almost periodic
function:}
\begin{equation}
   support_{\omega}  M[ \exp( - i \omega x) f(x)
   ] \; \leq \; \aleph_{0}
\end{equation}

Given a generic function $ f : \mathbb{R} \mapsto \mathbb{C} $:
\begin{definition} \label{def:Fourier transform of a function}
\end{definition}
\emph{Fourier transform of f:}
\begin{equation}
    \mathcal{F}[f] ( \omega ) \; := \; \left\{%
\begin{array}{ll}
    \lim_{T \rightarrow + \infty} \int_{- T}^{+ T} d \omega \exp ( - i \omega x ) f(x) , & \hbox{if the limit exists;} \\
    + \infty, & \hbox{otherwise} \\
\end{array}%
\right.
\end{equation}

The proposition \ref{prop:property of the mean value} implies
that:

\begin{proposition} \label{prop:Fourier spectrum of almost periodic functions}
\end{proposition}
\emph{Fourier spectrum of almost periodic functions:}

\begin{hypothesis}
\end{hypothesis}
\begin{center}
  f almost periodic function
\end{center}
\begin{thesis}
\end{thesis}
\begin{enumerate}
    \item
\begin{equation}
  \exists \{ A_{n} \} \in \mathbb{C}^{\mathbb{Z}} , \exists \{ \omega_{n}
  \} \in (0 , + \infty )^{\mathbb{Z}} \; : \;  \mathcal{F}[f] ( \omega )
  \, = \; \sum_{n = - \infty}^{+ \infty} A_{n} \delta ( \omega -
  \omega_{n} )
\end{equation}
    \item if  f is periodic of fundamental period $\frac{2 \pi}{\omega} $  then $ \omega_{n} \; =
    \; n \omega \; \forall n \in \mathbb{Z} $
\end{enumerate}

\begin{proof}
\begin{enumerate}
    \item
The thesis immediately follows combining the definition
\ref{def:Fourier transform of a function} and the proposition
\ref{prop:property of the mean value}.
    \item The thesis is nothing but the basic theorem concerning
    Fourier series of periodic functions
\end{enumerate}
\end{proof}

As to the existence of the Fourier transform of almost periodic
functions there exists a very powerful result
\cite{Corduneanu-89}:

\begin{proposition} \label{prop:About the specification of an almost periodic function
through its Fourier transform}
\end{proposition}
\emph{About the specification of an almost periodic function
through its Fourier transform:}

\begin{hypothesis}
\end{hypothesis}
\begin{equation*}
    \{ \omega_{n} \} \in (0 , + \infty )^{\mathbb{Z}}
\end{equation*}
\begin{equation}
   \{ A_{n} \} \in \mathbb{C}^{\mathbb{Z}} \; : \; \sum_{n= - \infty}^{+
   \infty} |  A_{n} | \in \mathbb{R}
\end{equation}

\begin{thesis}
\end{thesis}
\begin{equation*}
    \exists \; f \in A-PER( \mathbb{R} , \mathbb{C} ) \; : \;
    \mathcal{F}[f] ( \omega ) \, = \, \sum_{n= - \infty}^{+
   \infty} A_{n} \delta( \omega - \omega_{n} )
\end{equation*}

Let us now consider a generic measure $ \mu : \mathcal{B} (
\mathbb{R} ) \mapsto [ 0 , + \infty ) $ on the real line and let
us introduce the following:
\begin{definition} \label{def:Fourier transform of a measure}
\end{definition}
\emph{Fourier transform of $ \mu $:}
\begin{equation}
    \mathcal{F} [ \mu ] (\omega) \; := \; \int_{\mathbb{R}} e^{ - i \omega
    x} d \mu(x)
\end{equation}

Obviously:
\begin{proposition}
\end{proposition}

\begin{hypothesis}
\end{hypothesis}
\begin{equation*}
    \mu \; \prec \; \mu_{Id}
\end{equation*}

\begin{thesis}
\end{thesis}
\begin{equation}
  \mathcal{F} [ \mu ] \; = \; \mathcal{F} [ \frac{ d \mu }{ d
  \mu_{Id}} ]
\end{equation}
where the Fourier transform of the left hand side is intended in
the sense of the definition \ref{def:Fourier transform of a
measure} while the Fourier transform of the right hand side is
intended in the sense of the definition \ref{def:Fourier transform
of a function}.

\begin{proof}
It is sufficient to observe that:
\begin{equation}
   \mathcal{F} [ \mu ] \; = \; \int_{\mathbb{R}} e^{-i \omega x}  \frac{ d \mu }{ d \mu_{Id}
   }(x)  d \mu_{Id} (x) \; = \; \int_{- \infty}^{+ \infty}   e^{-i \omega x} \frac{ d \mu }{ d \mu_{Id}
   }(x) d x \; = \; \mathcal{F} [ \frac{ d \mu }{ d
  \mu_{Id}} ]
\end{equation}
\end{proof}

The situation is, instead, more complicated in the case in which
the measure $ \mu $ is singular with respect to the Lebesgue
measure $ d \mu_{Id} $.

Let us consider in particular the case in which $ \mu =
\mu_{\tilde{f}_{A,a}} $ is the singular measure associated to a
generalized Cantor function $ f_{A,a} $.

Though we have not yet a proof, we think that Eberlein's Theorem
(see the $ 11^{th} $ chapter "Almost periodicity of the
generalized Fourier transform " of
\cite{de-Lamadrid-Argabright-90}) strongly supports the following:
\begin{conjecture}
\end{conjecture}
\emph{About the Fourier transform of the Lebesgue-Stieltjes
measure associated to a generalized Cantor function:}
\begin{equation}
   \mathcal{F} [\mu_{\tilde{f}_{A,a}} ] \text{ is almost periodic}
\end{equation}
\newpage
\section{Quasicrystals} \label{sec:Quasicrystals}
 Given $ n \in \mathbb{N}_{+} $ let us introduce the following:

\begin{definition}
\end{definition}
\emph{n-dimensional euclidean space:}

the Hilbert space $ \mathbb{E}^{n} \; := ( \mathbb{R}^{n} , \cdot
) $, where $ \vec{x} \cdot \vec{y} \; := \; \sum_{i=1}^{n} x_{i}
y_{i} $ is the usual euclidean inner product.

\smallskip
Given $ \vec{x} \in \mathbb{E}^{n} $ and $ r \in \mathbb{R}_{+} $:
\begin{definition}
\end{definition}
\emph{ball with center $ \vec{x} $ and radius r:}
\begin{equation}
    B_{r} ( \vec{x} ) \; := \; \{ \vec{y} \in  \mathbb{E}^{n} \, : \, | \vec{y} - \vec{x} | < r \}
\end{equation}

\smallskip

Given $ S \subset \mathbb{E}^{n} $ let us introduce the following
\cite{Senechal-95a}:

\begin{definition}
\end{definition}
\emph{S is a Delone set in $ \mathbb{E}^{n} $:}
\begin{enumerate}
    \item
\begin{equation*}
  \exists r_{0} \in \mathbb{R}_{+} \; : \; | \vec{x} - \vec{y} | >
  2 r_{0} \; \; \forall \vec{x} , \vec{y} \in S \, \vec{x} \neq \vec{y}
\end{equation*}
    \item
\begin{equation}
      \exists R_{0} \in \mathbb{R}_{+} \; : \;B_{r} ( \vec{x}
       ) \cap S \, \neq \emptyset \; \; \forall r > R_{0} \, , \,  \forall \vec{x} \in  \mathbb{E}^{n}
\end{equation}
\end{enumerate}

Following the generalization of the mathematical definition of a
crystal given in \cite{Senechal-95a}, \cite{Hof-95}, \cite{Hof-97}
on the score of the indications given in 1992 by the Commission on
Nonperiodic Crystals established by the International Unit of
Crystallography, we will characterize a crystal by the condition
that its diffraction pattern exhibits a countable infinity of
Bragg peaks.

Given a Delone set S:
\begin{definition}
\end{definition}
\emph{density distribution of S:}
\begin{equation*}
    \rho_{S} ( \vec{x} ) \; := \; \sum_{\vec{s} \in S} \delta (
    \vec{x} - \vec{s} )
\end{equation*}
\begin{definition}
\end{definition}
\emph{autocorrelation of S:}
\begin{equation}
    \gamma_{S}( \vec{x} ) := \rho (\vec{x}) \star
\overline{\rho(- \vec{x})}
\end{equation}
where $ \star $ denotes the \emph{convolution operator}:
\begin{equation}
    (f \star g) ( \vec{x} ) \; := \; \int_{\mathbb{R}^{n}} d
    \vec{y} f( \vec{x} ) g( \vec{x} - \vec{y} )
\end{equation}

\begin{definition} \label{def:spectrum of a Delone set}
\end{definition}
\emph{spectrum of S:}

\begin{center}

the measure $ \mu_{S} $ on $ ( \mathbb{R}^{n} , \mathcal{B} (
\mathbb{R}^{n} ) ) $ identified by $ \lim_{L \rightarrow + \infty}
\frac{1}{(2L)^{n}
 } \mathcal{F} [ \gamma_{S \cap [ - L , L]^{n} } ( \vec{x} ) ] $
\end{center}
where $ \mathcal{B} ( \mathbb{R}^{n} )  $ is the
Borel-$\sigma$-algebra of $  \mathbb{R}^{n} $ and where $
\mathcal{F} $ denotes the Fourier transform \cite{Reed-Simon-80},
\cite{Reed-Simon-75}.

By the Lebesgue Decomposition Theorem (i.e. the straigthforward
multidimensional generalization of the theorem \ref{th:Lebesgue
Decomposition Theorem}) we know that $ \mu_{S} $ may be decomposed
uniquely as:
\begin{equation}
    \mu_{S} \; = \; \mu_{S}^{(p.p.)} + \mu_{S}^{(a.c.)}+  \mu_{S}^{(sing)}
\end{equation}
where $ \mu_{S}^{(p.p.)} $ is a \emph{pure point measure}, $
\mu_{S}^{(a.c.)} $ is \emph{ absolutely continuous} w.r.t. the
Lebesgue measure $ \mu_{Lebesgue} $ while $  \mu_{S}^{(sing)} $ is
\emph{singular} w.r.t. $  \mu_{Lebesgue} $.

The mathematical characterization of crystals involves only
$\mu_{S}^{(p.p.)} $; introduced the following
\cite{Reed-Simon-80}:
\begin{definition}
\end{definition}
\emph{set of pure points of $ \mu_{S} $ }:
\begin{equation}
    PP( \mu_{S} ) \; := \; \{ \vec{x} \in \mathbb{R}^{n} \; : \;
    \mu_{S} ( \{ \vec{x} \}) \neq 0 \} \; = \; \{ \vec{x} \in \mathbb{R}^{n} \; : \;
     \mu_{S}^{(p.p.)} ( \{ \vec{x} \}) \neq 0 \}
\end{equation}
we can at last define:
\begin{definition} \label{def:crystal}
\end{definition}
\emph{S is a crystal:}
\begin{equation}
    |  PP( \mu_{S} ) | \; = \; \aleph_{0}
\end{equation}

\begin{remark}
\end{remark}
It is claimed in the book review \cite{Radin-96} of
\cite{Senechal-95a} that the definition \ref{def:spectrum of a
Delone set} is not mathematically rigorous.

It should be noticed, with this regard, that the author of such a
criticism hasn't proposed  \cite{Radin-99} an, according to him,
better definition.

Demanding to \cite{Hof-95} for further details let us observe that
since for every Delone set S:
\begin{equation}
    \frac{1}{(2L)^{n}
 } \mathcal{F} [ \gamma_{S \cap [ - L , L]^{n} } ( \vec{x} ) ]  \in
 \mathcal{S}' ( \mathbb{R}^{n} ) \; \; \forall L \in \mathbb{R}_{+}
\end{equation}
its limit for $ L \rightarrow \infty $ is perfectly well defined
with respect to the natural topology of the Schwartz space of
tempered distributions \cite{Reed-Simon-80}.

As to definition \ref{def:crystal} it is important to remind the
double nature of the Dirac delta as a tempered distribution and as
a Lebesgue-Stieltjes measure.

\smallskip

Let us recall that given $ k \in \mathbb{N}_{+} $ vectors $
\vec{a}_{1} , \cdots , \vec{a}_{k} \in \mathbb{R}^{n} $:
\begin{definition}
\end{definition}
\emph{$ \mathbb{Z} $-module generated by  $  \vec{a}_{1} , \cdots
, \vec{a}_{k} $:}
\begin{equation}
    \Gamma (  \vec{a}_{1} , \cdots
, \vec{a}_{k} )  \; := \; \{ \sum_{i=1}^{k}  m_{i} \vec{a}_{i} \;
\; m_{i}, \in \mathbb{Z} \; i= 1 , \cdots , k \}
\end{equation}

\begin{definition}
\end{definition}
\emph{lattice in $ \mathbb{E}^{n} $:}
\begin{center}
   a $ \mathbb{Z} $-module generated by a basis of $ \mathbb{E}^{n} $
\end{center}

Let us now consider the particular case in which S is a lattice $
\Gamma ( \vec{a}_{1}, \cdots , \vec{a}_{n} ) $ (where hence $ \{
\vec{a}_{1}, \cdots , \vec{a}_{n} \} $ is a basis of $
\mathbb{E}^{n} $) and let us  introduce the following:
\begin{definition}
\end{definition}
\emph{dual lattice of $  \Gamma ( \vec{a}_{1}, \cdots ,
\vec{a}_{n} ) $:}
\begin{equation*}
 \Gamma' ( \vec{a}_{1}, \cdots , \vec{a}_{n} ) \; := \; \Gamma (  \vec{a}_{1}' , \cdots ,
 \vec{a}_{n}')
\end{equation*}
where $ \{  \vec{a}_{1}' , \cdots ,
 \vec{a}_{n}' \} $ is called the basis dual to the basis  $ \{  \vec{a}_{1}, \cdots , \vec{a}_{n}
 \} $ and is defined by the condition:
\begin{equation}
   \vec{a}_{i} \cdot  \vec{a}_{j}' \; := \;  \delta_{i,j} \; \;
   i,j= 1 , \cdots , n
\end{equation}

A corner stone of Mathematical Crystallography (both classical and
quasi) is the following \cite{Senechal-95a}:
\begin{theorem}
\end{theorem}
\emph{Poisson's summation formula:}
\begin{equation}
    \mathcal{F}( \rho_{\Gamma ( \vec{a}_{1}, \cdots ,
\vec{a}_{n} )}) \; = \; \rho_{\Gamma' ( \vec{a}_{1}, \cdots ,
\vec{a}_{n} )}
\end{equation}
from which it follows that:
\begin{corollary}
\end{corollary}
\begin{center}
  $ \Gamma ( \vec{a}_{1}, \cdots , \vec{a}_{n} ) $ is a crystal
\end{center}

The problem of characterizing  which nonperiodic Delone sets  are
crystals is an extremely high and steep mountain that has been
challenged by some of the best minds in the scientific community
(see e.g. \cite{Bombieri-Taylor-86} and
\cite{Bombieri-Taylor-87}).

\smallskip

 Let us now introduce the following:
\begin{definition} \label{def:isometry of En}
\end{definition}
\emph{isometry in $ \mathbb{E}^{n} $:}

a map $ \phi :  \mathbb{E}^{n} \mapsto  \mathbb{E}^{n} $ such
that:
\begin{equation*}
  | \phi ( \vec{x} - \vec{y} ) | \; = \; | \vec{x} - \vec{y} |  \; \; \forall \vec{x} , \vec{y} \in \mathbb{E}^{n}
\end{equation*}

The isometries of $ \mathbb{E}^{n} $ form a group that we will
denote by $ Is( \mathbb{E}^{n} ) $.

\smallskip

Given a Delone set S:
\begin{definition}
\end{definition}
\emph{symmetry group of S:}
\begin{equation}
    Sim ( S ) \; := \; \{ \phi \in Is( \mathbb{E}^{n}  ) \; : \;
    \phi \circ \mu_{S} \; = \; \mu_{S} \}
\end{equation}

Given a group G and an element $ g \in G $:
\begin{definition}
\end{definition}
\emph{order of g:}
\begin{equation*}
    ord(g) \; := \; \min \{ n \in \mathbb{N}_{+} : g^{n} = 1 \}
\end{equation*}
\begin{definition} \label{def:Hiller's function}
\end{definition}
\emph{Hiller's function:}
\begin{equation*}
    Hil : \mathbb{N}_{+}  \mapsto  \mathbb{N} \; : \; Hil(k) \, :=
    \, \min \{ n \in \mathbb{N} :  g \in GL ( n , \mathbb{Z} )
    \wedge ord(g) = k \}
\end{equation*}
Introduced the well-known:
\begin{definition}
\end{definition}
\emph{Euler's $ \phi $ function:}
\begin{equation*}
    \phi : \mathbb{N}_{+} \mapsto \mathbb{N}_{+} \; : \; \phi (n)
    \, := \, | \{ k \in \mathbb{N}_{+} \, : \, k < n \, \wedge \,
    gcd( k , n) = 1 \} |
\end{equation*}
then \cite{Senechal-95b}:
\begin{theorem} \label{th:Hiller Theorem}
\end{theorem}
\emph{Hiller's Theorem:}
\begin{equation*}
    Hil ( \prod_{n=1}^{\infty} p_{n}^{\alpha_{n}} ) \; = \;
    \sum_{\{ n \in \mathbb{N}_{+} \, : \, p_{n}^{\alpha_{n}} \neq 2 \} } \phi ( p_{n}^{\alpha_{n}} )      \; \;
    \forall  \{ \alpha_{n} \in \mathbb{N} \}_{n \in \mathbb{N}} :
    \exists N \in \mathbb{N} :  \alpha_{n} = 0 \; \forall n > N
\end{equation*}
where $ p_{n} $ is the $ n^{th} $ prime number.

 The \emph{Hiller function} is algorithmically implemented in the
 section \ref{sec:Mathematica implementation of this paper}.

 It may be easily verified that:

 \smallskip

\begin{tabular}{|c|c|}
  \hline
  $ n  $ & $ Hil(n) $ \\
  \hline
  1 & 0 \\
  2 & 0 \\
  3 & 2 \\
  4 & 2 \\
  5 & 4 \\
  6 & 2 \\
  7 & 6 \\
  8 & 4 \\
  9 & 6 \\
  10 & 4 \\
  11 & 10 \\
  12 & 4 \\
  13 & 12 \\
  14 & 6 \\
  15 & 6 \\
  16 & 8 \\
  17 & 16 \\
  18 & 6 \\
  19 & 18 \\
  20 & 6 \\
  21 & 8 \\
  22 & 10 \\
  23 & 22 \\
  24 & 6 \\
  25 & 20 \\
  26 & 12 \\
  27 & 18 \\
  28 & 8 \\
  29 & 28 \\
  30 & 6 \\
  31 & 30 \\
  32 & 16 \\
  33 & 12 \\
  34 & 16 \\
  35 & 10 \\
  36 & 8 \\
  \hline
\end{tabular}

\smallskip

\begin{remark}
\end{remark}
Let us remark that the constraint $ p_{n}^{\alpha_{n}} \neq 2 $ in
theorem \ref{th:Hiller Theorem} is essential as it is proved in
\cite{Senechal-95b} where, contrary to \cite{Senechal-95a}, such a
theorem is reported correctly.

\smallskip

 It can be easily verified that:
\begin{corollary} \label{cor:crystallographic restriction in low dimension}
\end{corollary}
\begin{equation*}
    Hil(n) \leq 3 \; \Leftrightarrow \; n \in \{ 1,  2 , 3 , 4, 6 \}
\end{equation*}

\bigskip

Given a Delone set S in  $ \mathbb{E}^{n} $:
\begin{proposition}
\end{proposition}
\emph{Crystallographic restriction:}
\begin{equation*}
 S \text{ is a lattice} \; \Rightarrow \;   Hil ( ord ( g ) ) \; \leq \; n \; \; \forall g \in Sim(S)
\end{equation*}
\begin{proof}
 The thesis follows by the same definition \ref{def:Hiller's
 function} of Hiller's function.
\end{proof}

\begin{definition}
\end{definition}
\emph{S is a quasicrystal:}
\begin{equation*}
    S \text{ is a crystal } \; \wedge \; \exists g \in Sim (S) :
    Hil(ord(g)) > n
\end{equation*}

\begin{example}
\end{example}
By the corollary \ref{cor:crystallographic restriction in low
dimension}
 n-fold rotational symmetry axes, i.e. axes around which there is rotational symmetry for rotations by $ \frac{2 \pi}{n} $,
 can exist in two and three dimensions if and only if $ n \in \{
 2,3,4,6 \} $. Hence a bidimensional or three-dimensional
 crystal exhibiting an n-fold rotational symmetry axis with $ n \notin  \{
 2,3,4,6 \} $ is a quasicrystal.

\smallskip

\begin{example}
\end{example}
Passing from three-dimension to four-dimension the only symmetries
that become allowed are those of order $ 5 , 8 , 10 , 12 $.  Hence
a four-dimensional
 crystal exhibiting a symmetry  whose order  $ \notin \ \{
 2,3,4,5,6,8,10,12 \} $ is a quasicrystal.

\newpage
\section{Mathematica implementation of this paper} \label{sec:Mathematica implementation of this paper}
The notions introduced in this paper may be implemented
algorithmically through the following Mathematica 5
\cite{Wolfram-03}, \cite{Pemmaraju-Skiena-03} notebook:
\begin{verbatim}
Off[General::spell1];

$MaxExtraPrecision=\[Infinity];

<<DiscreteMath`Combinatorica`;

<<NumberTheory`AlgebraicNumberFields`

(***words[alphabet,
        n] is the list of all the words of length n in lexicographic ordering***)

words[alphabet_,n_]:=Strings[alphabet,n]

(***wordsupto[alphabet,
        n] is the list of all the words of length less or equal to n in \
lexicographic ordering***)

wordsupto[alphabet_,n_]:=
  If[n\[Equal]1,words[alphabet,1],
    Join[wordsupto[alphabet,n-1],words[alphabet,n]]]

(*** subwords[x] gives the list of all the sub-
    words of the word x in lexicographic ordering***)

subwords[x_]:=Flatten[Table[Take[x,{i,j}],{i,1,Length[x]},{j,i,Length[x]}],1]

(*** generalizedselect[l,predicate,
        y] picks out all elements e of the list l for which the binary \
predicate predicate[e,y] is True ***)

(*** WARNING:
      since the instruction is implemented throwing away elements of pattern \
String the list l must not to contain strings ***)

generalizedselect[l_,predicate_,y_]:=
  DeleteCases[
    Table[If[predicate[Part[l,i],y],Part[l,i],"throw away"],{i,1,Length[l]}],
    x_String]

havinggivenlengthQ[word_,n_]:=Equal[Length[word],n]

(*** language[word,n] is the language of length n of the word x
***)

language[word_,n_]:=generalizedselect[subwords[word],havinggivenlengthQ,n]

(*** topologicalentropy[alphabet,word,
        n] is the approximation at level n of the topological entropy of a \
sequence ***)

topologicalentropy[alphabet_,word_,n_]:=
  Divide[Log[Length[alphabet],Length[language[word,n]]],n]

occurences[x_,y_]:=If[AtomQ[x],Count[{x},y],Count[x,y]]

(*** substitution[rule,
        x] is the action of the substitution rule over the word x ***)

substitution[rule_,x_]:=Flatten[Map[rule,x]]

incidencematrix[alphabet_,rule_]:=
  Table[occurences[rule[Part[alphabet,j]],Part[alphabet,i]],{i,1,
      Length[alphabet]},{j,1,Length[alphabet]}]

(*** sequence[rule,x,n] is the n-
    the iterate of the action of the substitution rule over the word x ***)

sequence[rule_,x_,n_]:=
  If[n\[Equal]1,substitution[rule,x],substitution[rule,sequence[rule,x,n-1]]]

fibonacci[x_]:=If[x\[Equal]0,{0,1},0]

padovan[x_]:=If[x\[Equal]0,{1,2},If[x\[Equal]1,2,0]]

padovannumber[n_]:=If[n\[LessEqual]2,1,padovannumber[n-2]+padovannumber[n-3]]

procedure[rule_,x_,n_]:=Do[Print[sequence[rule,x,k]],{k,1,n}]

algebraicorder[x_]:=
  If[x\[Element]Algebraics,Exponent[MinimalPolynomial[x,y],y],\[Infinity]]

differentQ[x_,y_]:=Unequal[FullSimplify[x-y],0]

conjugates[x_]:=
  If[x\[Element]Algebraics,
    generalizedselect[
      Table[Root[MinimalPolynomial[x],k],{k,1,algebraicorder[x]}],differentQ,
      x],\[Infinity]]

absolutevaluelessthanoneQ[x_]:=Abs[x]<1

pvnumberQ[x_]:=
  And[AlgebraicIntegerQ[x],x>1,
    Length[Select[conjugates[x],absolutevaluelessthanoneQ]]\[Equal]
      Length[conjugates[x]]]

polynomial[x_,listofcofficients_]:=
  Sum[listofcoefficients[[i]]*x^i,{i,1,Length[listofcofficients]}]

\[Tau]=GoldenRatio;

\[Rho]=Root[(#^3-#-1)&,1];

pointofpolygon[k_, i_] := {Cos[(2*\[Pi]*i)\k], Sin[(2*\[Pi]*i)/k]}


lineofpolygon[k_,i_,j_]:=Line[{pointofpolygon[k,i],pointofpolygon[k,j]}]

polygonpentagram[k_]:=Flatten[Table[lineofpolygon[k,i,j]
,{i,1,k},{j,1,k}]]

polygonpicture[k_]:=Show[Graphics[polygonpentagram[k]]]

pointofpentagram[n_, i_] :=
    If[n \[Equal] 1, {Cos[(2*\[Pi]*i)/5],
        Sin[(2*\[Pi]*i)/5]}, - (GoldenRatio/(1 + 2*GoldenRatio))
        pointofpentagram[n - 1, i]]

lineofpentagram[n_,i_,j_]:=Line[{pointofpentagram[n,i],pointofpentagram[n,j]}]

pentagram[n_]:=Flatten[Table[lineofpentagram[n,i,j]
,{i,1,5},{j,1,5}]]

pentagrampicture[n_]:=Show[Graphics[Table[pentagram[i],{i,1,n}]]]

pvpoint[\[Lambda]_,n_]:=
  If[n\[Equal]1,{1,
      0},{Re[Exp[\[ImaginaryI]*Mod[ (2*\[Pi]) * (\[Lambda]^n),2\[Pi]]]],
      Im[Exp[\[ImaginaryI]*Mod[ (2*\[Pi])*(\[Lambda]^n),2\[Pi]]]]}]

pvside[\[Lambda]_,n_]:=Line[{pvpoint[\[Lambda],n],pvpoint[\[Lambda],n+1]}]

pvline[\[Lambda]_,i_,j_]:=Line[{pvpoint[\[Lambda],i],pvpoint[\[Lambda],j]}]

pvcurve[\[Lambda]_,n_]:=Table[pvside[\[Lambda],k],{k,1,n-1}]

pvcurvepicture[\[Lambda]_,n_]:=
  Show[Graphics[{Circle[{0,0},1],pvcurve[\[Lambda],n]}]]

pvpentagram[\[Lambda]_,n_]:=
  Flatten[Table[pvline[\[Lambda],i,j],{i,1,n},{j,1,n}]]

pvpentagrampicture[\[Lambda]_,n_]:=
  Show[Graphics[{Circle[{0,0},1],pvpentagram[\[Lambda],n]}]]

elementaryrotation[alphabet_,spacings_,letter_ ,initialpoint_]:=
  Mod[Part[spacings,Extract[Position[alphabet,letter],{1,1} ]]+initialpoint,
    2\[Pi]]

orbit[alphabet_,spacings_,word_ ,initialpoint_,n_]:=
  If[n>Length[word],"undefined",
    If[n\[Equal]1,
      elementaryrotation[alphabet,spacings,Part[word,1] ,initialpoint],
      elementaryrotation[alphabet,spacings,Part[word,n] ,
        orbit[alphabet,spacings,word ,initialpoint,n-1]]]]

pointofpisotsequence[alphabet_,spacings_,rule_,x_,
    n_]:={Re[Exp[\[ImaginaryI]*
          orbit[alphabet,spacings,sequence[rule,x,n] ,0,
            Length[sequence[rule,x,n]]]]],
    Im[Exp[\[ImaginaryI]*
          orbit[alphabet,spacings,sequence[rule,x,n] ,0,
            Length[sequence[rule,x,n]]]]]}

curveoffpisotsequence[alphabet_,spacings_,rule_,x_,n_]:=
  Flatten[Table[
      Line[{pointofpisotsequence[alphabet,spacings,rule,x,k],
          pointofpisotsequence[alphabet,spacings,rule,x,k+1]}] ,{k,1,n-1}]]

pisotsequencepicture[alphabet_,spacings_,rule_,x_,n_]:=
  Show[Graphics[{Circle[{0,0},1],
        curveoffpisotsequence[alphabet,spacings,rule,x,n]}]]

Hiller[n_]:=
  If[n==1,0,
    If[n==2,0,
      If[Length[FactorInteger[n]]==1,
        EulerPhi[FactorInteger[n][[1]][[1]]^FactorInteger[n][[1]][[2]]],
        Sum[Hiller[
            FactorInteger[n][[i]][[1]]^FactorInteger[n][[i]][[2]]],{i,1,
            Length[FactorInteger[n]]}]]]]

\end{verbatim}

\end{document}